%% file: 4197.tex
 \documentclass{aa}
  \usepackage{graphicx}
  \usepackage{natbib}
  \bibpunct{(}{)}{;}{a}{}{,} 
  \newcommand{\lsim}{\ \raise -2.truept\hbox{\rlap{\hbox{$\sim$}}\raise 5.truept\hbox{$<$}\ }}
  \newcommand{\gsim}{\ \raise -2.truept\hbox{\rlap{\hbox{$\sim$}}\raise 5.truept\hbox{$>$}\ }}
  \newcommand{\me}{M_{\mathrm e}}

\begin{document}

  \title{Monte Carlo Simulations of Metal-Poor Star Clusters\thanks{All
  the tables (i.e. for models with $M_V^{tot}=-4, -6, -8$)
  are available in electronic form at the CDS via anonymous
  ftp to cdsarc.u-strasbg.fr (130.79.128.5) or via
  http://cdsweb.u-strasbg.fr/cgi-bin/qcat?J/A+A/}$^,$\thanks{All the
  figures are available as colored figures in the electronic edition of the Journal.}}

   \author{
   Martina Fagiolini \inst{1,2}
    \and
   Gabriella Raimondo \inst{3}
    \and
   Scilla Degl'Innocenti \inst{1,2}
            }

   \offprints{Gabriella Raimondo, \email{raimondo@oa-teramo.inaf.it}}

   \institute{
     Dipartimento di Fisica, Universit\`a di Pisa,
     Largo B. Pontecorvo 3, I-56126 Pisa, Italy\\
     \and
     INFN, Sezione di Pisa,
         Largo B. Pontecorvo 3, I-56126 Pisa, Italy\\
     \and
     INAF, Osservatorio Astronomico di Teramo,
         Via M. Maggini, I-64100, Teramo, Italy}

     \date{Received date / Accepted date}
     \authorrunning{Fagiolini et al.}
     \titlerunning{Monte Carlo Simulations of Metal Poor Star Clusters}

\abstract
{Metal-poor globular clusters (GCs) can provide a probe of the
earliest epoch of star formation in the Universe, being the oldest
stellar systems observable. In addition, young and
intermediate-age low-metallicity GCs are present in external
galaxies. Nevertheless, inferring their evolutionary status by
using integrated properties may suffer from large \emph{intrinsic}
uncertainty caused by the discrete nature of stars in stellar
systems, especially in the case of faint objects. }
{In this paper, we evaluate the \emph{intrinsic} uncertainty (due
to statistical effects) affecting the integrated colours and
mass--to--light ratios as a function of the cluster integrated
visual magnitude ($M_V^{tot}$), which represents a quantity
directly measured. We investigate the case of metal-poor
single-burst stellar populations with age from a few million years
to a likely upper value for the Galactic globular cluster ages
($\sim$15 Gyr).}
{Our approach is based on Monte Carlo techniques for randomly
generating stars distributed according to the cluster's mass
function. }
{Integrated colours and mass--to--light ratios in different
photometric bands are checked to be in good agreement with the
observational values of low-metallicity Galactic clusters; the
effect of different assumptions on the Horizontal Branch (HB)
morphology is shown to be not relevant, at least for the
photometric bands explored here. We present integrated colours and
mass--to--light ratios as a function of age for different
assumptions on the cluster total $V$ magnitude. We find that the
\emph{intrinsic} uncertainty cannot be neglected. In particular,
in models with $M_V^{tot}=-4$ the broad-band colours show an
\emph{intrinsic} uncertainty so high as to prevent precise age
evaluation of the cluster. The effects of different assumptions on
the initial mass function and on the minimum mass for which carbon
burning is ignited on both integrated colours and mass--to--light
ratios are also analyzed. Finally, the present predictions are
compared with recent results available in the literature, showing
in some cases non-negligible differences.}
{}


\maketitle
%

  \section{Introduction}

  A key issue in astronomy is to determine ages and chemical
  compositions of those stars and stellar systems that are needed to
  reconstruct the formation and evolution of galaxies. Toward this
  goal, the analysis of the stellar clusters population in external
  galaxies and in the Milky Way is fundamental for tracing the history
  of the parent galaxy \citep[see, e.g.][]{Cote+98,West+04}.
  Contrarily to the situation in the Galaxy, in which massive
  ($\approx 10^5 \div 10^6$ stars) stellar clusters are old ($\gsim$10
  Gyr) and metal poor ($[Fe/H]\lsim -$0.7), observations indicate that
  in the Magellanic Clouds, in the Local Group galaxies, and even in
  galaxies beyond the Local Group, massive stellar clusters spread a large
  range of both metallicity and age; thus they are considered the main
  indicators of stellar formation events in the galaxies history
  \citep{Larsen00, Matteucci+02, Harris03}.  Hereinafter we will use
  the term ``globular clusters'' (GCs) to signify massive clusters of
  whatever age and chemical composition.

  GCs have the advantage to be bright objects, and then to be easily
  observed beyond the Local Group. In addition, they are thought to be
  simple stellar populations (SSPs), consisting of a gravitationally
  bound group of stars born at nearly the same moment, and with a
  nearly identical chemical composition.  The integrated broad-band
  colours, line indices, and mass-to-light ratios, we observe from
  those systems, are the unique observational tools to understand
  their properties. Since the pioneering work by Tinsley (1972),
  different groups have developed population synthesis models,
  e.g. \citep[][and references therein]{Brocato+90b,
  Charlot&Bruzual91, Buzzoni93, Bressan+94, Worthey94, Maraston98,
  Kurth+99, Brocato+00, Vazdekis99, Girardi+00, Anders+03,
  Bruzual&Charlot03, Maraston05} in order to interpret such
  observables.

  Besides the \emph{systematic} uncertainties due to different sets of
  stellar evolutionary tracks and different spectral libraries--used
  to transform the models from luminosity and effective temperature to
  observable quantities-- \citep[see e.g.][]{Charlot+96, Maraston98,
  Brocato+00, Bruzual&Charlot03, Yi03}, broad-band colours may suffer
  from large \emph{intrinsic} fluctuations caused by the discrete
  nature of the number of stars in the system.  The first studies were
  carried out in the optical by \citet{Barbaro&Bertelli77} - for
  population I clusters - and \citet{Chiosi+88}- for intermediate
  ($Z$=0.001) and solar metallicity -, while \citet{Santos&Frogel97}
  analyzed the case for near infrared (NIR) bands. Among other
  results, the quoted authors concluded that it is necessary to
  include stochastic effects when deriving ages and metallicities from
  integrated broad--band colours.

  Most studies normalized their theoretical predictions to the total
  number of stars or total mass in the cluster, while
  \citet{Brocato+99,Brocato+00} derived the mean broad-band colours
  and the corresponding dispersions as a function of the cluster
  visual magnitude ($M_V^{tot}$) for selected values of ages. This
  approach directly links theory and observations and it is crucial
  when the cluster age and metallicity are inferred from the observed
  broad--band colours, especially for clusters at the faint end of the
  GC luminosity function.

  In this paper we extend the investigations by \citet{Brocato+99} and
  \citet{Brocato+00} and we analyze stochastic effects not only on
  broad--band colours, but also on mass--to--light ratios as a
  function of the adopted cluster visual magnitude, $M_V^{tot}$.  We
  provide a comprehensive study on this problem by investigating the
  time--evolution of both integrated colours and mass--to--light
  ratios for a fine grid of stellar ages and for three different
  values for $M_V^{tot}$. The analysis is carried out using new
  single-burst low-metallicity models ($Z$=0.0002) based on the
  updated stellar models database by \citet{Cariulo+04}.  We choose
  this metallicity because the metal--poor GCs can provide a probe of
  the earliest epoch of star formation in the Universe, being the
  oldest stellar systems observable. In addition, young and
  intermediate age low metallicity GCs are present in external
  galaxies (Larsen \& Richtler 1999, 2000).  For these reasons, and
  also because an analysis of young metal--poor clusters gives an idea
  on how old GCs appeared when they formed, we explore a wide range of
  ages (7.7 $\leq \log [age (yr)] \leq 10.3$). We note that we did not
  take into account redshift effects thus our calculations can be used
  only for objects with a redshift lower than about 0.1.

  The results are compared with a sample of low-metallicity clusters
  in the Galaxy with different HB morphology, chosen as prototypes of
  the old stellar populations studied in this work. The present
  theoretical predictions are also compared with recent results
  available in the literature showing in some cases non-negligible
  differences.

  We also discuss the influence of the adopted Initial Mass Function
  (IMF) on integrated colours and mass--to--light ratios and the
  effect of changing the maximum mass ($M_{up}$) for which carbon
  burning is not ignited, due to effects of the degenerate pressure
  and neutrino energy losses in the core.  The assumption of a fixed
  $M_V^{tot}$ can lead to a peculiar behaviour when varying the shape
  of the IMF, since an adjustment of the total number of stars might
  be required to keep the $M_V^{tot}$ value fixed.

  The layout of the paper is the following. In
  Section~\ref{section:thecode} the ingredients of the stellar
  population synthesis code are outlined, together with a brief
  description of the method adopted to derive the integrated
  quantities. In Section~\ref{section:GGC} we show the comparison of
  the theoretical results with selected observations of galactic
  globular clusters.  Then, we discuss the uncertainties affecting
  integrated colours (Section~\ref{section:colours}) and
  mass--to--light ratios (Section~\ref{section:ML}), together with a
  comparison with previous works.

  \section{Description of the code}
  \label{section:thecode}

  Synthetic CMDs and magnitudes presented in this paper are based on
  the stellar population synthesis code developed by
  \citet{Brocato+99,Brocato+00}, and \citet{Raimondo+05}\footnote{
  http://www.oa-teramo.inaf.it/SPoT}. In this section, we briefly
  describe the main ingredients and recall the method used to derive
  integrated magnitudes and colours, referring to the cited papers for
  more details.

  \subsection{Ingredients}

  The present SSP models rely on the evolutionary tracks of the ``Pisa
  Evolutionary library''\footnote{
  http://astro.df.unipi.it/SAA/PEL/Z0.html. Data files are also
  available at the CDS.} for masses $M\geq 0.6M_{\odot}$
  \citep{Cariulo+04}.  The input physics adopted in the models has
  already been discussed in \citet{Cariulo+04}. We only point out here
  that the models take into account atomic diffusion, including the
  effects of gravitational settling, and thermal diffusion with
  diffusion coefficients given by \citet{Thoul+94}; radiative
  acceleration \citep[see e.g.][]{Richer+98,Richard+02} is not
  included. The effects of rotation \citep[see
  e.g.][]{Maeder&Zahn98,Palacios+03} are also not included.

  Convective regions, identified following the Schwarzschild criterion, are
  treated with the mixing length formalism. We use throughout the canonical
  assumption of inefficient overshooting, so the He burning structures are
  calculated according to the prescriptions of canonical semiconvection
  induced by the penetration of convective elements in the radiative region
  \citep{Castellani+85}. The efficiency and presence of a mild
  overshooting are still open questions \citep[][]{Barmina+02,
  Brocato+03}; however, as discussed in \citet{Yi03}, a modest amount of
  overshooting (i.e.  $H_P\sim 0.2$, see also Brocato et al. 2003) influences
  only integrated colours with ages $\lsim 1.5$ Gyr for a maximum amount of
  $\approx$0.1 mag.  The models span the
  evolutionary phases from the main sequence up to C ignition or the onset of
  thermal pulses (TP) in the advanced Asymptotic Giant Branch (AGB) in the mass
  range $0.6 \div 11\, M_{\odot}$. This allows us to calculate stellar
  population models in the age range $\approx 50$ Myr $\div$ 20 Gyr.

  Beyond the early--AGB phase, thermally pulsating (TP) stars have
  been simulated in the synthesis code using the analytic formulas of
  \citet{Wagenhuber&Groenewegen98} which describes time evolution of
  the core mass, and luminosity of TP stars.  These formulae include
  three important effects: ($i$) the fact that the first pulses do not
  reach the full amplitude, ($ii$) the hot bottom burning process that
  occurs in massive stars, and ($iii$) the third dredge--up.  The
  effective temperature ($T_e$) of each TP--AGB star has been
  evaluated using prescriptions by \citet{Renzini&Voli81}, considering
  the appropriate slope $d\log (L/L_{\sun})/d\log (T_e)$ of the
  adopted evolutionary tracks. The analytic procedure ends up
  providing the time evolution of the temperature and luminosity for a
  given mass \citep[see for details][]{Raimondo+05}.

  Mass loss affecting Red Giant Branch (RGB) stars and early--AGB
  stars has been taken into account following prescriptions by
  \citet{Reimers75}:

  \begin{equation}
  \dot M_{R} = - 4\cdot 10^{-13}\eta_R \cdot LR/M,\label{eq:reimers}
  \end{equation}

  \noindent while during the TP phase we adopted the \citet{Baud&Habing83}
  mass--loss rate:

  \begin{equation}
  \dot M_{BH}= \mu LR/\me \label{eq:bh}.
  \end{equation}
  Here, $L, R, M, \me$ are, respectively, the star luminosity, radius,
  total mass, and envelope mass in solar units; $\mu = -4\cdot
  10^{-13}(M_{e,0}/M)$, being $ M_{e,0}$ the envelope mass at the
  first TP. Eq.~\ref{eq:bh} is a modification of the Reimers formula
  with $\eta_R =1$ which also includes a dependence on the actual mass
  envelope. The initial-final mass relation is applied by
  \citet{Dominguez+99}, and no white dwarfs (WD) more massive than
  $1.1\, M_{\sun}$ are accepted \citep{PradaMoroni05}.

  In this paper the TP phase is included but we do not adopt any
  separation between C-rich and O-rich TP-stars. All stars are
  oxygen-rich when they enter the AGB phase.  Whether or not they
  become C-stars depends primarily on the efficiency of the third
  dredge-up (TDU) occurring on the TP-AGB phase, and the extent and
  time-variation of the mass-loss \citep[e.g.\ ][]{Marigo+99,
  Straniero+03}. In low-metallicity stars ($Z < 0.004$), the amount of
  oxygen in the envelope is so low that a few thermal pulses are
  sufficient to convert an O-rich star into a C-star
  \citep[][]{Renzini&Voli81}.  In addition, the smaller the
  metallicity the smaller the minimum mass for the onset of TDU
  \citep[e.g.][]{Straniero+03}.  On the other hand, TP-AGB stars may
  experience episode of strong mass loss, that in the case of low mass
  stars may cause a reduction of the envelope mass that may delay or
  even prevent the TDU occurrence and the formation of C-rich stars
  \citep{Marigo+99}.  In conclusion, the presence of C-rich stars may
  affect NIR-bands luminosity and its uncertainty in the case of
  low-metallicity, intermediate-age massive clusters \citep[see
  e.g.][]{Maraston98}, while at the typical age of Galactic globular
  clusters their presence become more uncertain, as confirmed by the
  fact that AGB stars in GGC are all observed to be oxygen-rich, so
  that carbon does not appear to have been dredge-up into the envelops
  during thermal pulses \citep{Lattanzio&Wood03}.  In addition, our
  assumption is also expected to have marginal effect on integrated
  quantities of faint populations, as bright TP-AGB stars are
  statistically less frequent, or even absent.  Finally, it will be
  shown that the details and the treatment of the physical processes
  at work on the TP-AGB phase, as well as their impact on synthetic
  colours, is still uncertain (Section \ref{section:colours}).

  The adopted colour transformations for the standard $UBVRIJHK$ bands
  are from \citet{Castelli99}, see also \citet{Castelli+97}. To
  calculate integrated colours in the Hubble Space Telescope (HST)
  bands (WFPC2 and NICMOS systems)
  we adopted the colour transformations by \citet{Origlia&Leitherer00}
  based on the \citet{Bessel+98} stellar atmospheric
  models.

  As well known the mixing length parameter ($\alpha$) governs the
  efficiency of convection in the convective envelope of a stellar
  structure. The $\alpha$ parameter used for calculating the
  evolutionary tracks adopted in the present work has been calibrated in
  such a way that the isochrones reproduce, with the adopted colour
  transformations, the observed RG branch colour of GCs with the proper
  metallicity and age. This is evident from Fig.~\ref{fig:GGC} in which
  our synthetic models nicely reproduce the RGB colours of the selected
  GCs. The $\alpha$ parameter needed in the present evolutionary models
  to obtain this agreement is $\alpha\approx$2.0, but it's worth
  noticing that $\alpha$ is a free parameter, sensitively dependent on
  the chosen color transformations and on all the adopted physical
  inputs which affects the effective temperature of a
  model. \citet{Cariulo+04} showed that tracks with the same physical
  assumptions for metallicity up to Z$\approx$0.001 reproduces the RGB
  colours of galactic GCs for the same $\alpha$ values, while metal-rich
  (Z$>$0.001) and standard solar models require an $\alpha$ slightly
  lower \citep[see][]{Ciacio+97, Castellani+03}.

  The very low mass tracks (VLM, $M\leq 0.6\, M _{\odot}$) are taken
  from \citet{Baraffe+97}. The tracks have been already transformed by
  the authors to the observational plane in the Johnson-Cousins system
  adopting the colour transformations by \citet{Allard+97},
  particularly suitable for low mass stars. We checked that low mass
  models satisfactory match, in all the available colours, the higher
  mass models (Fagiolini 2004).  However, as already discussed by
  e.g. \citet{Brocato+00}, VLM stars do not contribute to the
  photometric indices, although their contribution to the cluster mass
  is fundamental. Masses lower than the minimum mass for the central H
  ignition $\approx 0.08 \, M_{\odot}$, \citep[see e.g.][]{Baraffe+97}
  have been estimated do not significantly contribute to the total
  mass of the cluster \citep[see e.g.][]{Chabrier&Mera97}, and thus
  they are not taken into account.

  Post--AGB evolution, until the entrance in the white dwarfs (WD)
  cooling sequence, are not considered because the evolutionary time
  is too short to have a significant influence \citep[see
  e.g][]{Blocker&Schon97}.  WDs have been included in the code using
  evolutionary models by \citet{Salaris+00}. They have been
  transformed by adopting the atmospheric models by
  \citet{Saumon&Jacobson99} for DA WDs, which include the treatment of
  the collision-induced absorption of $H_2$ molecules, for $T<4000\,
  K$. For higher temperatures the transformations of
  \citet{Bergeron+95} have been used.  As we will discuss in the
  following our {\em standard} model adopts $6.5\, M_{\odot}$ as
  $M_{up}$, \citep{Dominguez+99}. As expected, and as already noted by
  e.g. \citet{Angeletti+80}, the contribution of the WD population to
  optical and NIR photometric indices is negligible although they
  significantly contribute to the total mass of the cluster.

  The IMF of \citet{Kroupa02} is adopted in the mass interval $0.1
  \leq M/M_{\odot} \leq 11 $, unless explicitly stated otherwise. To
  simulate the mass distribution of stars in the synthetic CMD we use
  a Monte Carlo method: the position of each randomly created star in
  the $\log L/L_{\odot}$ vs. $\log T_e$ diagram is determined for each
  given age. As already discussed, the chemical composition is fixed
  to $Z=0.0002$ and $Y=0.23$.

\begin{figure}[t]
  \center
  \includegraphics[width=8cm]{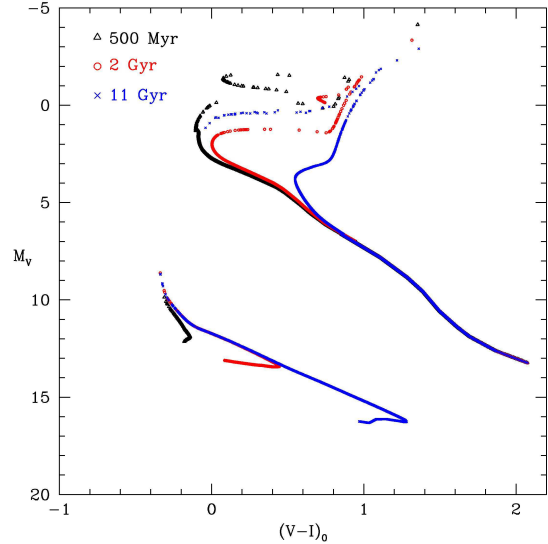}
   \caption{Synthetic CMDs for $Z=0.0002$, $Y=0.23$ and three different ages:
   500 Myr (black), 2 Gyr (red), and 11 Gyr (blue). For each population
   $M_V^{tot}=-6$ mag is adopted. The simulation of photometric errors is not
   included.}
  \label{fig:500211}
  \end{figure}

\begin{figure}[t]
  \center
  \includegraphics[width=6cm]{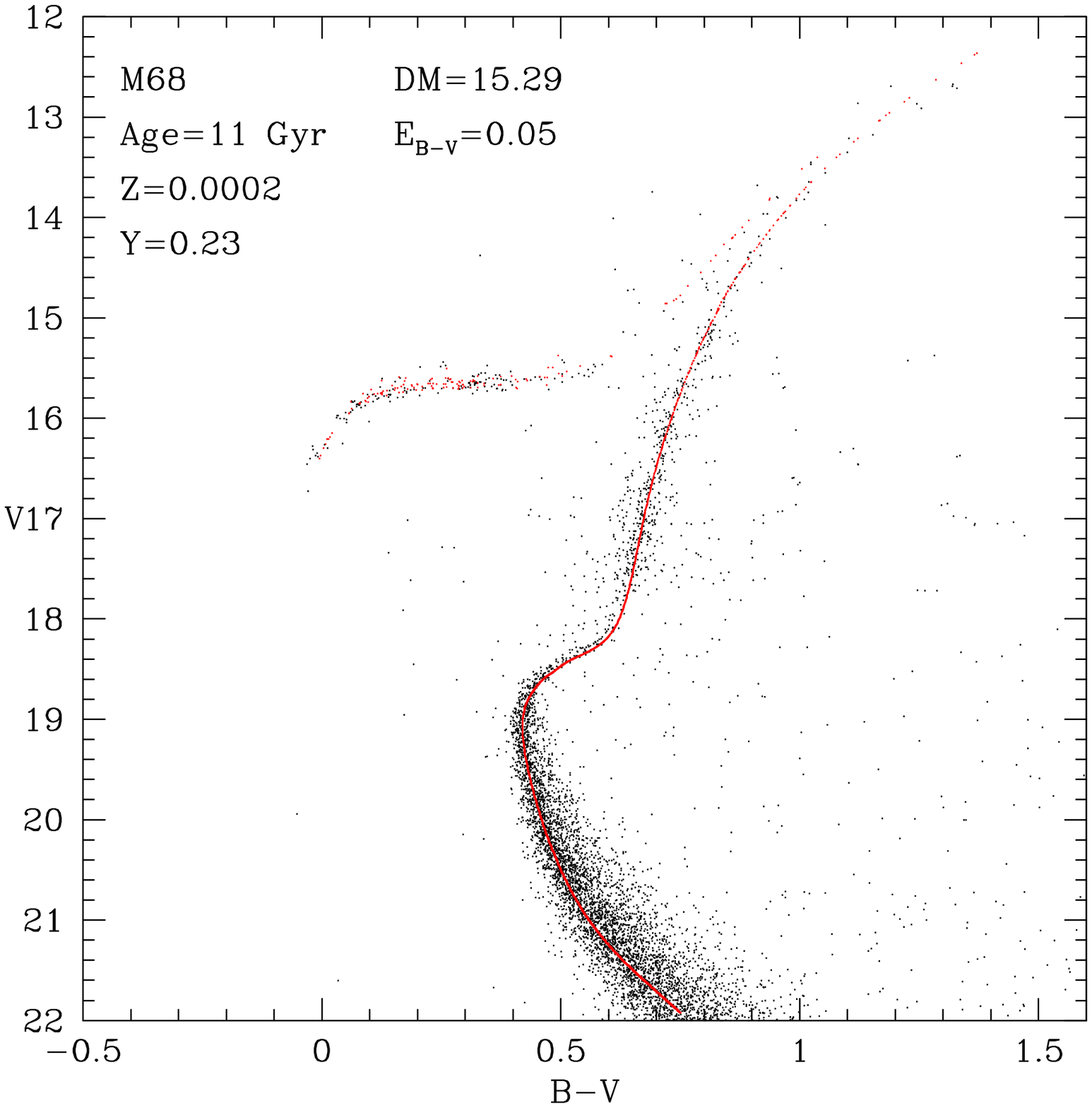}
  \includegraphics[width=6cm]{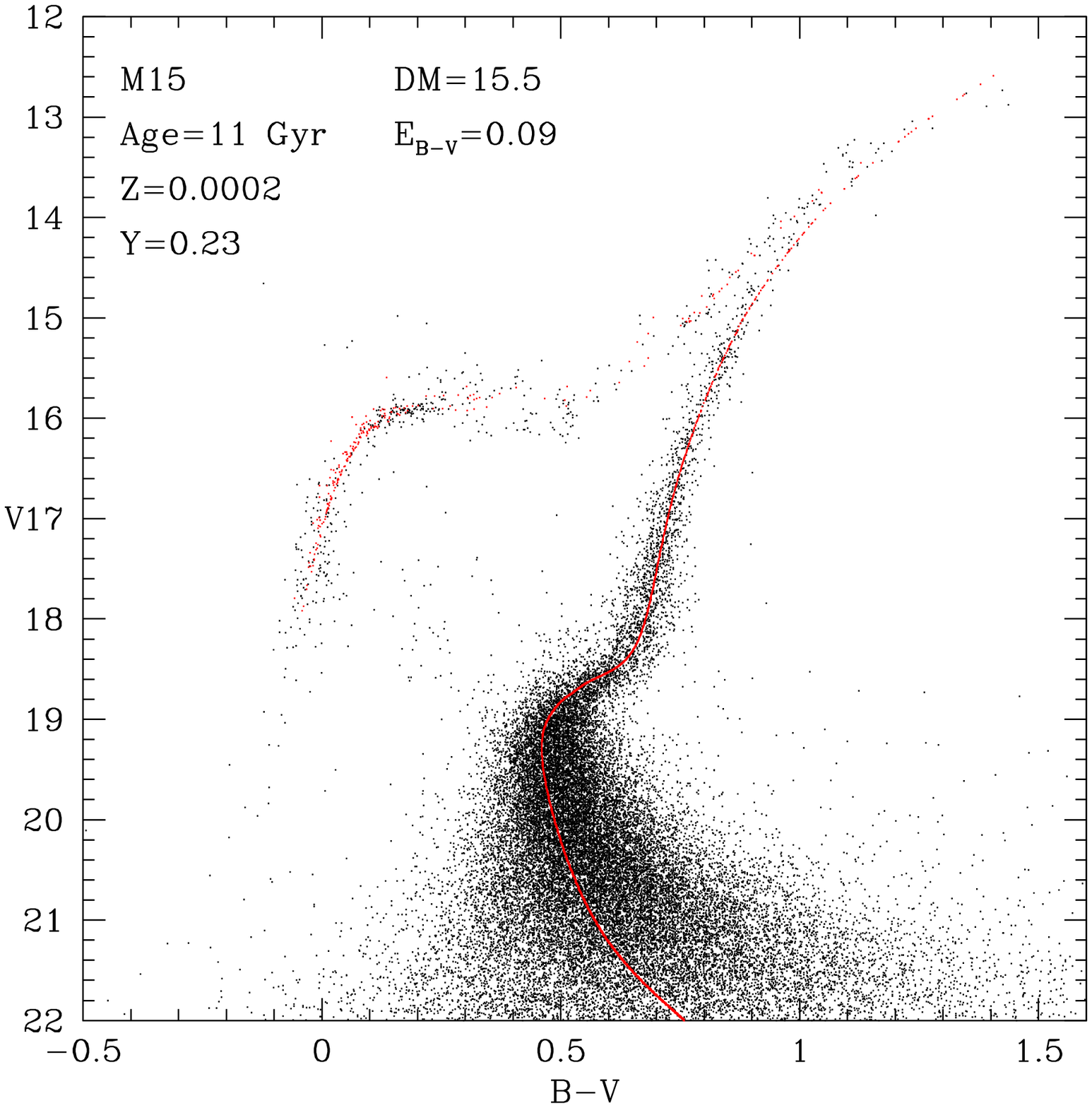}
  \includegraphics[width=6cm]{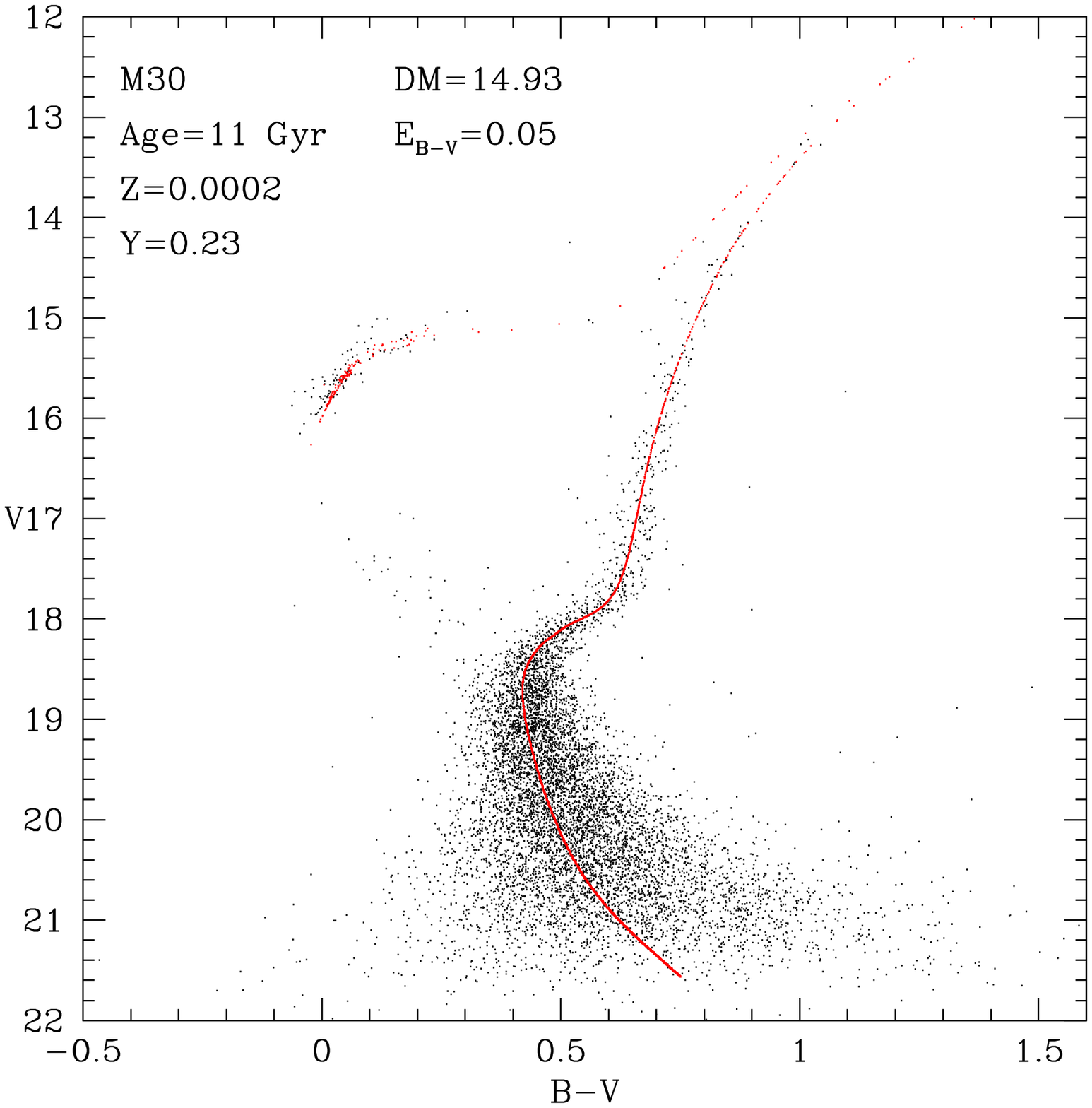}
  \caption{Comparison between present synthetic (red dots), without the
     inclusion of simulated photometric errors, and the
     observed CMDs of the globular clusters \emph{M68}, \emph{M15} and
     \emph{M30} (black dots). The assumed age is 11 Gyr, and the
     adopted chemical composition is $Z=0.0002$ and $Y=0.23$ (see
     text). The values of the distance modulus and the reddening
     obtained from the analysis are also indicated.}
  \label{fig:GGC}
  \end{figure}

  Fig.~\ref{fig:500211} shows, as an example, synthetic CMDs (without
  simulation of photometric errors) for the selected chemical
  composition, a total absolute visual magnitude $M_V^{tot}=-6$ mag,
  and three different ages (500 Myr, 2 Gyr, and 11 Gyr). All the
  evolutionary phases described above are clearly visible.

 \subsection{Integrated magnitudes and colours}

 To compute integrated fluxes and magnitudes we assume that the
 integrated light from the stellar population is dominated by light
 emitted by its stellar component. This implies that i) no source of
 non-thermal emission are at work, ii) the thermal emission by
 interstellar gas gives no sizeable contribution to the integrated
 flux, and iii) the absorption from dust and gas are negligible. On
 this basis, the integrated flux in each photometric filter mainly
 depend on two quantities. The first is the flux $f_{i}$ emitted by the
 $i$-th star of mass $M$, age $t$ (stellar system age), luminosity $l$,
 effective temperature $T_{eff}$, and chemical composition ($Y$, $Z$):

 \begin{equation}
 f_{i}[l(M,t,Y,Z),T_{eff}(M,t,Y,Z),Y,Z].
 \end{equation}
 $f_{i}$ is defined by the stellar evolutionary-tracks library, and
 by the adopted temperature-colours transformation tables.

 The second quantity is $\Phi(M,N)$ which describes the number of
 stars with mass $M$ in a population globally constituted of $N$
 stars with age $t$ and chemical composition ($Y$, $Z$). It is
 strictly related to the IMF.

 A fundamental point of our code is that the mass of each generated
 star is obtained by using Monte Carlo techniques, while the mass
 distribution is ruled by the IMF. The mass of each star is
 generated randomly, and its proper evolutionary line is computed
 by interpolating the available tracks in the mass grid. This
 method is crucial for poorly populated stellar systems, as we
 shall study in the following. It ensures to treat undersampled
 evolutionary phases properly; for instance NIR colours may be
 dominated by a handful of red giant stars
 \citep{Santos&Frogel97,Brocato+99,Cervino+03,Raimondo+05}.

 In each model, stars are added until to reach a given value of the
 absolute visual magnitude, $M_V$, at any age. The mass values are
 generated randomly and distributed according to the chosen IMF. It
 is relevant to note that the random extraction of masses is fully
 independent from model to model even if the same input quantities
 ($M_V$, $t$, $N$, $Y$, $Z$, IMF, etc.) are assumed.

 Both the previous quantities ($f_{i}$ and $\Phi(M,N)$) are
 combined and integrated by the stellar population code to derive
 the total integrated flux $F$ in a given photometric band:

 \begin{equation}
 F[N,t,Y,Z] = \sum _{i=1}^N f_{i},
 \end{equation}
 and the corresponding magnitudes.

  \input{4197.tab1.tex}

  \section{Comparison with Galactic Globular Clusters}
  \label{section:GGC}

  Before studying the general behaviour of colours and mass-to-light
  ratios as a function of $M_V$, the natural test for SSP models is to
  compare their predictions to the observed properties of Galactic
  star clusters.  Three old galactic globular clusters (GGCs) have
  been selected for checking our capability to reproduce both the CMD
  morphology of each cluster and the integrated magnitudes: NGC4590
  (M68), NGC7078 (M15) and NGC7099 (M30).  They are all metal--poor,
  span a relativity wide range of the visual magnitude
  (Table~\ref{table:GGC}), and show different Horizontal Branch
  morphologies. This allows us to check the effect of HB morphologies
  on the integrated colours in the selected bands. The $[Fe/H]$ values
  estimated for the three clusters are, respectively,
  $[Fe/H]=  -2.06,-2.26,-2.12$
  \citep{Harris96}\footnote{www.physics.mcmaster.ca/resources/globular.html.
  February 2003 version.}, while the $\alpha$-elements enhancement can
  be evaluated as $[\alpha/Fe] \approx 0.3$ \citep[see, e.g,
  the discussion in][]{Ferraro+99}; thus we can assume $Z \approx
  2\cdot 10^{-4}$.  We include M15 due to its brightness ($M_V\sim
  -9$), even if it is recognized as a cluster which should have
  undergone a gravothermal catastrophe. This results in a contraction
  of the cluster core while the external regions expand, with some
  stars escape the system. As a possible consequence of such a
  dynamical evolution, observational evidence of mass segregation, and
  of radial color gradients have been found \citep{Bailyn+89,
  DeMarchi+94, Stetson94}.

  Fig.~\ref{fig:GGC} shows the observed CMDs for the selected
  clusters.  $B$ and $V$ photometry comes from the HST--Snapshot
  Catalog by \citet{Piotto+02}, except for M68 whose data are from
  \citet{Walker94}.  The synthetic CMD which better reproduces the
  observational data is over-plotted as black dots in each panel. For
  each cluster an age of $11\,Gyr$ is assumed; we are not interested
  in a detailed calibration of the cluster ages which is far from our
  purpose. Results for each comparison between synthetic and observed
  cluster are, briefly:

  \vspace{.15cm}

  -- \emph{M68} (NGC4590): We find $(m-M)_V = 15.29$ and $E_{B-V} =
  0.05$ in agreement, within the uncertainties, with the current
  distance modulus and reddening determinations
  \citep[e.g.][]{Harris96, Carretta+00,Dicriscienzo+04}.  The HB
  morphology is reproduced using $\eta_{R} = 0.30$ with a dispersion
  $\sigma_{\eta} =0.08$. Due to the present uncertainties in the
  factors that influence the HB morphology and in the precise
  treatment of mass loss in the RG branch \citep[see e.g.][]{Lee+94,
  Rey+01}, the tuning of the adopted value of $\eta_{R}$ is just a way
  to obtain the observed HB morphology but it should not be
  interpreted in terms of physical parameters of the stellar cluster.

  \vspace{.15cm}

  -- \emph{M15} (NGC7078): The best result is obtained by assuming
  $(m-M)_V = 15.50$ and $E_{B-V} = 0.09$, in agreement with current
  determinations on the cluster distance modulus and reddening
  \citep{Harris96, Dicriscienzo+04, McNamara+04}.  The HB morphology
  is reproduced using $\eta_{R} = 0.44$ and $\sigma_{\eta} = 0.1$. The
  cluster contains one of the few planetary nebulae (PN) known in
  GGCs: the star K648 identified as a PN by \citet{Pease28}, for which
  Alves et al. (2000) measured an apparent magnitude of $V=14.73$.

  \vspace{.15cm}

  -- \emph{M30} (NGC7099): Assuming $E_{B-V} =0.05$ we obtain
  $(m-M)_V=14.93$ in agreement with \citet{Sandquist+99}.  The HB
  morphology is reproduced using the same parameters as in
  \citet{Brocato+00}, i.e. $\eta_{R} = 0.40$, and $\sigma_{\eta} =
  0.2$.

  Table~\ref{table:GGC} gives the observed integrated colours for the
  selected clusters, taken from the \emph{Catalog of Parameters for
  Milky Way Globular Clusters}
  \citep{Harris96}\footnote{www.physics.mcmaster.ca/resources/globular.html. February
  2003 version. The integrated colours $U-B$ and $B-V$ are on the
  standard Johnson system, and $V-R$, $V-I$ on the Kron-Cousins
  system. The values are the simply average of results from Peterson
  (1993) and Reed (1985).}  and synthetic integrated colours obtained
  from the present models by reproducing, within the errors, the
  observed total visual magnitude of each cluster. Optical
  observational data are de-reddened according to \citet{Reed+88} with
  reddening values taken from Harris (1996). To give an idea of
  uncertainties for the colours we report the 'residual' values
  computed by \citet{Reed85} which result from his homogenization
  procedure based on measurements of clusters colours by various
  authors. In some cases the availability of only one measurement
  prevents the determination of an uncertainty. NIR colours are from
  \citet{Brocato+90}, de-reddened according to \citet{Cardelli+89}
  with reddening values taken from Harris (1996).

  Theoretical errors correspond to $1\, \sigma$ dispersion evaluated
  from 10 independent simulations. Table~\ref{table:GGC} lists:
  cluster NGC number (Col. 1), total absolute $V$ magnitude
  ($M_V^{tot}$, Col. 2), integrated $U-B$, $B-V$, $V-R$, $V-I$, $V-J$
  and $V-K$ colours (Cols. 3--7), and ${\cal M}/L_V$ (Col. 8) from
  \citet{Pryor&Meylan93}. The last values are strongly
  model--dependent, as stressed by the authors themselves.
  Hereinafter we will indicate dereddened colours as, e.g., $U-B$
  instead of $(U-B)_0$.

  For each cluster, numerical simulations were performed by populating
  the synthetic CMD until the observed $M_V^{tot}$ of the cluster is
  reproduced (Sect. 2.1). Synthetic colours agree with data of all the
  clusters to within the uncertainties and theoretical statistical
  fluctuations.  The only exception is the $V-I$ colour of M15 that is
  found to be redder than observations.  The observed $V-I$ colour of
  this cluster is significantly bluer than that of other clusters with
  similar metallicity \citep{Harris96}.  On this regard, we recall
  that the $V-I$ measure available is based on one photoelectric
  measurement only \citep{Kron+60}, and it is not possible to estimate
  the uncertainty (Reed 1985).  To investigate this issue, we use $VI$
  stellar photometry by \citet{Rosenberg+00}, which contains more than
  90\% of the cluster light ($M_V\simeq -9.1$). By adding the flux of
  individual stars, we derive an integrated colour $V-I \simeq 0.85$,
  in fair agreement with our prediction.  This value is larger than
  the integrated value by \citet{Kron+60}, even if it would be
  affected by incompleteness at the faint end of the observed
  luminosity function.  To deeply investigate the origin of the latter
  inconsistency is well beyond the purpose of this paper, what we wish
  to point out here is to notice that the observed integrated $V-I$
  colour of such a cluster, reported in Table~\ref{table:GGC}, could
  be peculiar and then not representative of metal-poor clusters.

  Additional indications may come from the analysis of M15 NIR
  colours. $J-K$ prediction well agrees with the value by
  \citet{Brocato+90} ($J-K=0.67$, reddened), and is also consistent
  with data by \citet{Burstein+84} who found $J-K=0.62$ (reddened).
  Interestingly, \citet{Burstein+84} also provide the observed
  $(V-K)=2.14$ (reddened), that becomes $(V-K)=1.86$ if corrected
  according to our reddening assumptions. By comparing this value with
  theory ($V-K=2.05$), it appears that our models predict $V-K$ colour
  about 0.2 mag redder than observations.  Unfortunately, a similar
  comparison cannot be done for M30 and M68, since they are not listed
  by \citet{Burstein+84}. Instead the authors observed M92 (NGC6341),
  whose metallicity is estimated to be close to M15, being
  $[Fe/H]=-2.29$. Since M92 is as bright as $M_V=-$8.20, its colours
  data can be easily compared with our predictions as reported in
  Table~\ref{standardcol}, at age $\sim 11-13\, Gyr$.  Interestingly,
  the observed optical-NIR colour $V-K=2.13$ well agrees with our
  prediction, together with all the optical colours $U-B=0.01$,
  $B-V=0.61$, and $V-I= 0.86$ (from Harris's catalog).

  In conclusion, except for the peculiarity concerning M15, our models
  are able to properly predict the integrated colours of let's say
  'normal' metal-poor clusters, as confirmed by the comparison with
  observational data of M30, M68 (Table~\ref{table:GGC}), and the
  additional cluster M92.

  As noted above, ${\cal M}/L$ values reported in the last column of
  Table~\ref{table:GGC} (upper section), derived using isotropic King
  models by \citet{Pryor&Meylan93}, are strongly model--dependent and
  thus the reported values are only indicative. Moreover dynamical
  processes such as star evaporation
  \citep[e.g.][]{Spitzer87,Vesperini&Heggie97}, may affect the total
  mass and the inferred IMF shape as a function of time.  Observed
  integrated colours of the selected clusters, as well as theoretical
  results, indicate that they are not influenced by the HB morphology,
  at least in our modeled photometric bands. Shorter wavelength
  ultraviolet colours are expected to be more sensitive to the
  presence of an extended blue HB.

  \input{4197.tab2.tex}
  \input{4197.tab3.tex}

  \begin{figure*}[ht]
  \center
   \includegraphics[width=6cm]{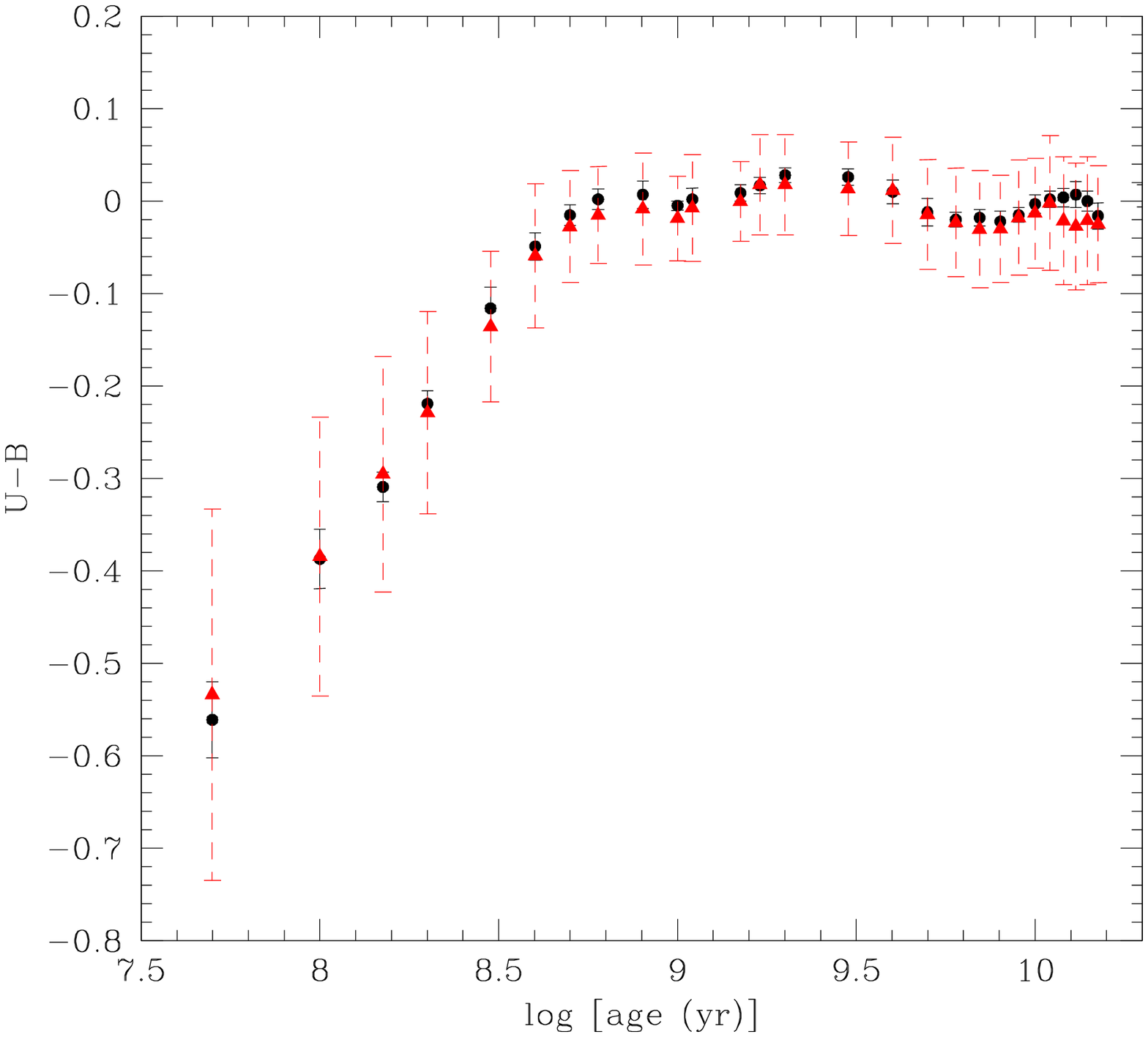}
   \includegraphics[width=6cm]{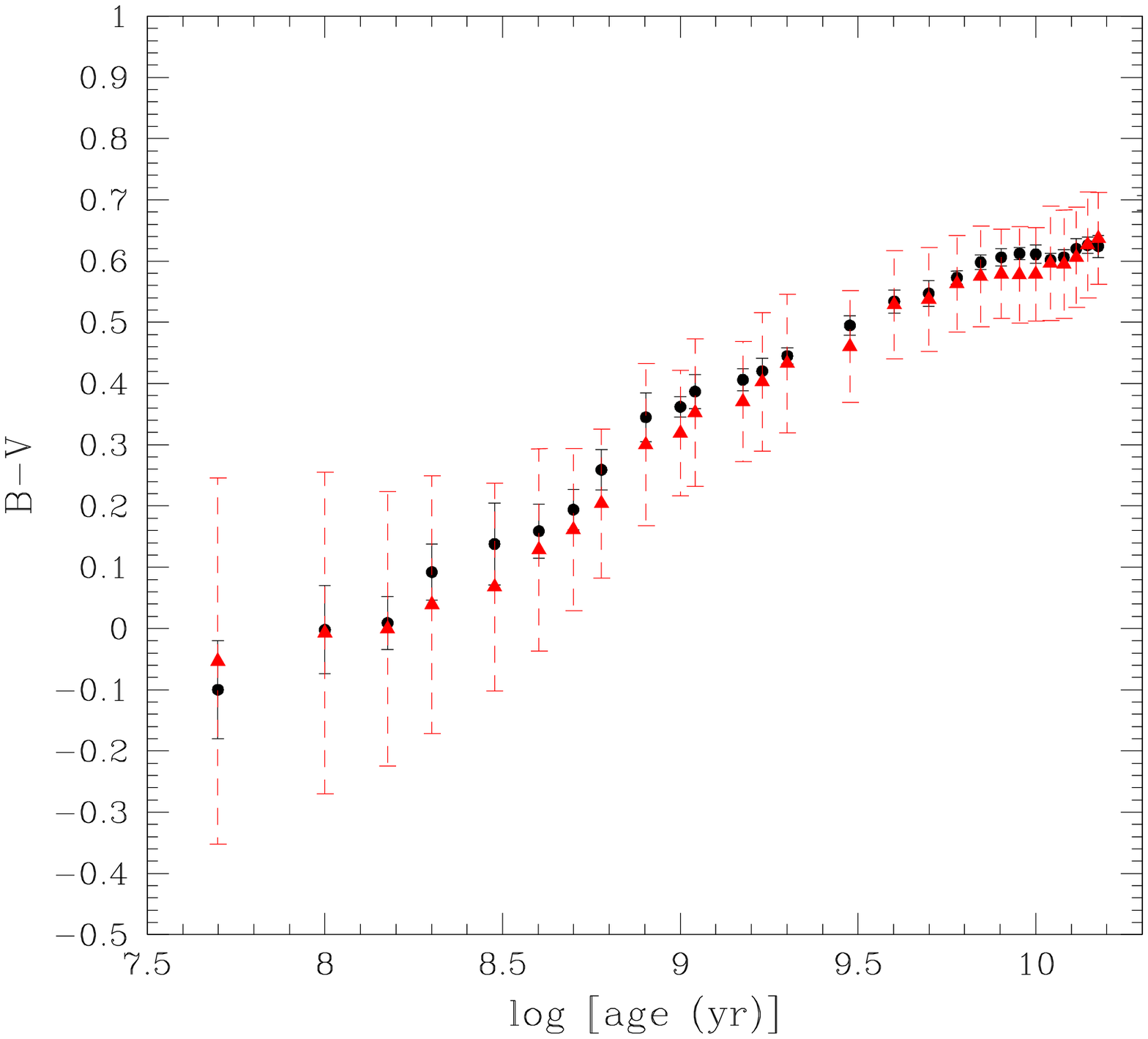}
   \includegraphics[width=6cm]{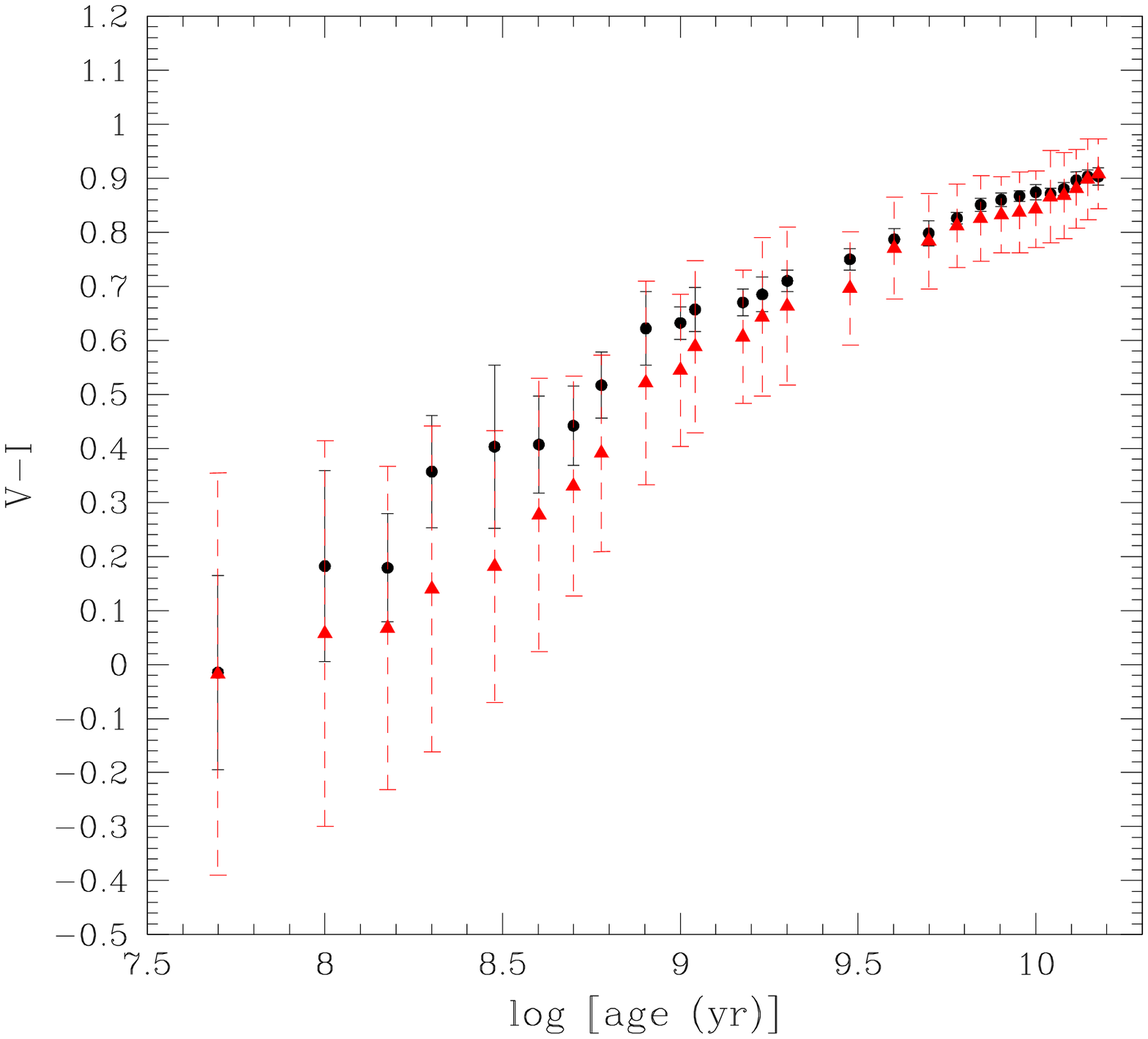}
   \includegraphics[width=6cm]{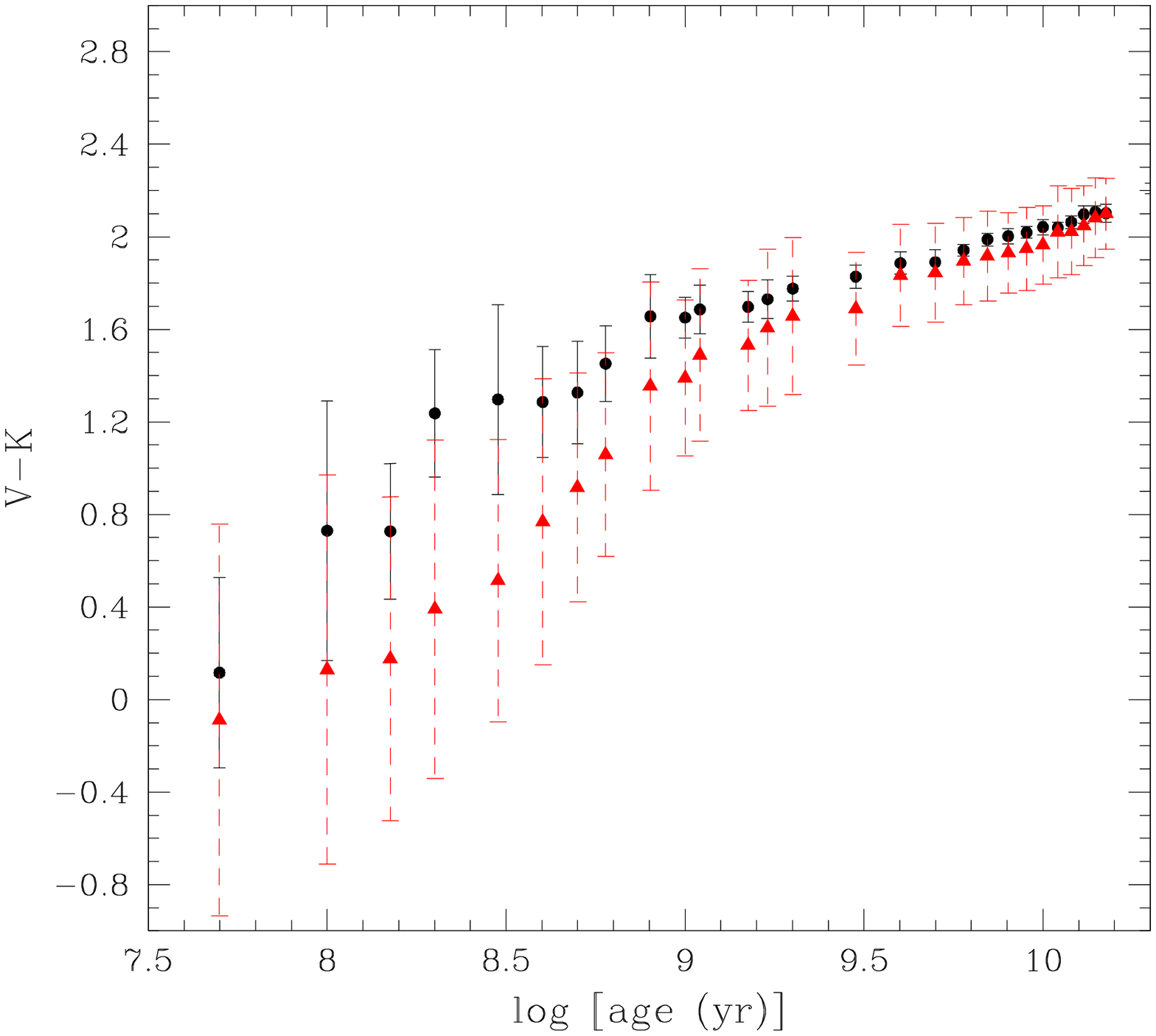}
   \caption{Time evolution of selected integrated colours in the
   standard \emph{UBVRIJHK} filters
   for models with $M_V^{tot}=-4$ (red triangles, dashed lines) and
   $M_V^{tot}=-8$ (black points, solid lines). For each model $ 1 \,
   \sigma$ colour dispersion is reported, see text. }
  \label{fig:coloursV}
  \end{figure*}
  \begin{figure}[ht]
  \center
   \includegraphics[width=8cm]{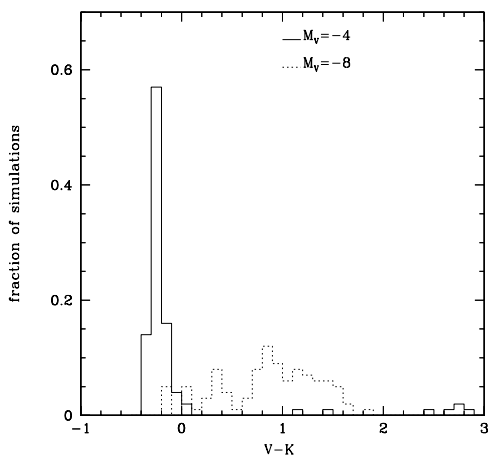}
   \caption{$V-K$ colour distribution resulting from 100
   simulations for a model with 100 Myr, $Z=0.0002$, and
   $M_V^{tot}=-4$ (solid) and -8 mag (dotted).}
  \label{fig:histogram}
  \end{figure}

  \section{Integrated colours}
  \label{section:colours}

  In this section we examine the effects on integrated colours of stochastic
  fluctuations for different values of the assumed total absolute visual
  magnitude and the effects of the still present uncertainty on the IMF
  shape. The \citet{Kroupa02} IMF is adopted for the \emph{standard} model.

  Results are presented at fixed absolute visual magnitude to aid
  comparisons with observational data. For these calculations the
  evaluation of the statistical fluctuations is
  fundamental. Statistical fluctuations of broad-band colours as well
  as mass-to-light ratios have been evaluated by computing a series of
  10 independent simulations for a fixed set of population's
  parameters ($Z$, $age$, IMF, ...)  at fixed $ M_V^{tot}$, and
  assuming a 1 $\sigma$ error.  When fluctuations become large, and
  the value above is not fully representative of colours variations,
  we extend the analysis up to 100 runs.
  This is especially the case of very
  poorly populated clusters ($M_V=-4$) and young ages.

  The \emph{standard} models are computed assuming: $\eta_{R} = 0.3$,
  $\sigma_{\eta} = 0.08$ and Kroupa's IMF.  As an example, Table
  \ref{standardcol} and \ref{standardHST} reports integrated colours
  as function of age for models with $M_V^{tot}=-8$; Johnson-Cousins
  colours in Tab.\ref{standardcol} and HST colours in
  Tab.\ref{standardHST}. Table \ref{standardcol} lists from left to
  right: age, $M_V^{tot}$, integrated $(U-B)$, $(B-V)$, $(V-R)$,
  $(V-I)$,$(V-J)$ and $(V-K)$ colours. Table \ref{standardHST} reports
  from from left to right: age, $V-F439W2$, $V-F555W2$, $V-F814W2$ for
  the $WFPC2$ and $V-F110W1$, $V-F160W1$ for the $NICMOS$ camera.

  \subsection{$ M_V^{tot}$ sensitivity}

  Our standard synthetic models are calculated for three different
  values of the total magnitude: $M_V^{tot}=-8,-6$ and $-4$. (Note,
  that all the results for both broad-band colours and mass--to--light
  ratios for $M_V^{tot}=-6$, and $-4$ models are available as
  electronic tables at CDS).

  In Fig.~\ref{fig:coloursV} the time evolution of selected integrated
  colours are shown for the two extreme cases ($M_V^{tot}=-8$, black
  circles, and $-4$, red triangles).  In general, broad--band colours
  become redder with age, because of the increasing fraction of cool
  stars in the populations. For the same reason, the $(U-B)$ colour is
  quite age-insensitive for age $\gsim 1\, Gyr$.

  Within the statistical uncertainties, the mean colours at different
  $M_V^{tot}$ do not show significant differences, except in a few
  cases at young ages, and for optical--NIR colours. This is due to
  the strong variation of the number of very bright red stars
  affecting significantly the total luminosity of the cluster.
  In a SSP with a total brightness as faint as
  $M_V^{tot}=-4$,
  the most of stars occupy the MS phase, and only a few (or none)
  red giants are present, including stars burning He in the core and stars
  in the double shells phase
  \citep[see for example Fig. 14 in][]{Raimondo+05}. Hence, the mean $V-K$ colour
  (and to a less extent $V-I$) is generally
  bluer than in models with $M_V^{tot}=-8$, where post-MS phase is
  always widely populated. This effect is large in young clusters, for which
  colours are more sensitive to the cluster luminosity  [see also Fig. 5 in \citet{Santos&Frogel97},
  and Fig. 8 in \citet{Brocato+99}].

  To clarify this point, Fig.~\ref{fig:histogram} illustrates the $V-K$ colour distribution for a
  SSP model with age 100 Myr; for each choice on $M_V^{tot}$, 100 simulations are computed.
  The $V-K$ colour distribution of faint clusters (solid line) shows a blue populous peak, due to the
  large fraction of simulations containing mainly
  \emph{blue} stars, and a few sparse simulations at redder colours
  containing a few red giants. Consequently, the average colour
  is bluer than that found for brighter clusters (Fig.~\ref{fig:coloursV}), which
  show a broadened colour distribution shifted towards the red side,
  since the probability to have a large number of red stars is higher
  (dotted line in Fig.~\ref{fig:histogram}).
  We emphasize that in the present paper
  for a given cluster age, the integrated colours at different cluster brightness
  are obtained by properly adding stars in all
  evolutionary phases according to the IMF and evolutionary
  time-scales (i.e. keeping fixed the proportion between the number of stars in
  different stages). So that,  by
  decreasing their absolute magnitude (mass), the \emph{natural} trend is that the cluster progressively
  misses post-MS stars, and due to the discreteness of stars
  "\emph{if an SSP is far less luminous, then there would be no post-MS stars and the integrated fluxes
  would be dominated by upper MS stars with little spread in color, which results
  in smaller color fluctuations}" \citep{Santos&Frogel97}.
  The colour behaviour described above is in agreement with \citet{Santos&Frogel97} and
  \citet{Brocato+99}, while opposite to what found by \citet{Bruzual&Charlot03}.
  A discussion on the nature of such a discrepancy is beyond the purpose of this paper, since it would required a
  deep comparison between all the assumptions adopted in all the
  codes (evolutionary tracks, atmospheres, etc.).

  In general, for a further increase of the cluster brightness, the stochastic
  effects decrease, the colour distribution becomes more regular,
  and the standard deviation squeezes (small spread in colour).

  At a given age, the size of errorbars increases as the cluster
  luminosity decreases; this is due to the small number of stars
  present in faint clusters ($M_V^{tot}=-4$, red symbols) with respect
  to bright ones ($M_V^{tot}=-8$, black symbols).  From the figure it
  is evident that for all ages the statistical errors at
  $M_V^{tot}=-4$ are so high as to prevent precise quantitative
  evaluations.

  The errorbars also increase from $B-V$ to $V-K$, as shown by
  \citet{Brocato+99}.  This last finding is related again to the
  expected small number of post-MS stars shining in the infrared
  wavelengths, especially in case of low-luminosity clusters
  ($M_V^{tot}=-4$). Moreover, in faint clusters it may happen that
  giant stars are not present in each of the simulations we computed
  at fixed age, being their appearance highly driven by stochastic
  phenomena.  At fixed total magnitude the \emph{intrinsic}
  uncertainty lowers with age, also because a larger number of stars
  is needed to reach the given total magnitude.

  \begin{figure*}[th]
  \center
   \includegraphics[width=6cm]{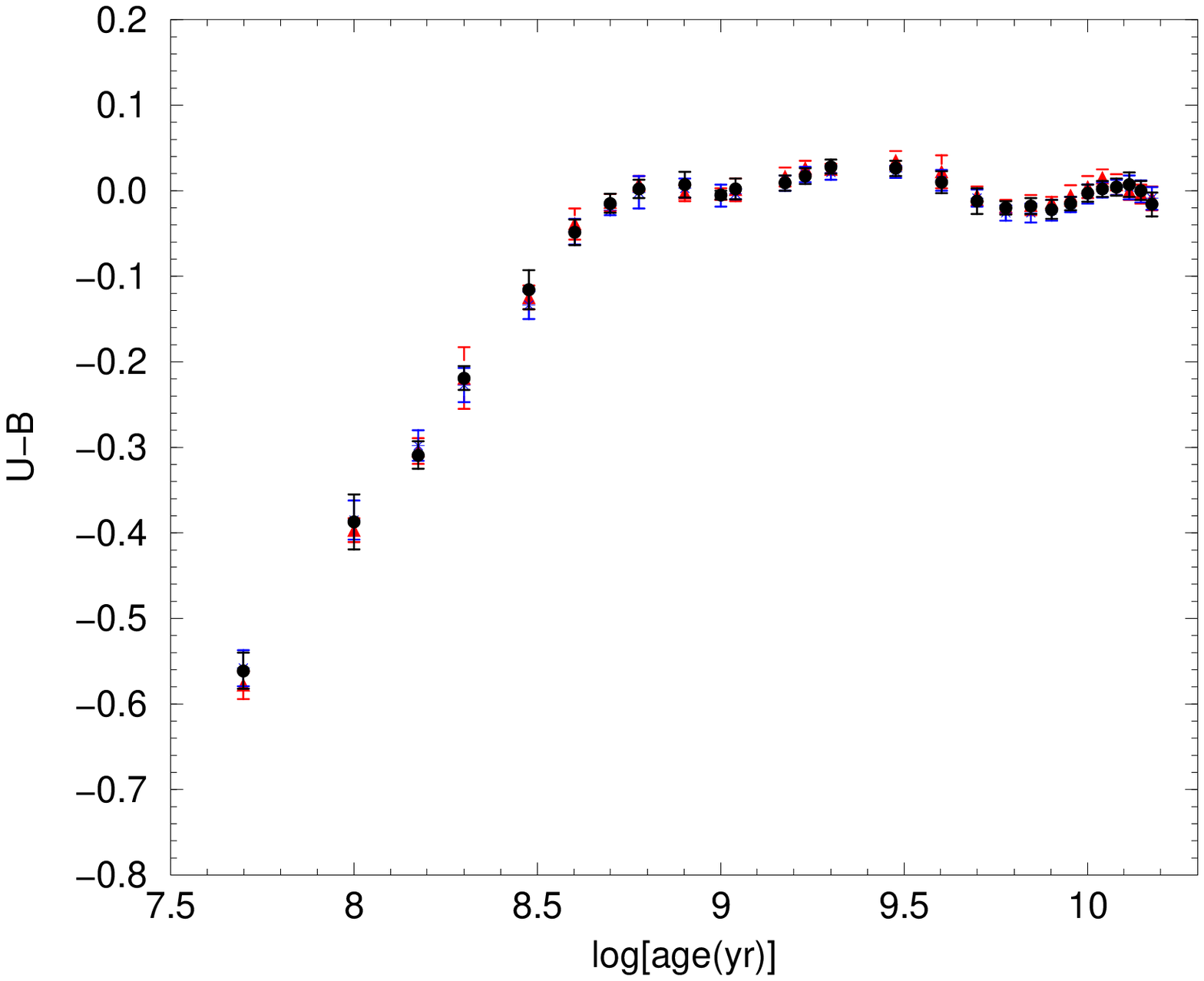}
   \includegraphics[width=6cm]{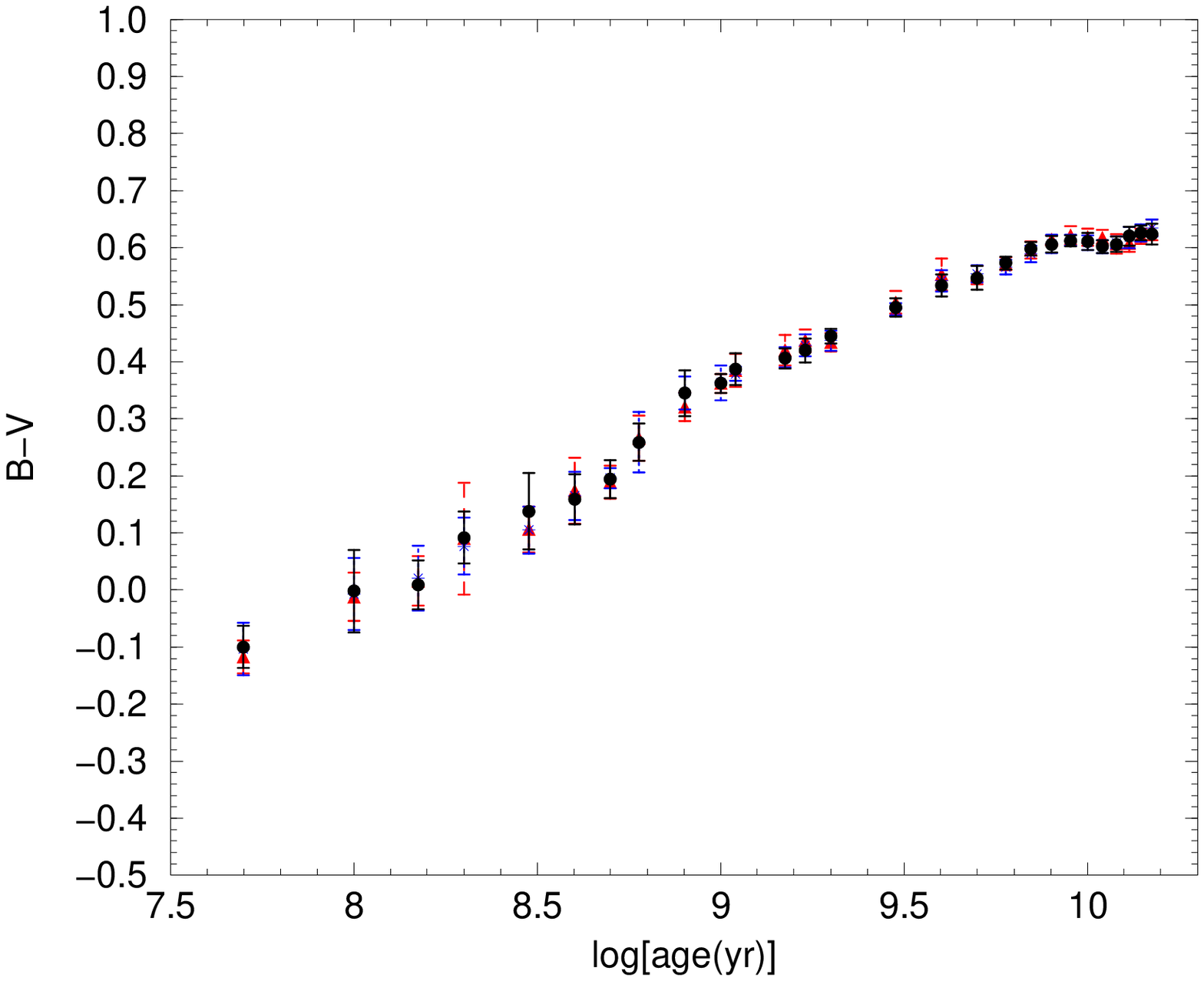}
   \includegraphics[width=6cm]{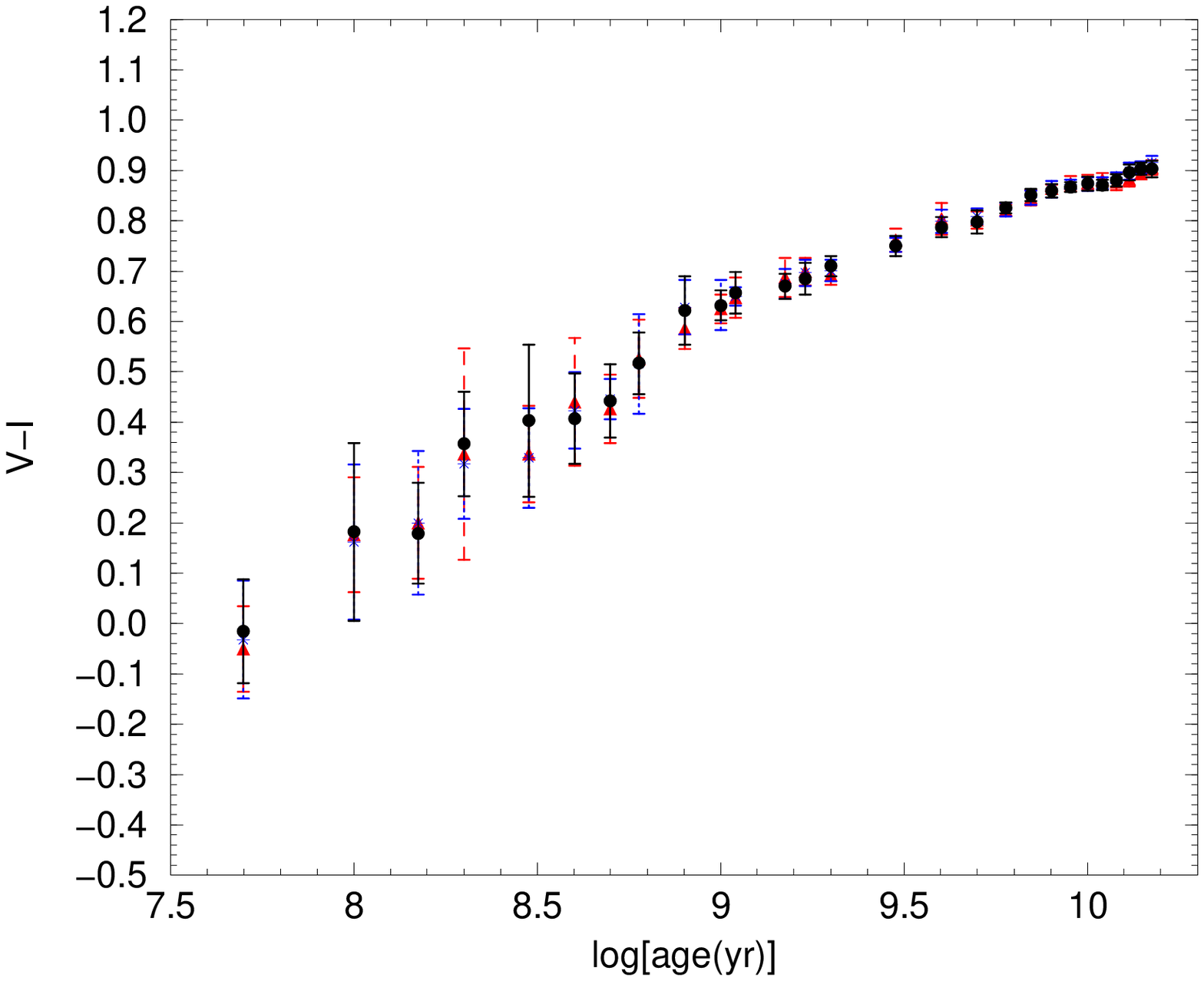}
   \includegraphics[width=6cm]{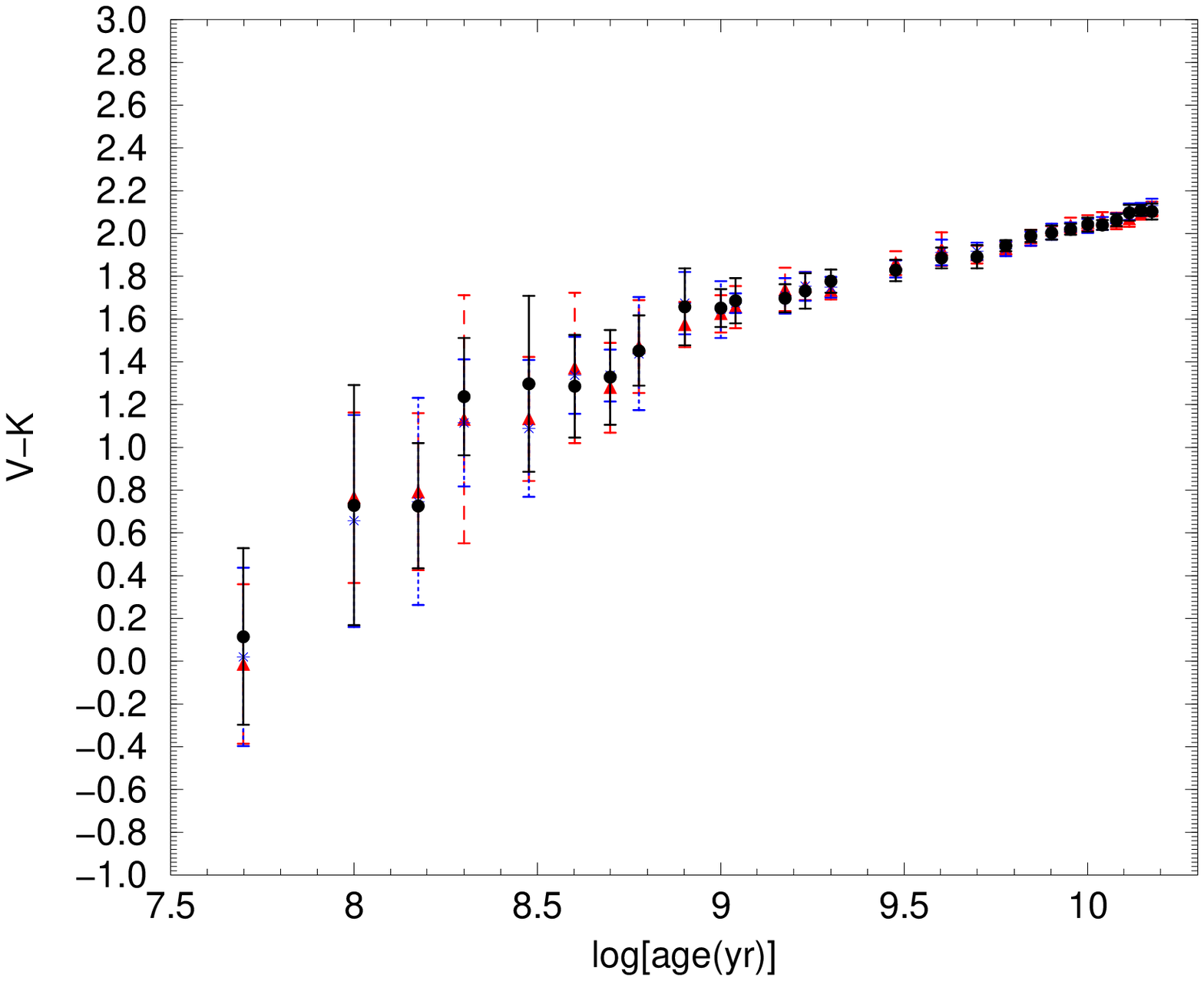}
   \caption{Time evolution of
     selected integrated colours for SSPs with $M_V^{tot}=-8$
     calculated with the lower, central and upper value for the
     \citet{Kroupa02} IMF exponent ($\alpha$) for $M \geq 0.5 M_{\odot}$.
     Black dots indicate our reference model ($\alpha =2.3$), red
     triangles indicate models with $\alpha =2.0$ and blue stars
     models with $\alpha =2.6$.}
  \label{fig:colorsIMF1}
  \end{figure*}

  \input{4197.tab4.tex}

  \begin{figure*}[th]
  \center
   \includegraphics[width=6cm]{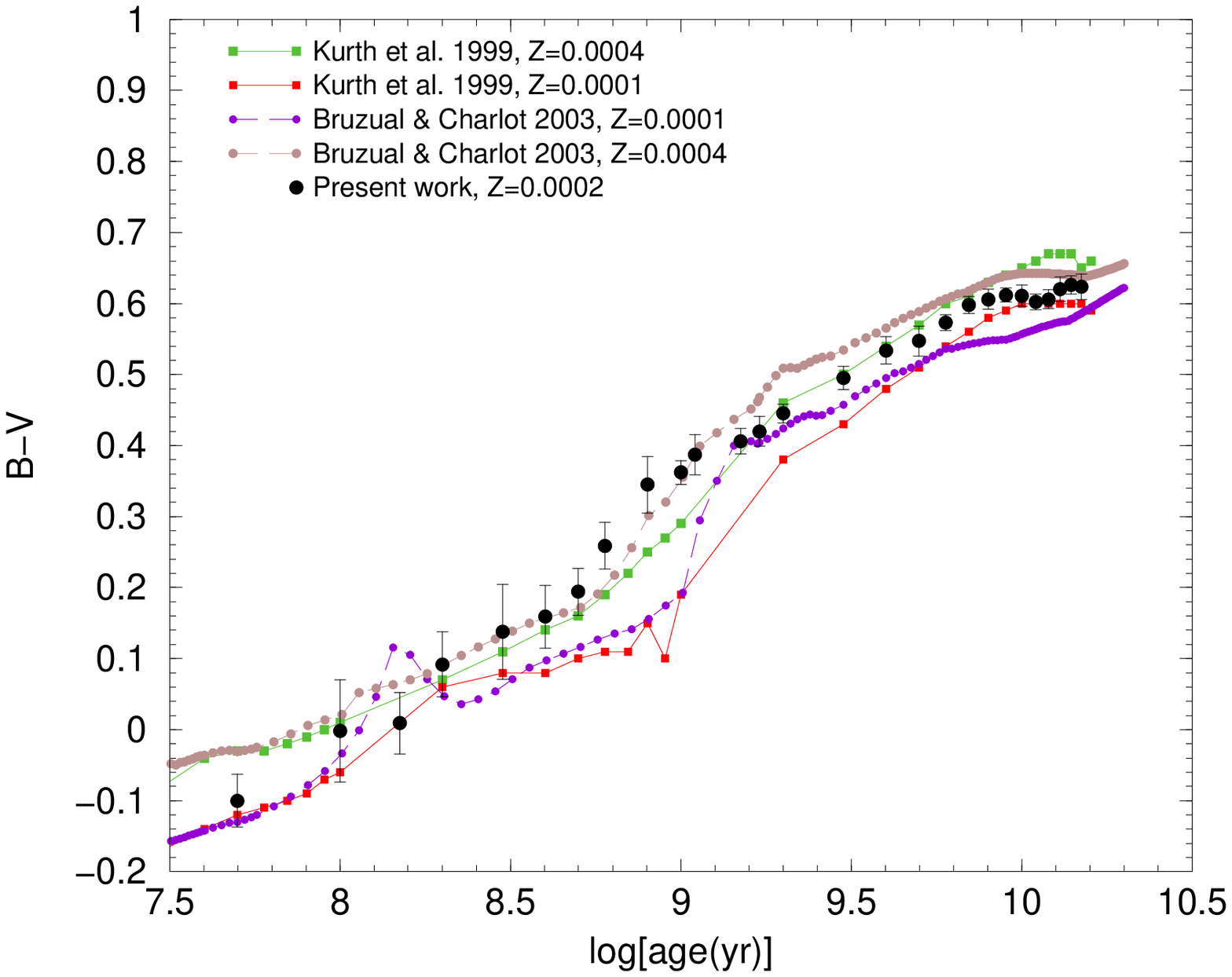}
   \includegraphics[width=6cm]{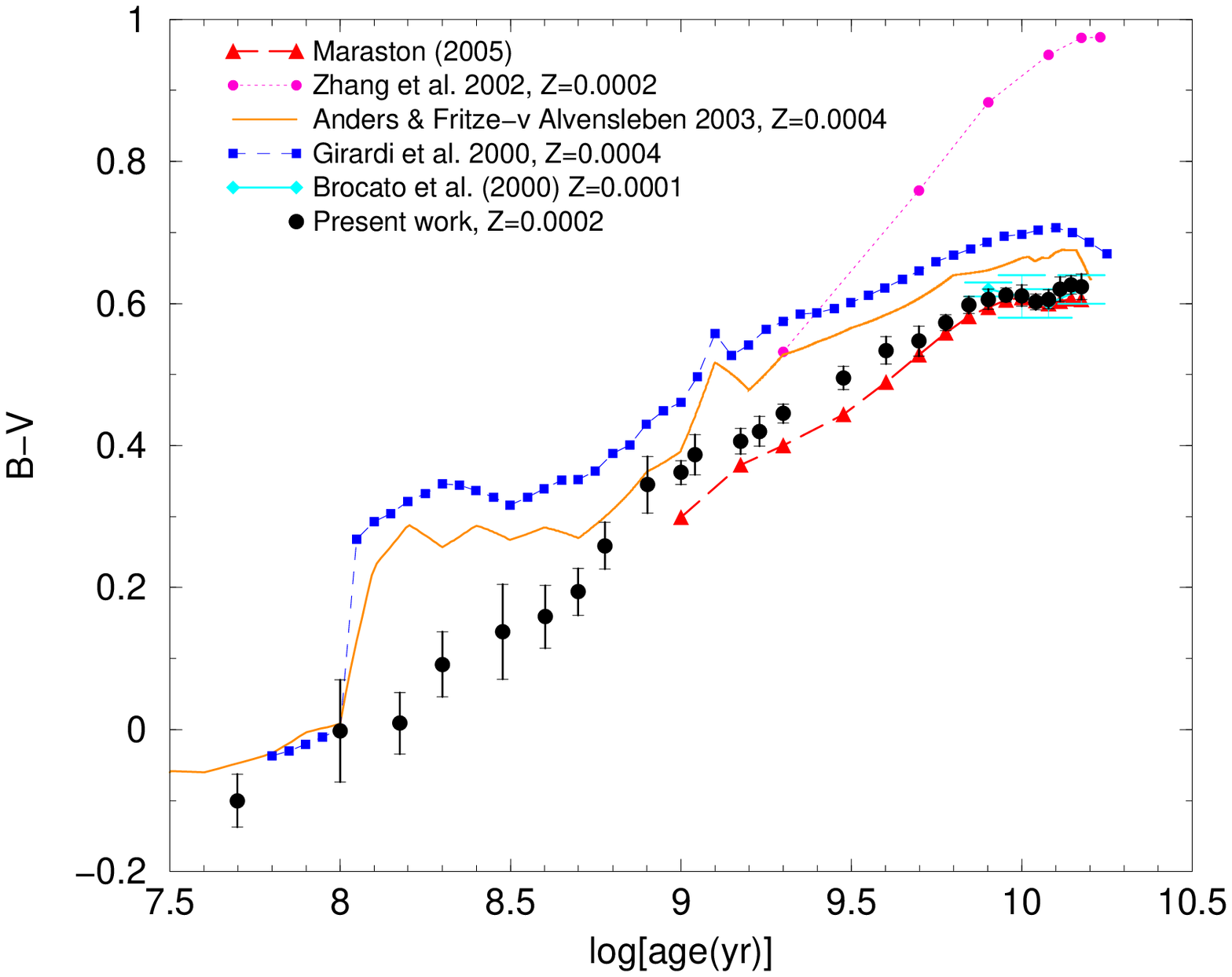}
   \includegraphics[width=6cm]{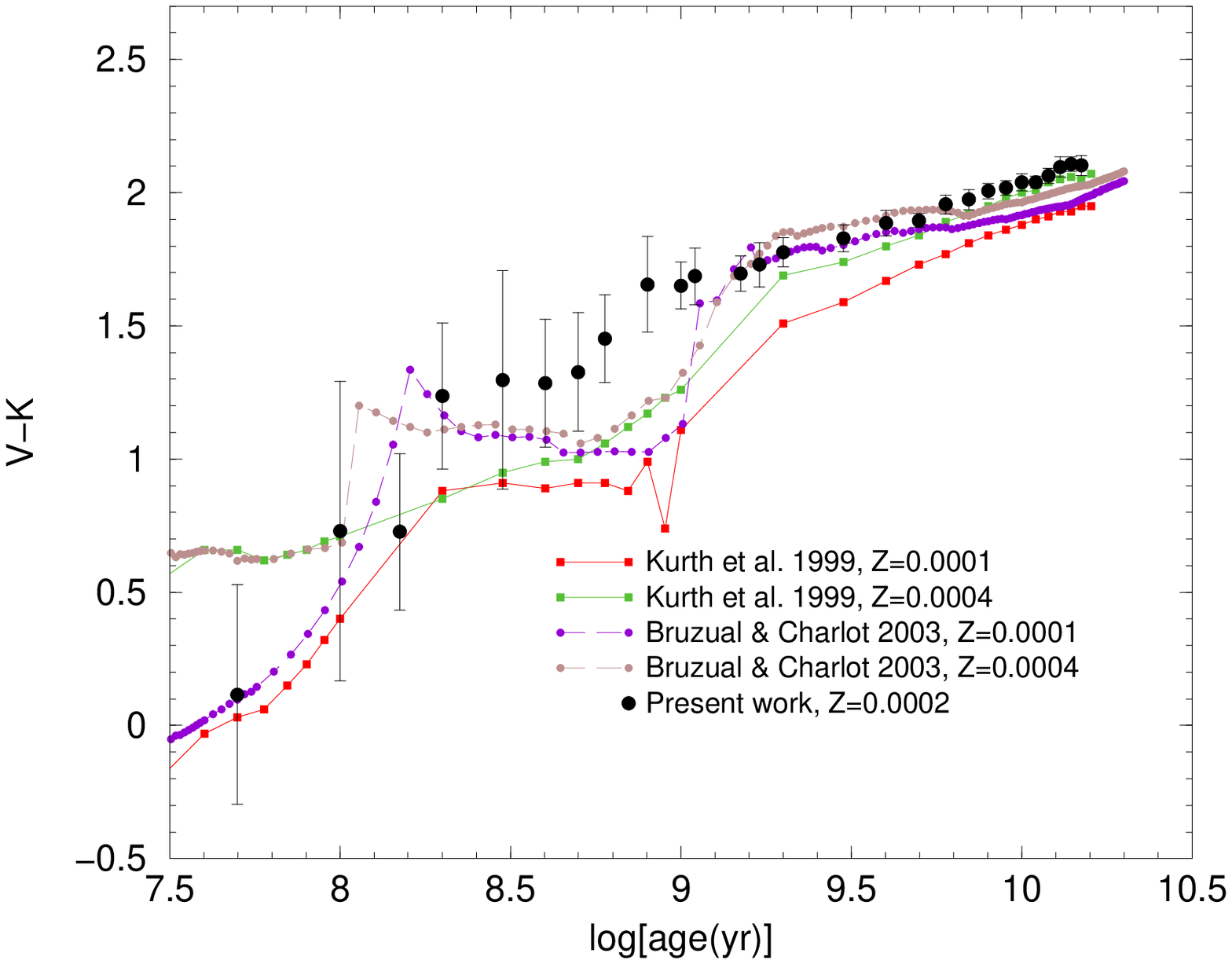}
   \includegraphics[width=6cm]{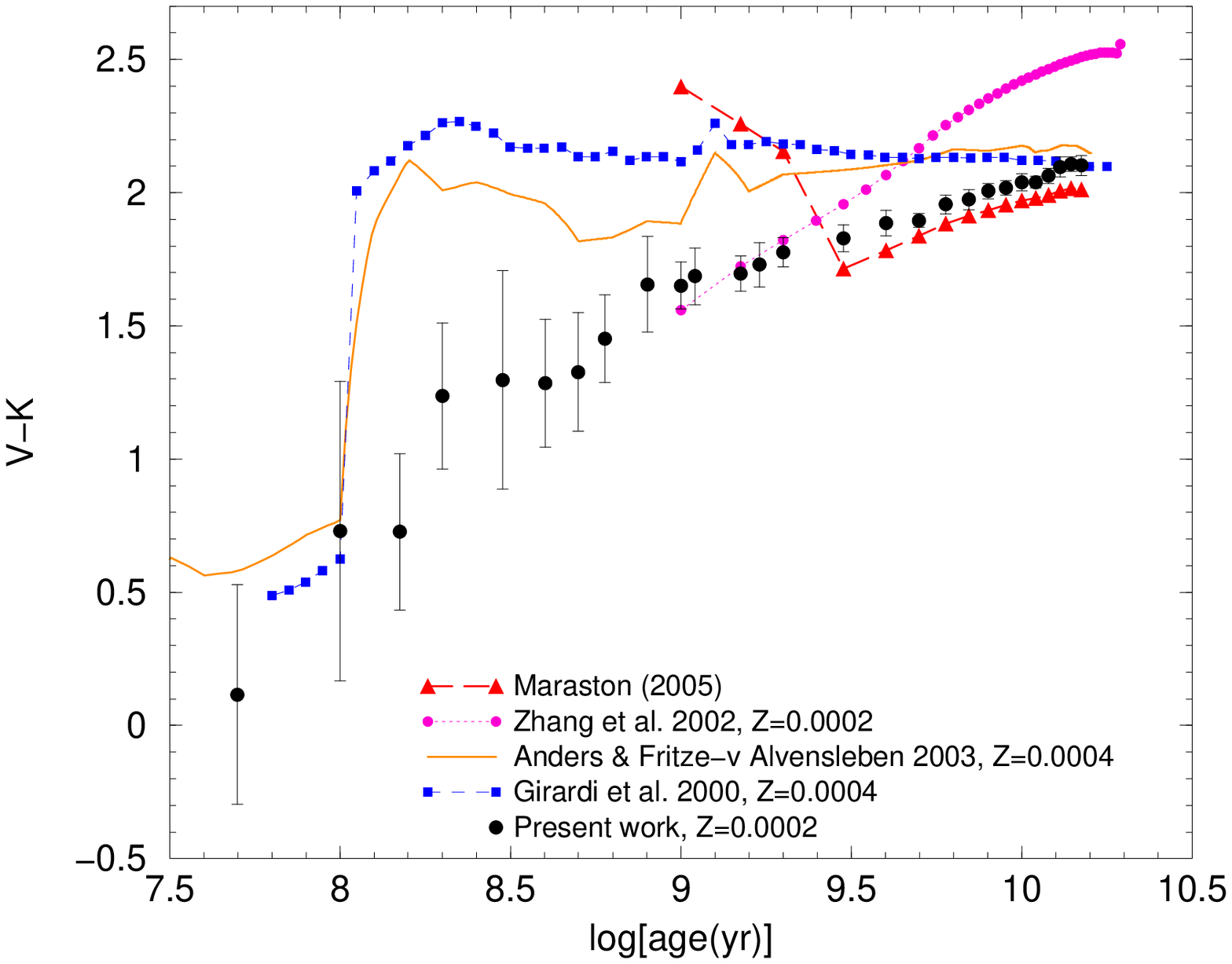}
    \caption{Comparison between theoretical integrated colours ($B-V$, upper panels;
    $V-K$, lower panels) from recent papers:
    present models (black dots);
    \citet[][red Z=0.0001, green Z=0.0004 squares, solid line]{Kurth+99},
    \citet[][cyan dots]{Brocato+00},
    \citet[][orange solid line]{Anders+03},
    \citet[][blue squares, short-dashed line]{Girardi+00},
    \citet[][pink circles, dotted line]{Zhang+02},
    \citet[][violet Z=0.0001, brown Z=0.0004 circles, dashed line]{Bruzual&Charlot03}.
    \citet[][red triangles, dashed line]{Maraston05}.}
    \label{fig:comparison}
   \end{figure*}

  \subsection{IMF variations}

  To evaluate the effect of IMF variations on integrated colours we
  changed the Kroupa (2002) IMF exponent (x) within the estimated
  uncertainty for masses $M > 0.5 \, M_{\odot}$. In
  Fig.~\ref{fig:colorsIMF1} models with $M_V^{tot}=-8$ calculated with
  the lower, central and upper value of the IMF exponent are plotted
  in selected pass-bands; the corresponding data tables are available
  as electronic tables at CDS.

  As obtained for metal-rich models investigated in previous papers
  \citep[see e.g.][]{Maraston98, Brocato+00, Yi03, Bruzual&Charlot03},
  we find that colours variations are negligible also for very
  low-metallicity models, at least in the range of exponent
  investigated here, being the colours variations well within the
  intrinsic uncertainty, for both values of the brightness of the
  cluster ($M_V^{tot}=-4$ mag and $-$8).  The errorbars behavior is
  similar to what described above: colour fluctuations increase from
  optical--optical to optical--NIR colours and for ages $\lsim 1$ Gyr.

  Changing the distribution of very low mass stars ($M \leq 0.5\,
  M_{\odot}$) has negligible effects on photometric indices whatever
  the total visual magnitude, as already shown by \citet{Brocato+00}
  for ages greater than 5 Gyr.  In poorly populated clusters
  ($M_V^{tot}=-4$ mag) as in the richest ones ($M_V^{tot}=-8$ mag),
  reducing the number of stars with $M < 0.5 \, M_{\odot}$ does not
  affect the integrated colours since the contribution of these stars
  to the total $V$ light is only of the order of few percent
  \citep{Brocato+00}.  The contribution of such stars slightly
  increases with age, but it remains comparable to the statistical
  errors even at the oldest ages (at least in the range explored
  here).

  \subsection{Comparison with previous works}

  Table~\ref{table:ingredients} summarizes the main ingredients used
  by various authors in their synthesis code, and
  Fig.~\ref{fig:comparison} compares the time--evolution of $B-V$ and
  $V-K$ colours of the present work with similar recent results
  available in the literature.  In the figure our models are plotted
  with $1 \sigma$ intrinsic error, while other authors do not estimate
  the colours statistical errors, except for \citet{Brocato+00}.  If
  colours are not available for exactly the same metallicity value
  adopted in the present work ($Z=0.0002$), the two closest $Z$ values
  are plotted. The models referred as Anders \& Fritze-v. Alvensleben
  (2003) are the ones of the quoted paper, updated concerning emission
  lines by private communication \citep[see also][]{Schulz+02}.

  The present predictions well agree with those by \citet{Brocato+00}.
  There is also a general good agreement with \citet{Kurth+99}, and
  \citet{Bruzual&Charlot03}, both based on \emph{Padua 1994} stellar
  evolutionary tracks (see the quoted papers for the detailed list of
  references).  Differences can be found in the age range $8.6 \lsim
  \log \, age(yr) \lsim 9$ for optical--NIR colours, possibly due to
  the treatment of the AGB and TP--AGB phases.  \citet{Girardi+00} and
  Anders \& Fritze-v. Alvensleben (2003) predict redder colours than
  all the other authors for $age \gsim 100$ Myr, as well as
  \citet{Zhang+02} for $age \gsim 3$ Gyr.  Differently from the
  others, both \citet{Maraston05} and our models are based on
  no-overshooting tracks. Nevertheless, while the present $B-V$
  colours well agree with Maraston's models at all ages in common,
  $V-K$ predictions are bluer than those by \citet{Maraston05} at ages
  younger than $\simeq 3\, Gyr$.

  Fig.~\ref{fig:comparison} shows that the $B-V$ and $V-K$ predictions
  from different authors agree within 0.1 mag at $\log age \gsim 9.3$.
  Large model-to-model differences arise at young and intermediate
  ages, when the contribution of AGB and TP-AGB stars to the total
  flux is high, especially in the NIR bands \citep[see for a
  discussion][]{Maraston98,Gir&Ber98}.  This has a direct consequence
  on predicted colours, since a large variety of stellar ingredients
  and prescriptions are used to simulate these phases.
  Temperature-colours relations, mass-loss prescriptions, and
  analytical description of the TP evolution all affect the
  photometric properties of such kind of stars, thus it is not
  straightforward to individuate a single origin (or multiple ones) to
  explain differences shown in Fig.~\ref{fig:comparison} in the
  intermediate-age range.

  For instance, mass-loss processes affecting TP--AGB stars are one of
  the physical mechanisms largely affecting the TP-AGB stars
  evolution.  However, the effect of mass loss on the observational
  properties of TP stars are largely unknown to date. To give an
  indication on how mass-loss may affects colour predictions, we made
  numerical experiments by computed integrated colours in the extreme
  case when no TP--AGB stars are present in the population, to mimic
  in such a way a very efficient mass-loss rate \citep[see
  also][]{Maraston98}.  As expected, the effect is larger for
  intermediate-age populations, and in the optical-NIR colours.  The
  $V-K$ colour becomes bluer than the $standard$ value by $\sim 0.5$
  mag at $age=500$ Myr, and the $B-V$ colour decreases only by 0.06
  mag at the same age. This effect tends to vanish by increasing the
  age.  On the other hand, if we lower the mass-loss efficiency with
  respect to our \emph{standard} assumption (BH) by using a Reimers'
  law with $\eta$ of the order of 1, and $V-K$ colours redder than our
  'standard' models are obtained. As a consequence the colour decline
  at age corresponding to the appearance of He-core degenerate stars
  is more evident ($t\sim 1\, Gyr$), similarly to what obtained by
  \citet{Maraston05}, and by \citet{Raimondo+05} for surface
  brightness fluctuation colours.

  \section{Mass-to-light ratio}
  \label{section:ML}

  Table \ref{standardML} lists our predictions for the mass-to-light
  ratio (${\cal M}/L$) in various photometric bands as function of age
  for $M_V^{tot}=-8$; similar tables for $M_V^{tot}=-6$ and
  $M_V^{tot}=-4$ are available at CDS.

  The total mass $\cal M$ includes all the evolving stars up to the
  AGB tip ($M \geq 0.1 \, M_{\odot}$) and WDs.  Stars with mass higher
  than $M_{up}=7\, M_{\odot}$ are assumed to leave a neutron star (NS)
  as a remnant.  To evaluate the maximum mass fraction of NS we set
  the neutron star mass at its upper limit \citep[$\approx
  2\,M_{\odot}$, see e.g.][]{Bombaci+04}, and we choose high cluster
  age ($\approx 12$ Gyr); we found that the NS mass fraction
  constitutes at maximum few percent of the cluster mass in agreement
  with the calculations by \citet{Vesperini&Heggie97} who found an
  upper value of $\approx$ 1\%.  The contribution of massive
  black holes to the total mass depends on assumption on the mass of
  the remnant and IMF \citep{Maraston05}. By using the same
  prescriptions on black holes mass as \citet{Maraston05}, we
  estimated that for a Kroupa IMF the mass fraction of both neutron
  stars and black holes does not exceed 5-10\%.

  In this paper, the ${\cal M}/L$ values are calculated without taking
  into account any dynamical processes, e.g. evaporation of stars due
  to two body relaxation or disk shocking \citep[see
  e.g.][]{Spitzer87, Vesperini&Heggie97, Boily+05}.  A treatment of
  these phenomena is beyond the scope of this work; N-body simulations
  showing the effects of this processes on the total mass and IMF
  shape as a function of time, and on the galactocentric distance of
  the cluster can be found in e.g. \citet{Vesperini&Heggie97}. We only
  note that since stellar evaporation is more efficient for low mass
  stars, one expects a flattening of the IMF as the dynamical
  evolution of a cluster proceeds. Evaluations of the mass fraction of
  WDs expected to be lost from clusters due to dynamical evaporation
  can be found in \citet{Vesperini&Heggie97} \citep[see
  also][]{Fellhauer+03, Hurley&Shara03}.  The presence of binary stars
  is not taken into account as well; the lack of information about the
  binary frequency, the distribution of binaries of different mass
  ratios, the separation of the components, and the occurrence of
  explosion of novae and supernovae prevents a quantitative treatment
  of this phenomenon.  An attempt to include binary systems in
  population synthesis models has recently been presented by
  \citet{Zhang+05}.

\begin{figure*}[t]
  \center
  \includegraphics[width=6cm]{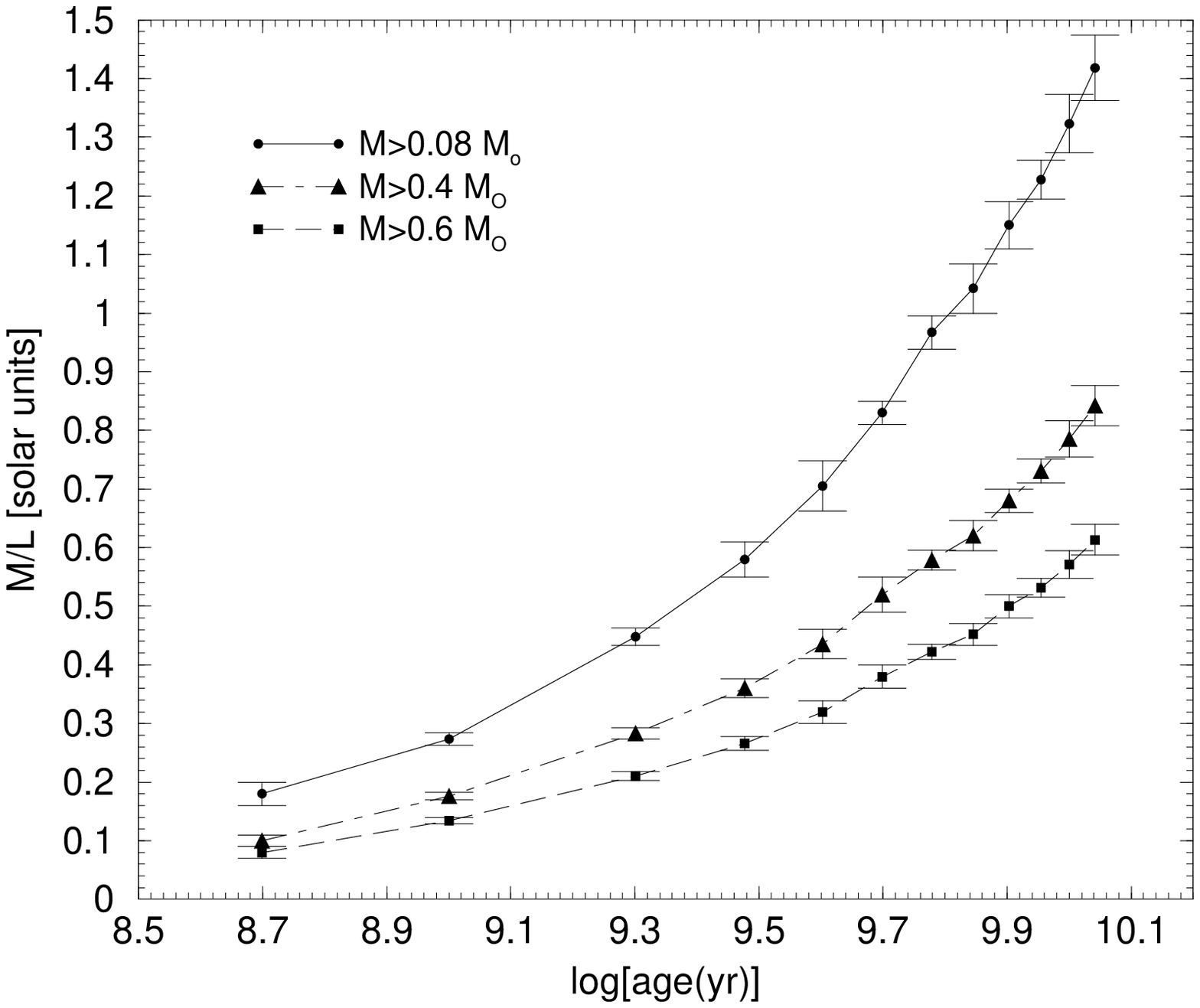}
  \includegraphics[width=6cm]{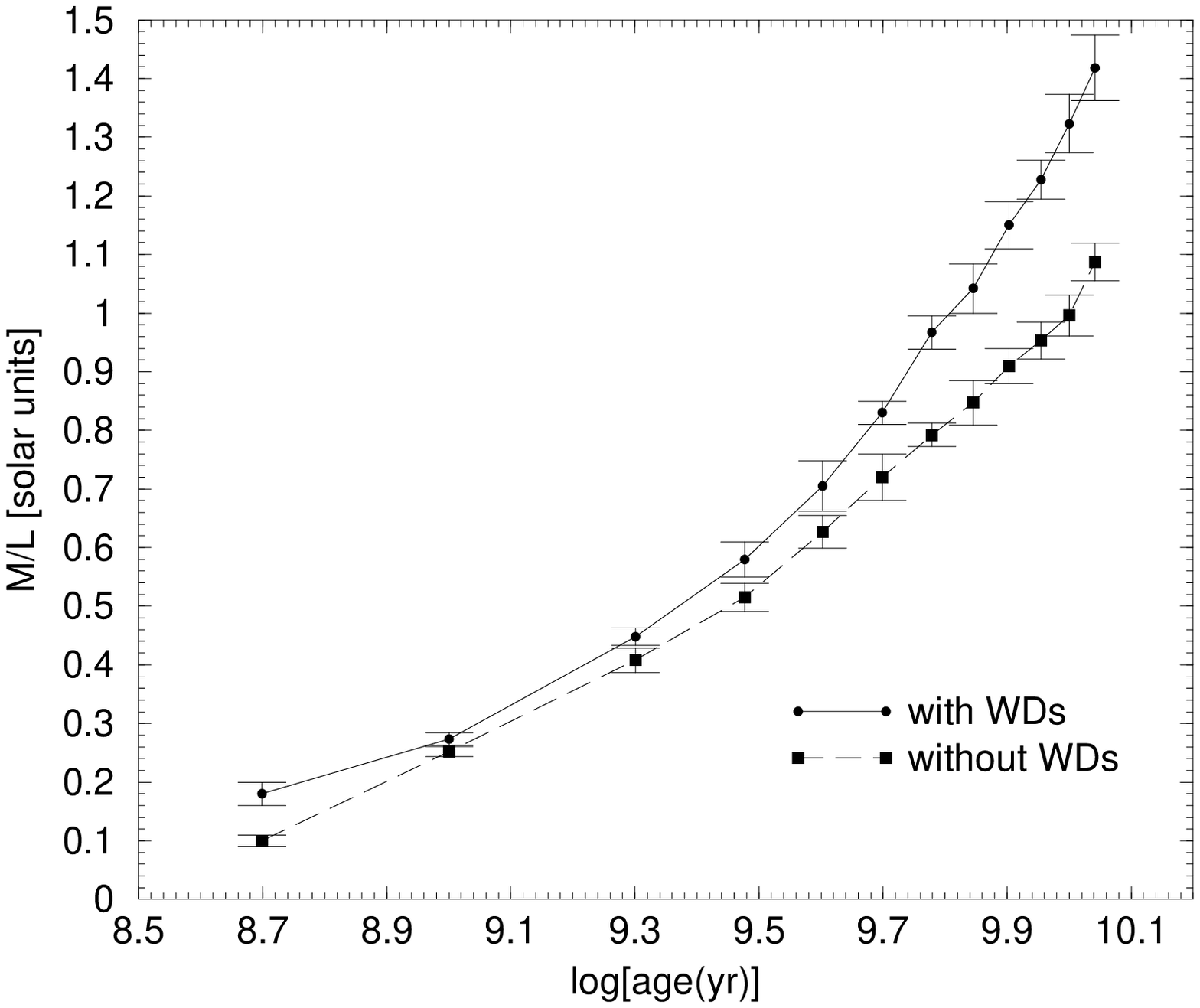}
  \caption{Left panel: Fraction of very low mass stars ($M \leq 0.6\, M_{\odot}$) that contribute to
  the mass--luminosity ratio (${\cal M}/L$) (in solar units) as a function of
  the age for $M_V^{tot}=-8$. Filled circles represent our {\em standard} model in
  which all masses with $M \geq 0.08\, M_{\odot}$ are included;
  the effect of considering only stars with $M \geq 0.4\, M_{\odot}$
  (filled triangles) or $M \geq 0.6\, M_{\odot}$ (filled
  squares) is analyzed separately. Right panel: as in the left panel, but
  when the WD population is (filled circles, {\em standard} model) or is not (filled squares)
  included.}
  \label{fig:VLMWD}
  \end{figure*}

\begin{figure}[ht]
  \center
  \includegraphics[width=6cm]{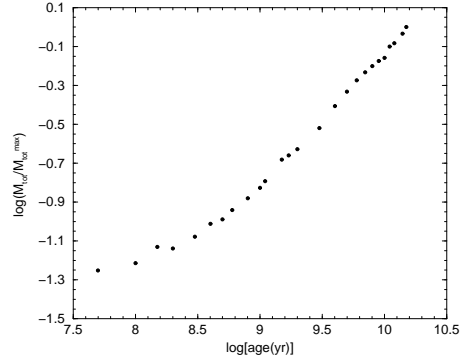}
  \label{fig:MLstd} \caption{Time behaviour of the logarithm of the cluster total mass
  needed to reach $ M_V^{tot}=-8$ normalized to the maximum value,
  as a function of age.}
  \label{fig:Mvsage}
  \end{figure}

  In Fig. \ref{fig:VLMWD} we show the influence of VLM stars ($M \leq
  0.6\, M_{\odot}$) and WDs on the ${\cal M }/ L$, $L$ is the
  bolometric luminosity in solar units, at fixed absolute visual
  magnitude ($M_V=-8$).  As already known, VLM and WDs provide a
  nearly negligible contribution to the total luminosity \citep[see
  for example][]{Maraston98}, while give a significative contribution
  to the total mass. Fig. \ref{fig:VLMWD} shows that the fraction of
  low mass stars and white dwarfs increases with the cluster age
  \citep[see also][]{Maraston98}.  Note that, since the simulations
  are performed at fixed absolute visual magnitude, the total mass of
  the cluster varies with age, as shown in Fig. \ref{fig:Mvsage},
  where we plot the case of $M_V=-8$. This is because as the age
  increases the mass of the typical star in the cluster decreases.

  Changing $M_{up}$ affects the number of
  stars cooling as WDs, or in other words the mass fraction locked as WDs,
  and thus the cluster total mass.
  Different input ingredients of stellar evolutionary tracks, as e.g. the
  neutrino energy losses in the core, lead to different predictions
  of $M_{up}$ at fixed metallicity, see e.g.
  \citet[][$M_{up}=5.7 \, M_{\sun}$]{Tornambe&Chieffi86},
  \citet[][$M_{up}=5\, M_{\sun}$]{Pols+98},
  \citet[][$M_{up}=6.5\, M_{\sun}$]{Dominguez+99},
  \citet[][$M_{up}=5 \div 6\, M_{\sun}$]{Cariulo+04}. We explore
  the effect of changing $M_{up}$. We find that increasing $M_{up}$ from 5
  up to 7 $M_{\sun}$ does not cause
  variations in the mass--to--light ratios, at least for the adopted IMF. Even
  pushing $M_{up}$ up to 10$M_{\sun}$ does not cause sensitive change in
  the mass-to-light ratios (Table \ref{tab:Mup}).

  \input{4197.tab5.tex}

  \subsection{$ M_V^{tot}$ sensitivity}

  Fig. \ref{fig:MLstd} shows the time evolution of ${\cal M}/L$ ratio
  in selected passbands for \emph{standard} models with $M_V^{tot}=-8$
  (Table \ref{standardML}) and $-4$ mag. The corresponding data tables
  for other passbands and for the case of $M_V^{tot}=-6$ are available
  at CDS.  Note that the mass-to-light ratios generally increase with
  age independently of the photometric band, because the total mass
  increases. This result is similar to what obtained by other authors,
  who do not keep constant the luminosity $L_V$ \citep[see
  e.g.][]{Bruzual&Charlot03,Maraston05}.

  The only effect of changing the absolute magnitude from $M_V=-8$ to
  $-4$ on mass-to-light ratios in the optical bands is to increase the
  intrinsic uncertainties. On the contrary, other than this effect
  mass-to-light ratios in the NIR bands suffer a larger scatter due to
  the decrease of red giants contributors to the total NIR luminosity,
  as shown in the right lower panel of the figure for $K$-band.

  Since $M_V^{tot}$ is fixed, the behaviour of (${\cal M}/L_{V}$) as a
  function of the age reflects the trend of the cluster mass
  (Fig. \ref{fig:Mvsage}); it is quite linear for ages $\ge 1\, Gyr$
  in all passbands indicating it is driven by the grown of the total
  mass. For ages $\leq 1\, Gyr$ NIR colours are flatter than the
  optical ($U,B,V$) ones due to the increase of the infrared
  luminosity following the develop of an extended asymptotic giant
  branch. We also find a dimming in ${\cal M}/L_K$ at the age
  corresponding to the AGB phase-transition confirming the result
  found by \citet{Maraston05} for higher metallicity.

  \begin{figure*}[t]
  \center
  \includegraphics[width=6cm]{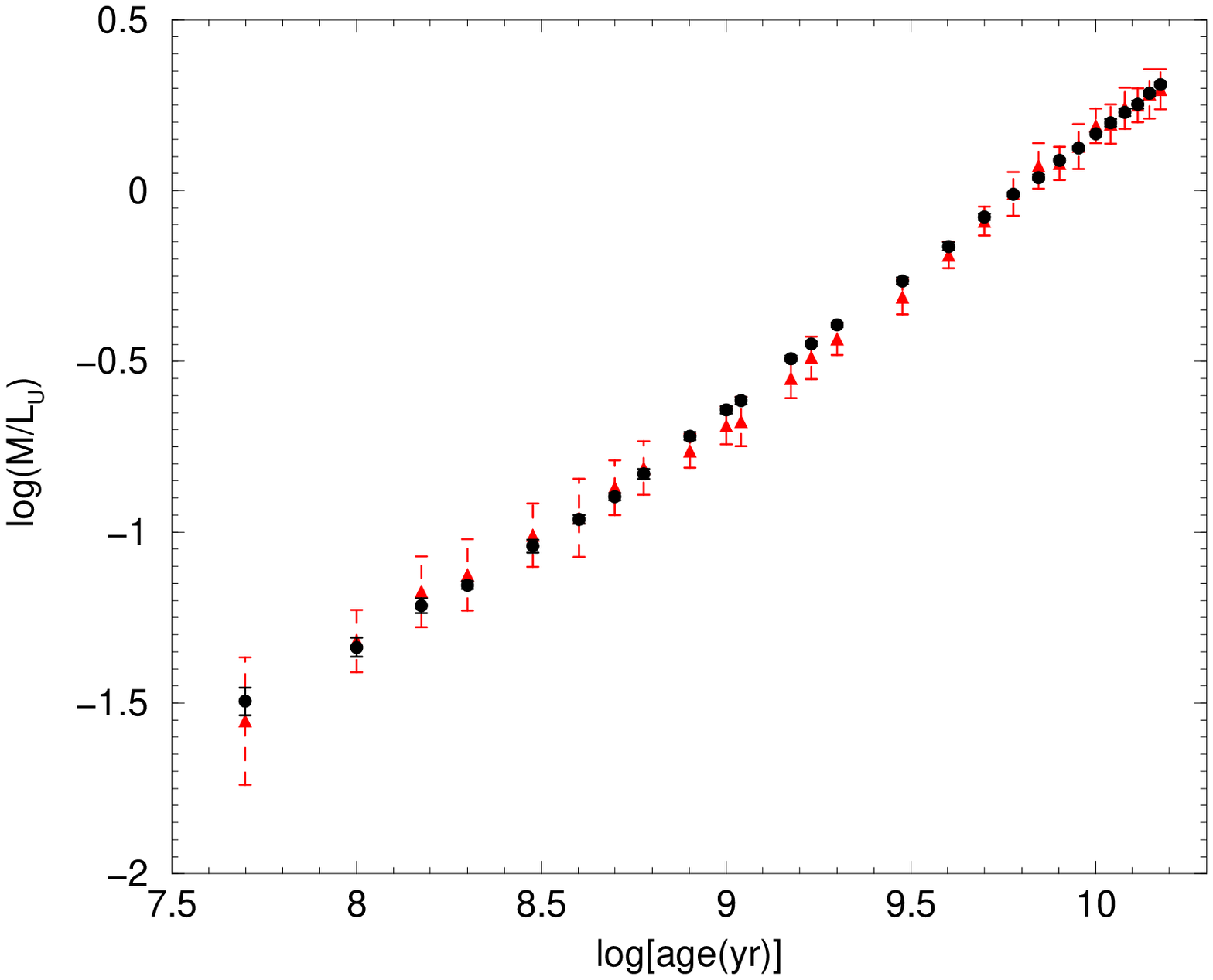}
  \includegraphics[width=6cm]{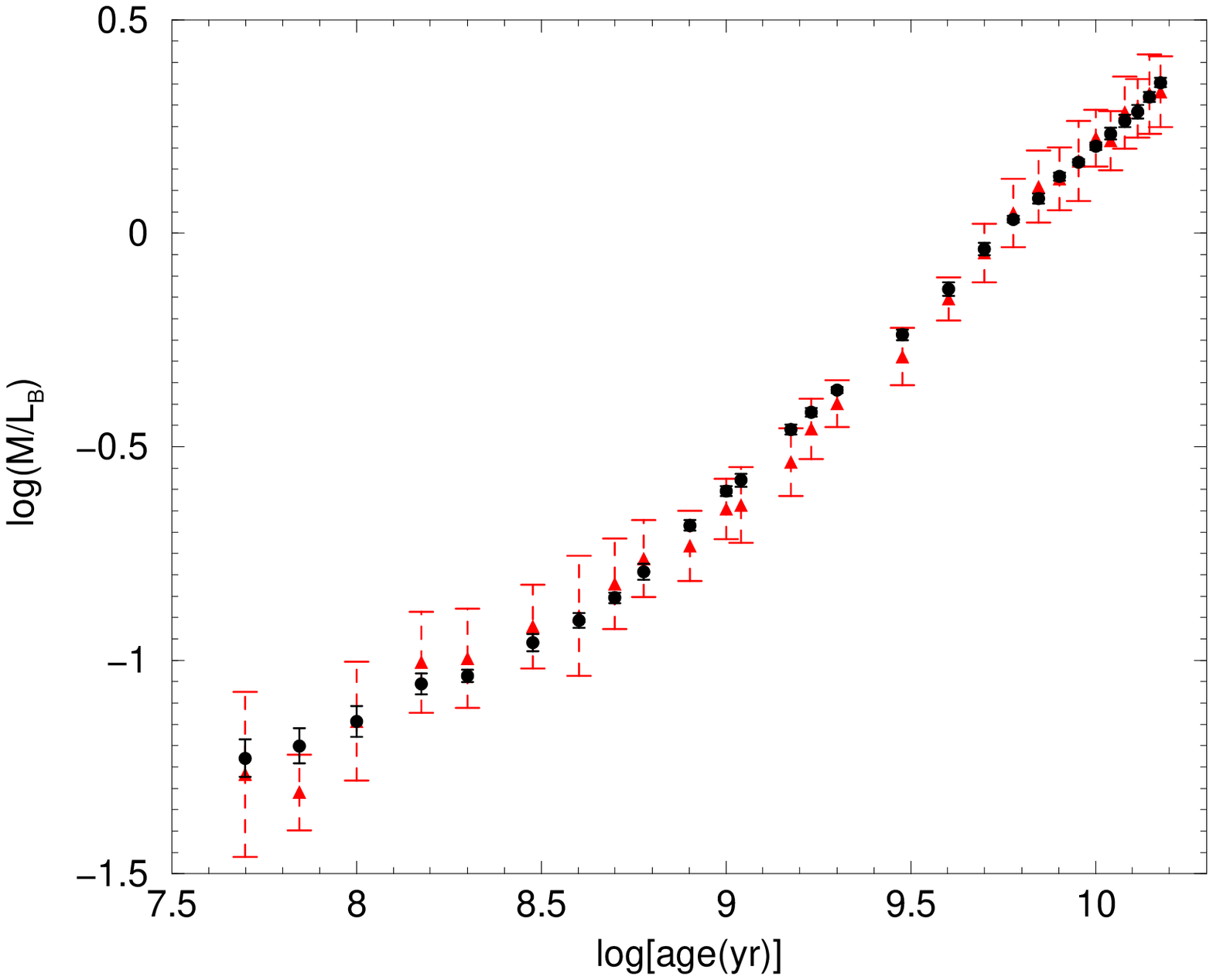}
  \includegraphics[width=6cm]{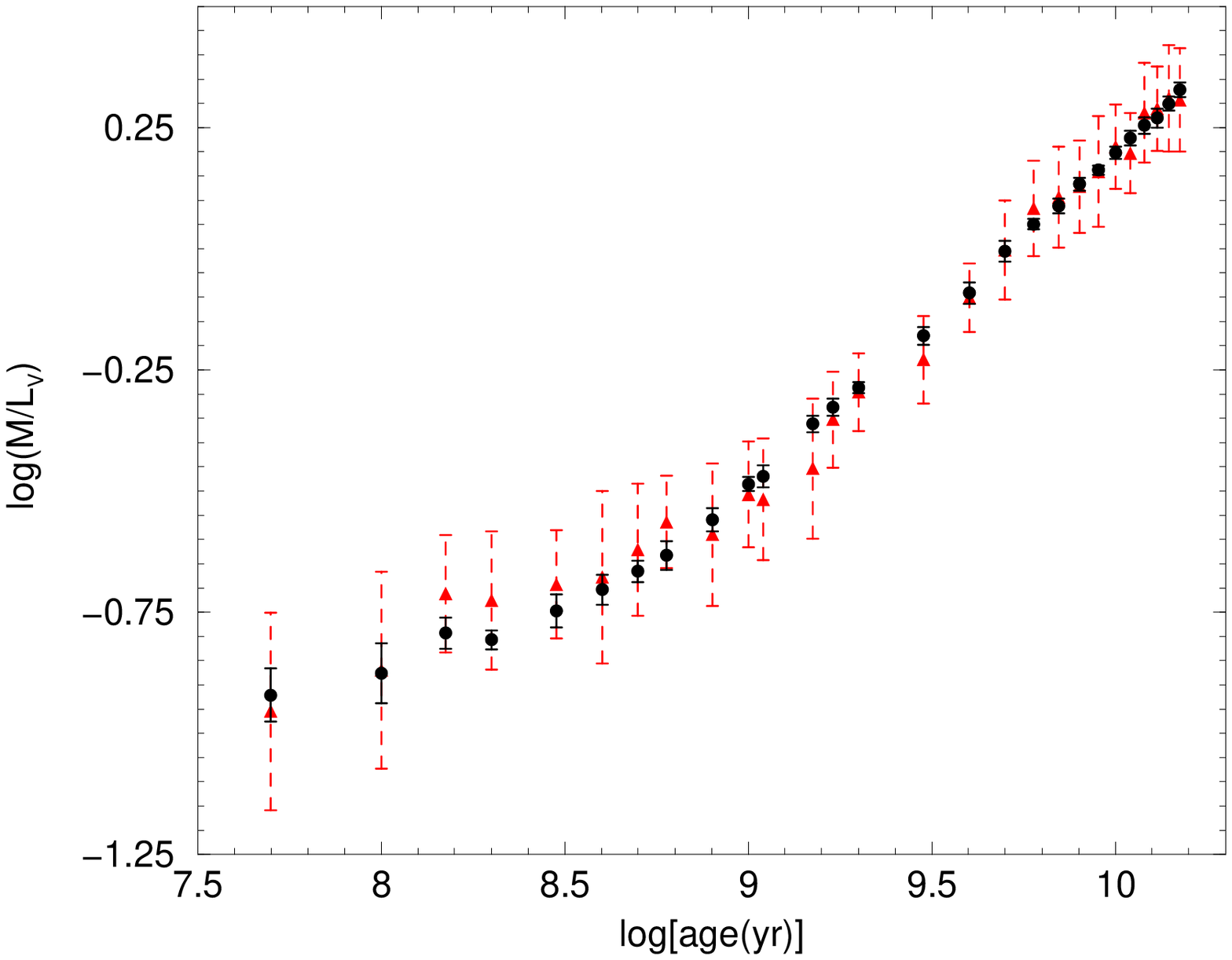}
  \includegraphics[width=6cm]{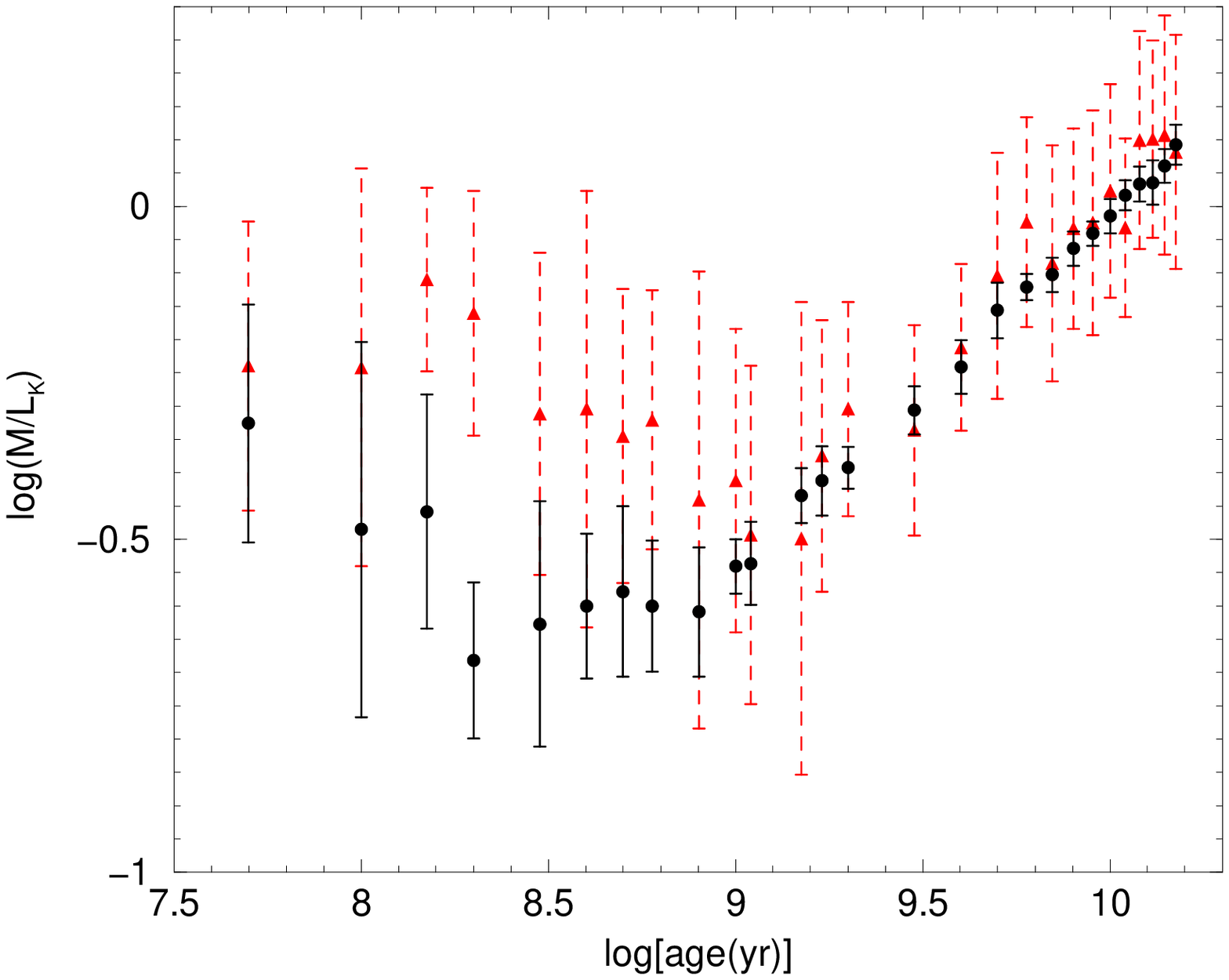}
  \label{fig:MLstd} \caption{Time behaviour of the ${\cal M}/L$ ratio (in solar units) for selected
  passbands ($U,B,V,K$). For each model $1\, \sigma$ dispersion is shown.
  Red triangles (dashed line) indicate models with $ M_V^{tot}=-4$, and black dots (continuos line)
  models with $ M_V^{tot}-8$.  }
  \label{fig:MLstd}
  \end{figure*}

  \begin{figure*}[ht]
  \center
  \includegraphics[width=8cm]{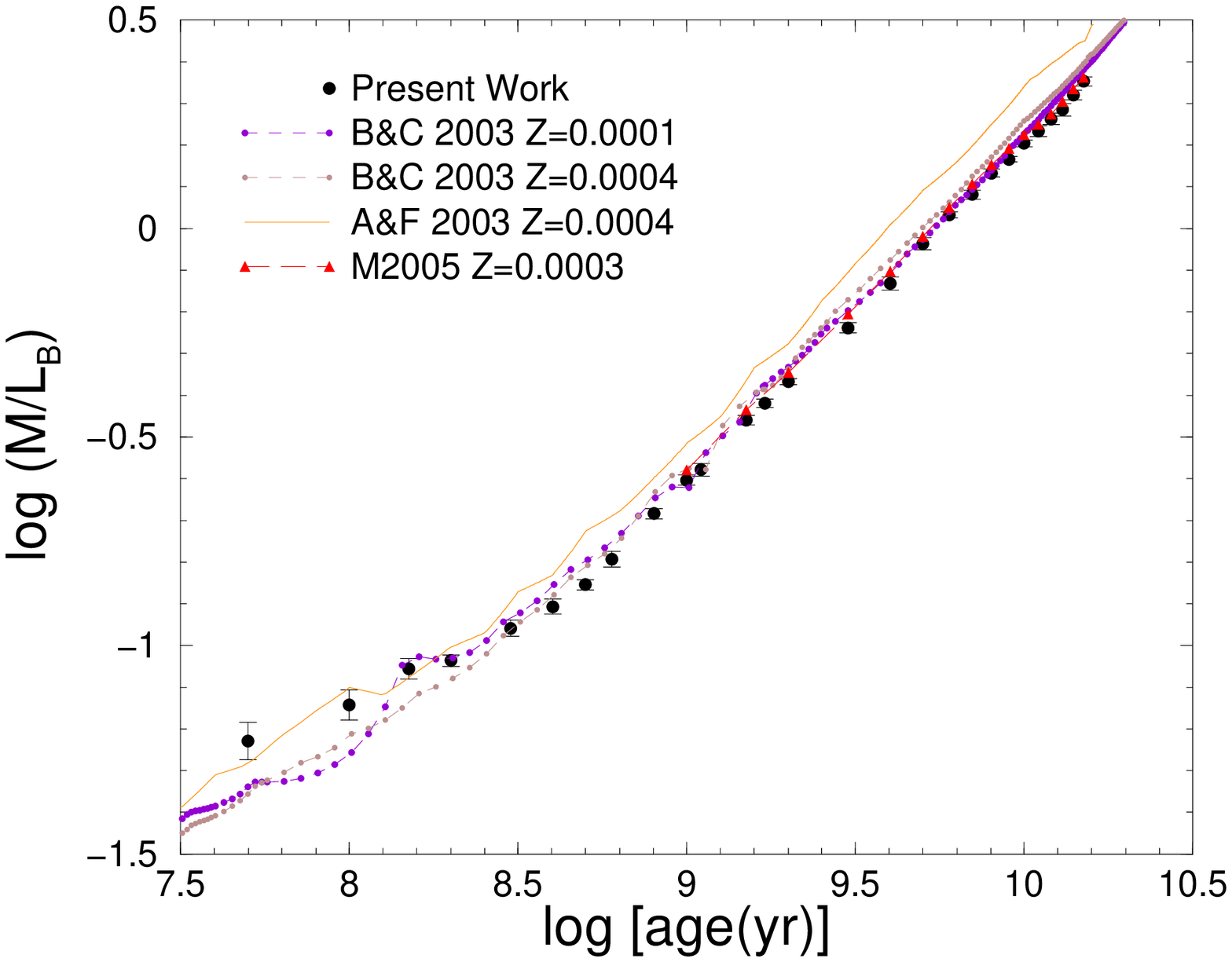}
  \includegraphics[width=8cm]{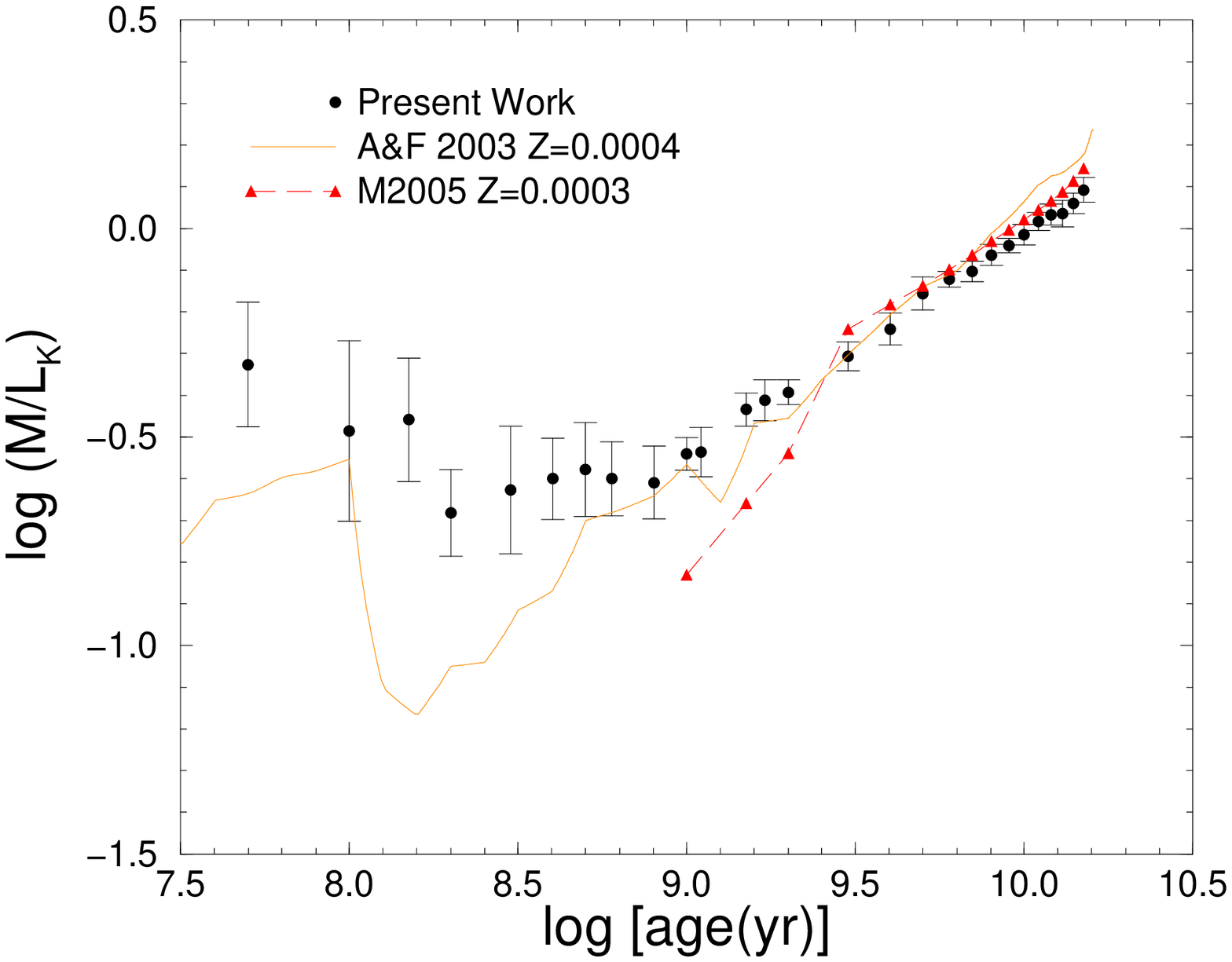}
  \label{fig:BruLM} \caption{Time evolution of ${\cal M}/L_B$
  (left panel) and ${\cal M}/L_K$ (right panel)
  in our reference model (black dots) compared  with
  the \citet[][B\&C 2003]{Bruzual&Charlot03} for models with $Z=0.0001$ (violet dashed
  line), and $Z=0.0004$ (brown dashed line), Anders \& Fritze-v. Alvensleben
  (2003, A\&F 2003) models (orange solid
  line), and \citet[][M2005]{Maraston05} for $Z=0.0003$ (red triangles, dashed line).
  The IMF adopted by various authors are
  reported in Table~\ref{table:ingredients}. }
  \label{fig:cfrBC03}
  \end{figure*}

  \input{4197.tab6.tex}

  Finally, Fig. \ref{fig:cfrBC03} compares the present values with the
  ones by \citet{Bruzual&Charlot03}, \citet{Anders+03}, and
  \citet{Maraston05} in two photometric bands, namely $B$ and $K$.
  Concerning the $B$-band, the agreement with
  \citet{Bruzual&Charlot03} and \citet{Maraston05} is very
  satisfactory, while \citet{Anders+03} predict slightly higher values
  at $\log age \gsim 8.5$.

  A more complex behaviour is shown by ${\cal M}/L_K$: at $\log age
  \gsim 9.3$ the ${\cal M}/L_K$ is independent of the model, while
  model-dependent at younger ages. This reflects the great
  uncertainties in simulating the AGB phase we have already discussed
  for colours.

  \begin{figure*}[t]
  \center
  \includegraphics[width=6cm]{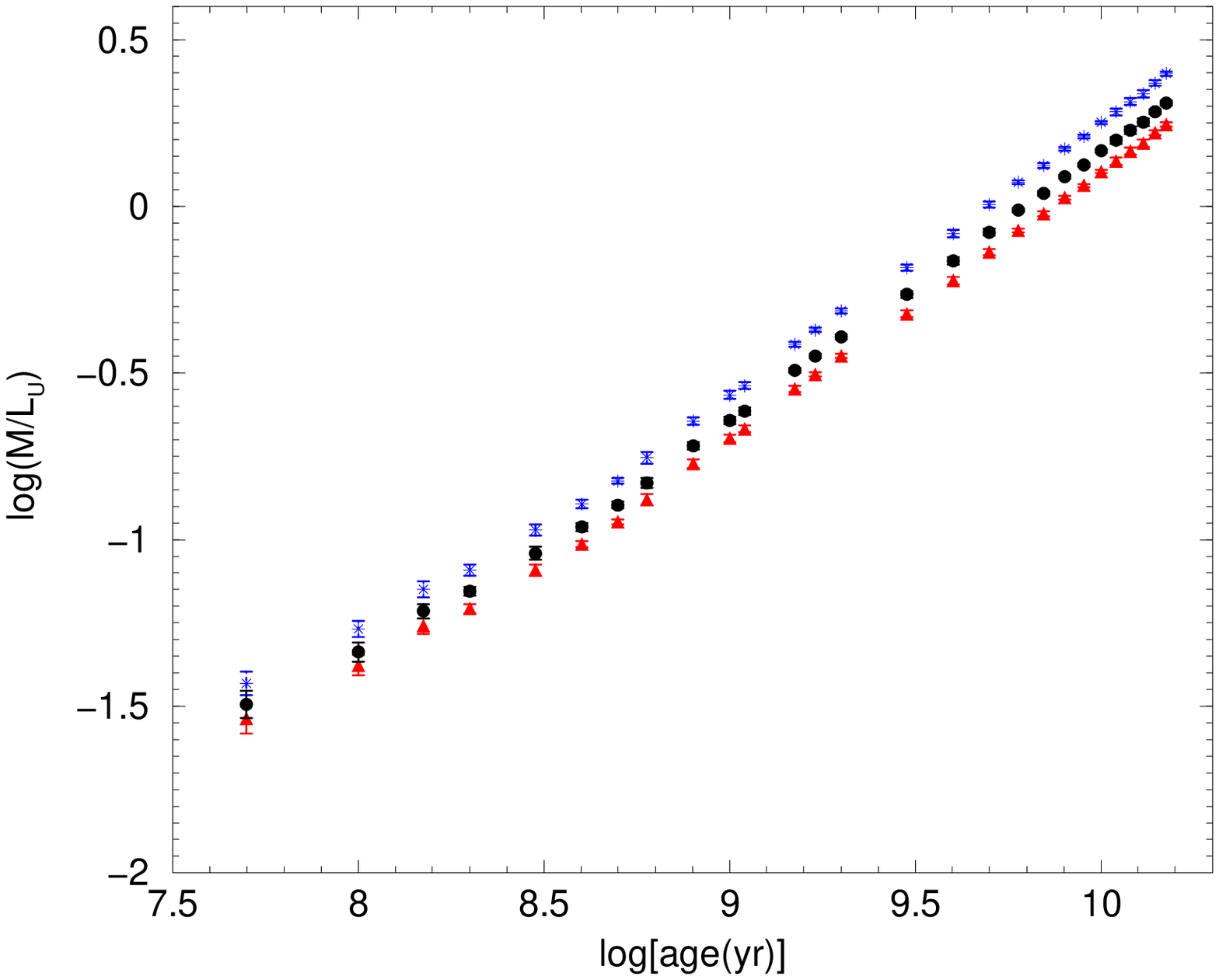}
  \includegraphics[width=6cm]{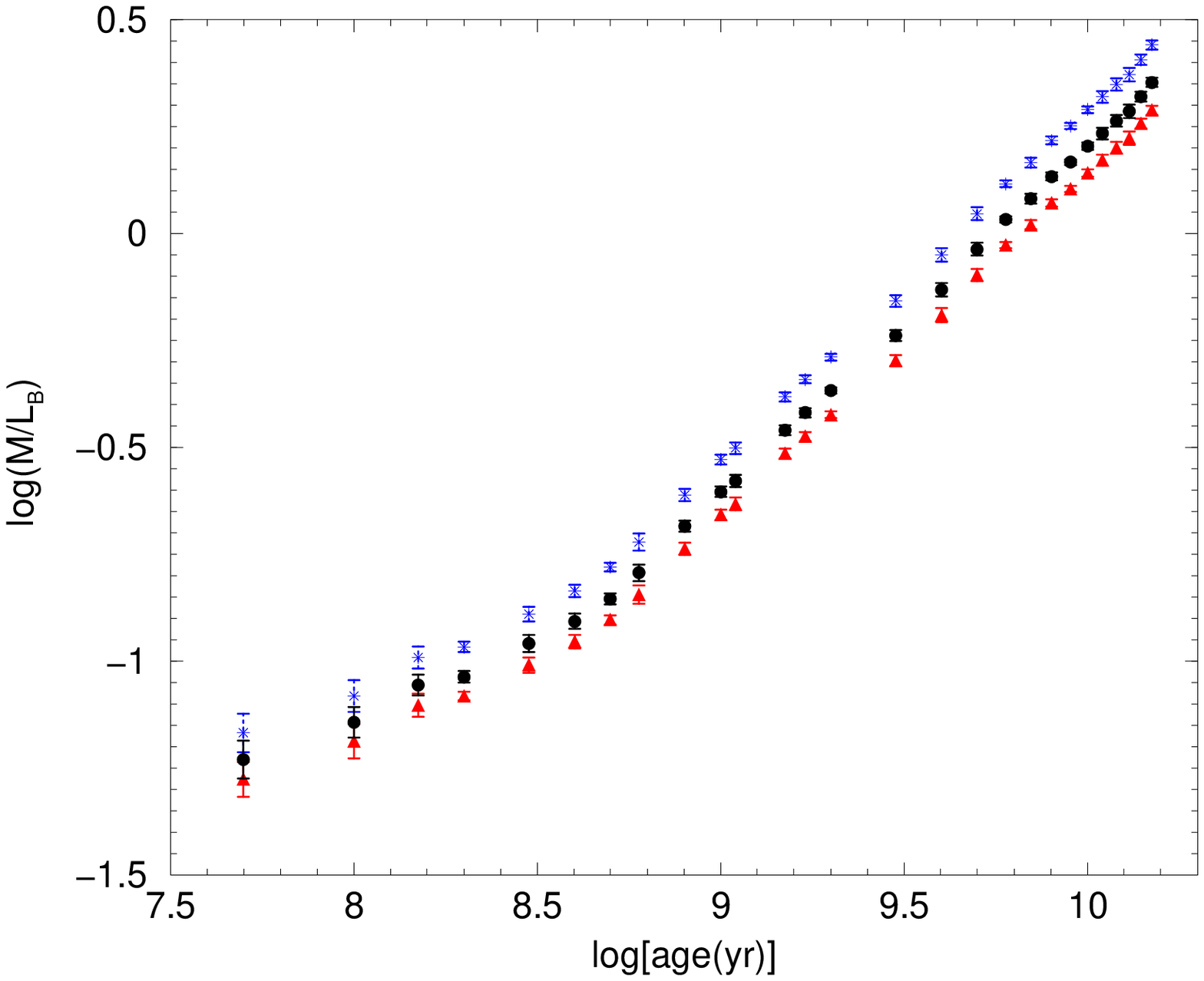}
  \includegraphics[width=6cm]{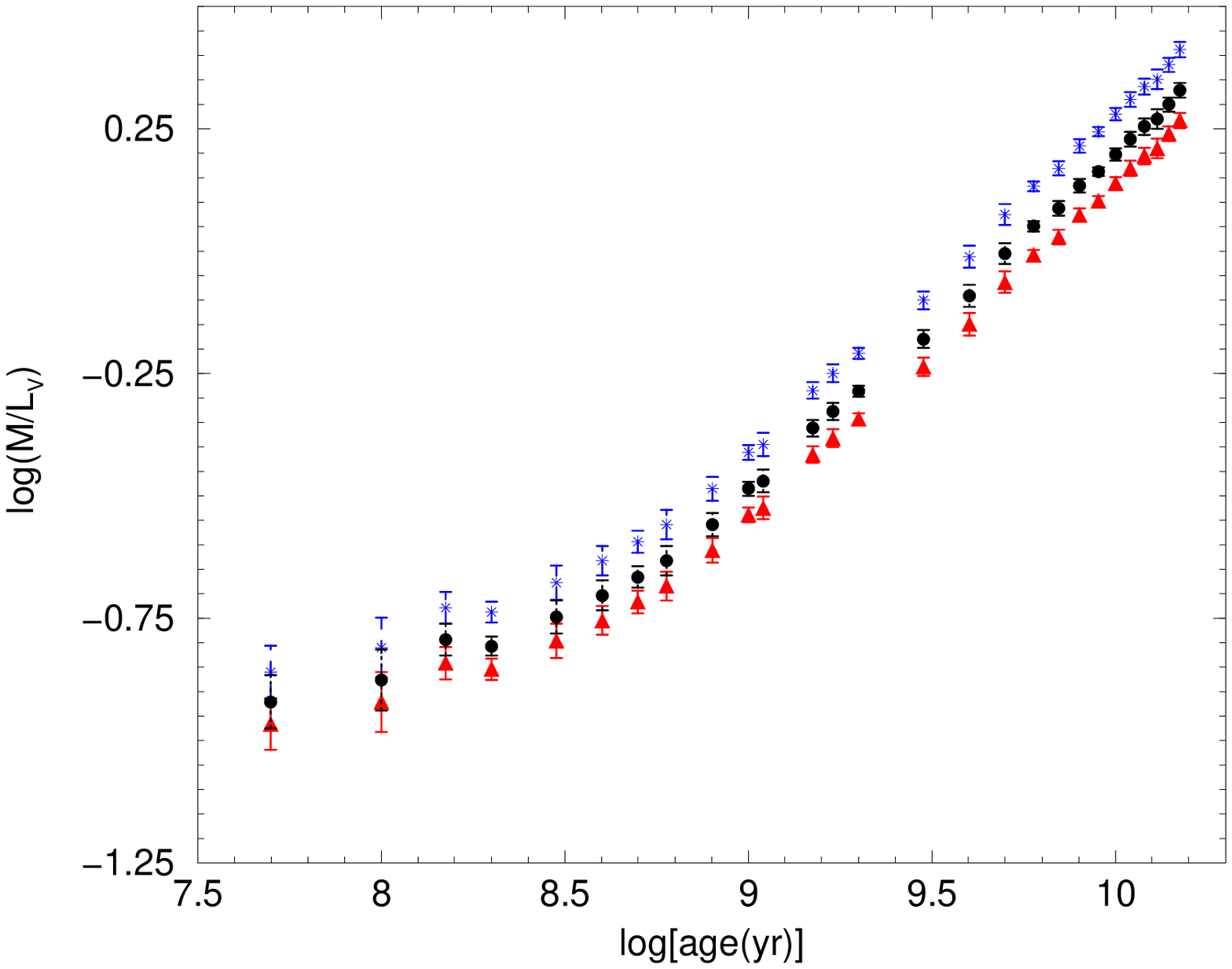}
  \includegraphics[width=6cm]{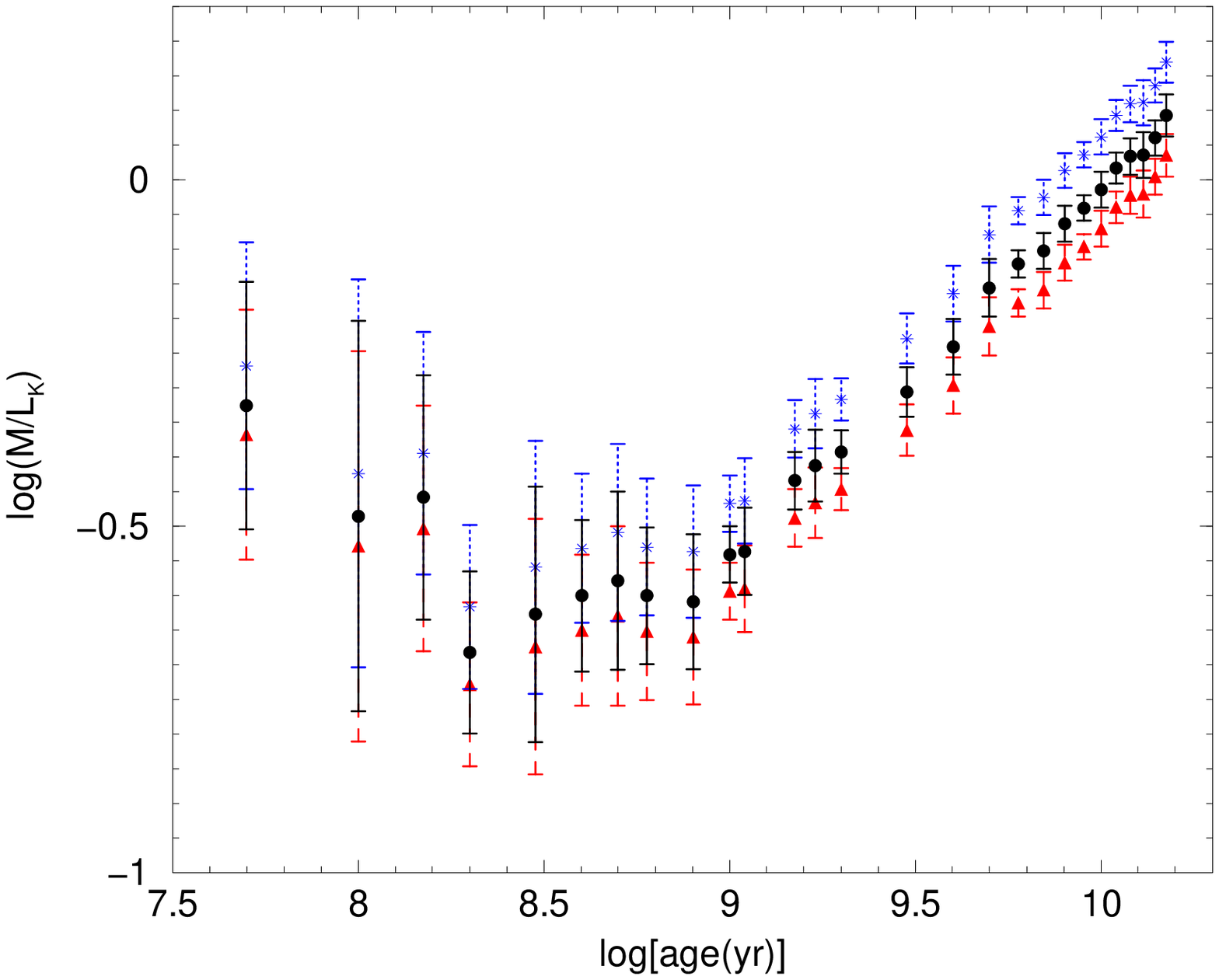}
  \caption{Time evolution of ${\cal M}/L$ (in
    solar units) in selected passbands ($U, B, V, K$) for models with $M_V^{tot}=-8$ calculated with the lower,
    central and upper value for the \citet{Kroupa02} IMF exponent for $M \leq
    0.5 M_{\odot}$. Black dots indicate our reference model ($\alpha=1.3$), red triangles
    indicate models with x=0.8 and blue stars models with $\alpha =1.8$.
    }
  \label{fig:MLlowmass}
  \end{figure*}

  \begin{figure*}[th]
  \center
  \includegraphics[width=6cm]{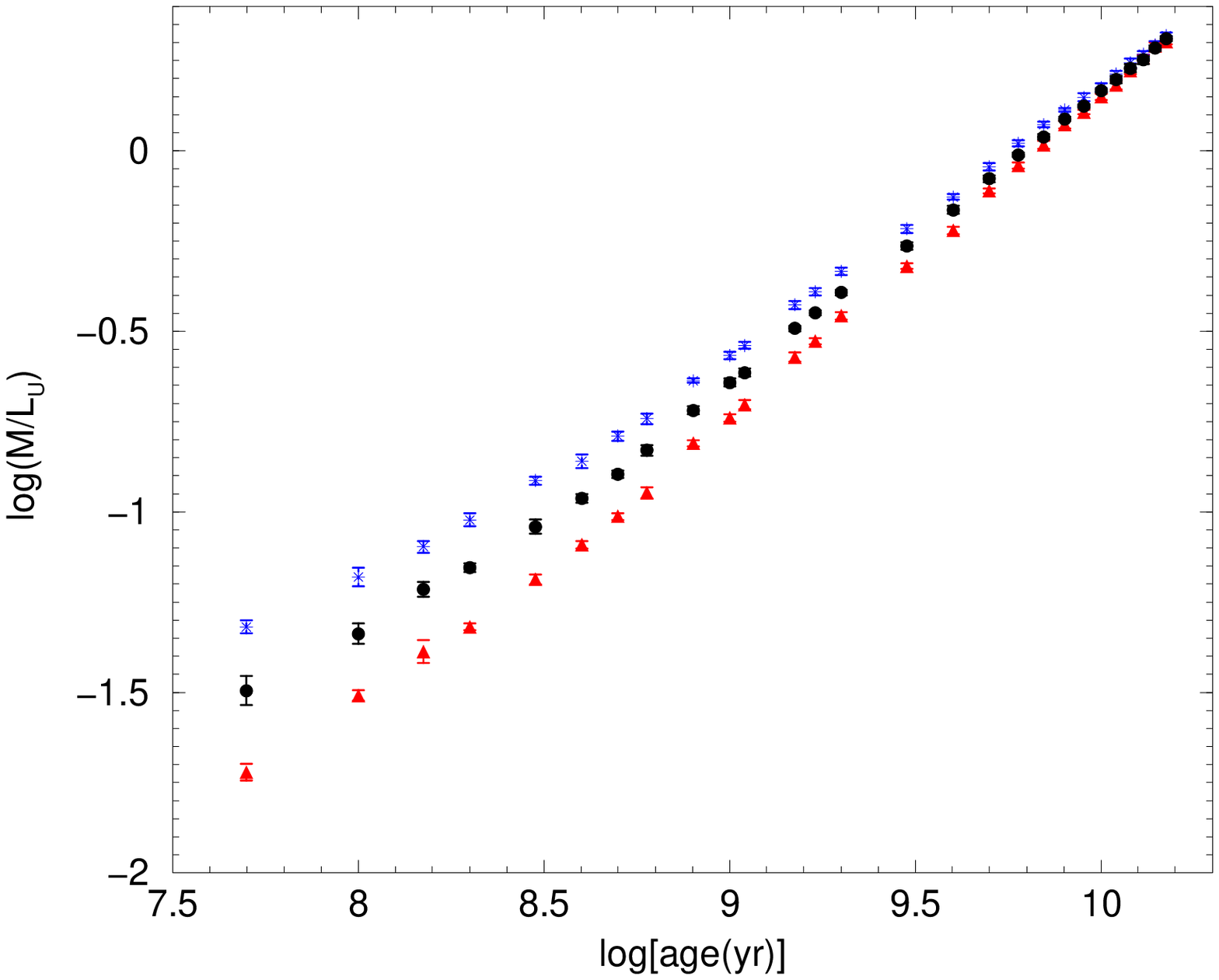}
  \includegraphics[width=6cm]{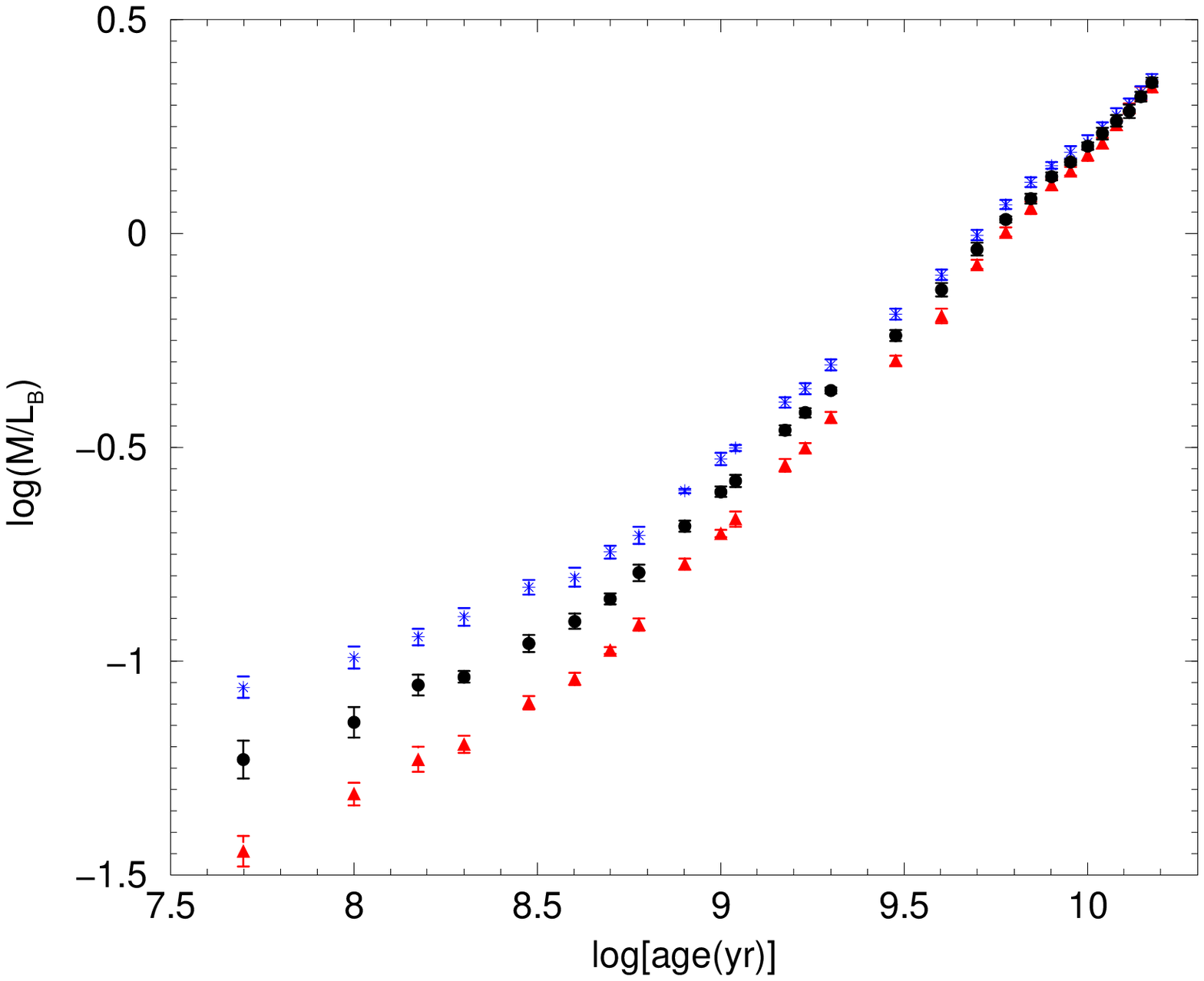}
  \includegraphics[width=6cm]{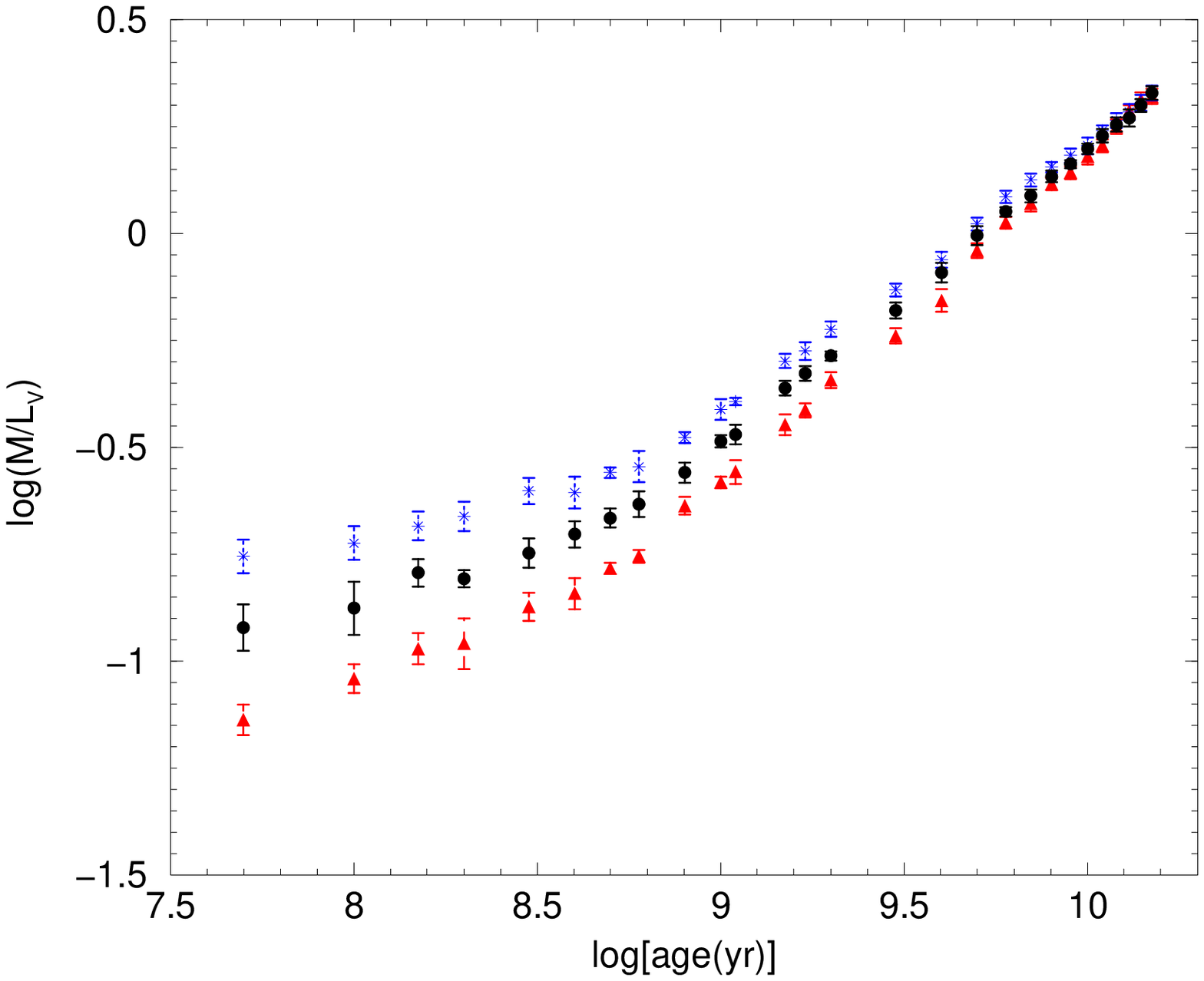}
  \includegraphics[width=6cm]{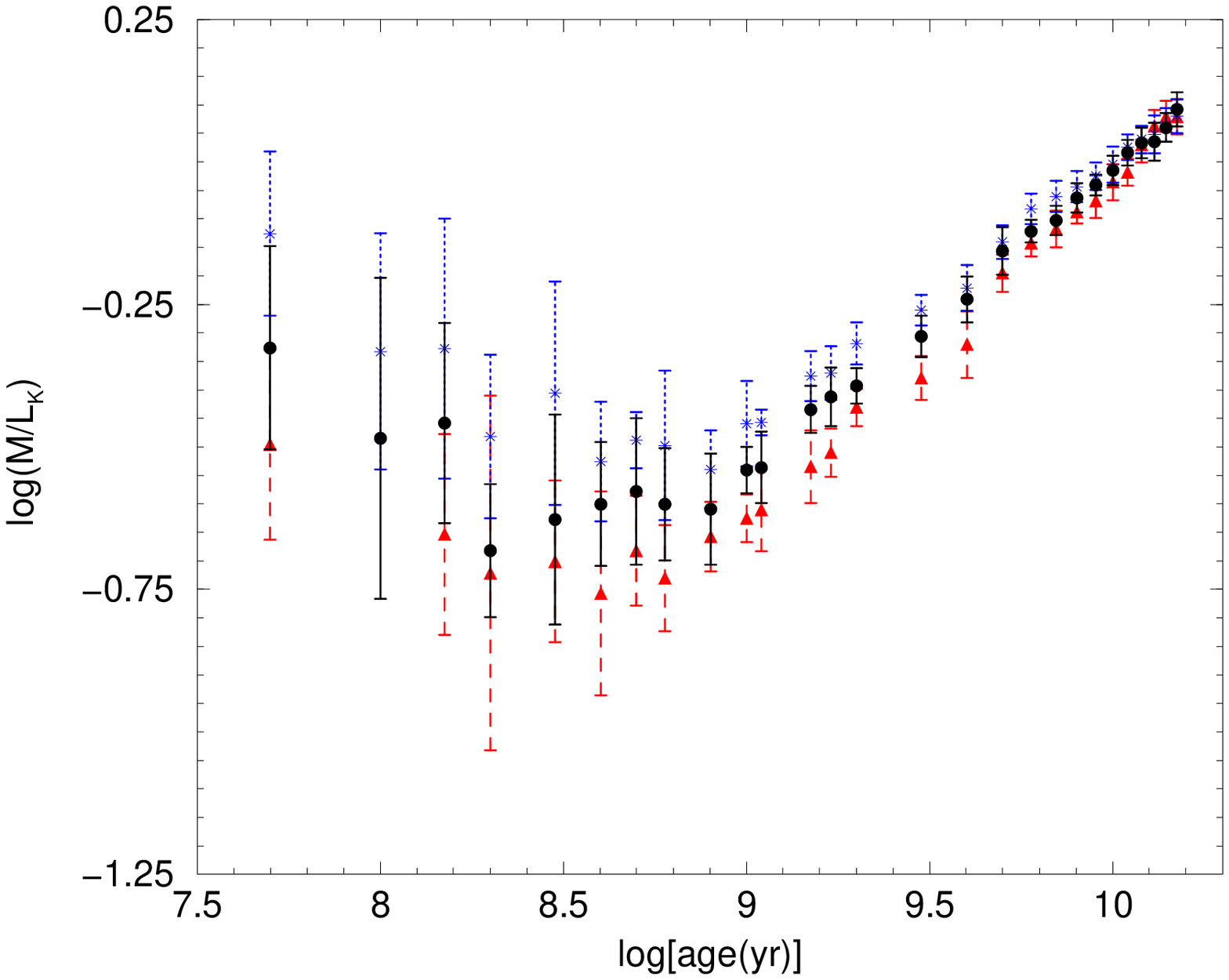}
  \caption{Time evolution of ${\cal M}/L$ (in
    solar units) in selected passbands ($U, B, V, K$)
    for models with $M_V^{tot}=-8$ calculated with the lower,
    central and upper value for the \citet{Kroupa02} IMF exponent for $M \geq
    0.5 M_{\odot}$. Black dots indicate our reference model ($\alpha =2.3$), red triangles
    indicate models with $\alpha =2.0$, and blue stars indicate models with $\alpha =2.6$.}
  \label{fig:MLhighmass}
  \end{figure*}

  \subsection{IMF variations}

  To evaluate the effect of the IMF shape on ${\cal M}/L$ we changed
  the IMF exponents within the estimated uncertainty \citep{Kroupa02}.
  Fig. \ref{fig:MLlowmass} and Fig. \ref{fig:MLhighmass} show the
  effect of changing the IMF exponents with respect to the Kroupa's
  formulation for masses higher and lower than 0.5 M$_{\odot}$,
  respectively. Models with $M_V^{tot}=-8$ mag are plotted in selected
  passbands while results for other passbands and for $M_V^{tot}=-4$
  mag are available at CDS.

  The ${\cal M}/L$ ratio is more sensitive than integrated colours to
  the IMF shape, as already noted by \citet{Maraston98,
  Bruzual&Charlot03, Maraston05}.  From Fig. \ref{fig:MLlowmass} it is
  evident that, by decreasing the IMF slope at $M\leq 0.5M_{\sun}$,
  the mass of the system decreases, while the luminosity remains
  almost constant. The final effect is to have smaller ${\cal M}/L$
  ratios than in the \emph{standard} models ($\alpha = 1. 3$).  At age
  lower than few billion years, the uncertainty due to stochastic
  effects is comparable or larger than the effects of exponent
  variations. Indeed, the IMF variations can be appreciated at ages of
  the order of $10^{10}$ yr.  We also find that the uncertainty
  predicted for $M_V^{tot}=-4$ models prevents to detect significant
  variations at any age.

  A different situation is found in Fig.  \ref{fig:MLhighmass}. The
  mass-to-light ratios slightly depend on IMF for age younger than a
  few billion years, while they are nearly insensitive at old
  age. This is because an IMF flatter than $\alpha = 2.3$ at $M \geq
  0.5$ (red triangles) requires a total mass smaller than in the
  \emph{standard} case ($\alpha = 2.3$, black circles) to reach the
  given luminosity, thus predicting low ${\cal M}/L_V$.  The effect is
  enhanced in young stellar populations in which the total mass is
  dominated by massive stars. On the other hand, by increasing the
  exponent up to 2.6 (blue stars), the cluster mass needed to have
  $M_V=-8$ increases, leading to high ${\cal M}/L$ values.  Again,
  mass-to-light ratios in the NIR bands are more affected by
  stochastic phenomena.

  Finally, note that in the case of small variations ($\alpha = 2 \div
  2.6$) explored here, the contribution of massive remnants to the
  total mass does not significantly change, while it is shown to be
  effective if large variations ($\alpha = 0.5 \div 3.5$) are adopted,
  see for a discussion \citet{Maraston98}.

  \section{Summary and conclusions}
  \label{section:conclusion}

  In this work we have analyzed the intrinsic uncertainties due to
  stochastic effects on integrated colours and mass--to--light ratios
  of metal-poor stellar clusters as a function of the total visual
  magnitude.  The calculations are performed for three different
  values of $M_V^{tot}$ and for a fine grid of stellar ages.
  Statistical errors are shown to be crucial, especially in the
  extreme case $M_V^{tot}=-4$ they are so high as to prevent precise
  quantitative evaluations for all ages. Calculations are done in the
  standard UBVRIJHK photometric passbands and in the Hubble Space
  Telescope bands (WFPC2 and NICMOS systems).

  We checked the consistency of our models on observational properties
  of three metal poor clusters, namely M68, M15, and M30, which differ
  mainly in the absolute visual magnitude, and HB morphology.  For
  each cluster we were able to reproduce both the features of the
  observed CMD and the integrated colours, showing that the HB
  morphology does not influence the photometric indices taken into
  account.

  A comparisons with recent results available in the literature shows,
  in some cases, non-negligible differences due to the large variety
  of prescriptions used in the model calculations.

  The uncertainties in the results on both colours and mass--to--light
  ratios due to the still present uncertainty on the IMF shape have
  been quantitatively estimated; while the colours fluctuations
  remains within theoretical uncertainties the ${\cal M}/L$ ratio is
  more sensitive to the IMF shape.  We also showed that the influence
  on ${\cal M}/L$ of the adopted value for the minimum mass for which
  carbon burning is ignite is quite negligible.

  \begin{acknowledgements}

  This work is dedicated to the memory of Vittorio Castellani,
  whose advise always supported us, and whose enthusiasm and
  dedication in challenging the astrophysics questions are an
  invaluable example for all of us. We warmly
  thank E. Brocato, P.G. Prada Moroni and S.N. Shore for useful
  discussions and for a careful reading of the manuscript.  We are
  grateful to the anonymous referee for her/his suggestions and
  comments that greatly improved the paper.  Financial support for
  this work was provided by MIUR--COFIN 2003.  This work made use of
  computational resources granted by the Consorzio di Ricerca del Gran
  Sasso according to the Progetto 6 {\it 'Calcolo Evoluto e sue
  Applicazioni (RSV6)'}--Cluster C11/B.  This paper utilizes the
  HST--snapshot database by the Globular Cluster Group of the Padova
  Astronomy Department, and the Catalog of parameters for Milky Way
  Globular Clusters by W.E. Harris.

 \end{acknowledgements}

 \end{document}

%% file: 4197.tab1.tex
\begin{table*}[t]
 \begin{center}
 \small
 \caption{Observed and theoretical integrated colours for the selected globular clusters. Values from
 the Harris' catalogue are dereddened according to Reed et al.
 (1988). The $J-K$ near--IR colour is from Brocato et al.
 (1990), dereddened according to Cardelli et al. (1989). ${\cal
 M}/L_V$ is from the compilation by Pryor \& Meylan (1993).}
 \label{table:GGC}
 \begin{footnotesize}
 \begin{tabular}{cccccccc}
 \hline \hline
 Identity & $M_V^{tot}$ & $U-B$ & $B-V$ & $V-R$& $V-I$ & $J-K$ & ${\cal M}/L_V$\\
 \hline \hline & \multicolumn{6}{c}{Observations} & \\
 \hline
 M68 & $-$7.35 & $-$0.01 $\pm$ 0.01 & 0.58 $\pm$ 0.03 & 0.43 & 0.88 $\pm$ 0.01 & $-$ & 1.6 \\
 M15 & $-$9.17 & $-$0.03 $\pm$ 0.02 & 0.58 $\pm$ 0.01 & $ -$ & 0.73 & 0.62 & 2.2 \\
 M30 & $-$7.43 & $-$0.00 $\pm$ 0.02 & 0.57 $\pm$ 0.04 & 0.39 $\pm$ 0.02 & 0.82 $\pm$ 0.04 & 0.54 & 2.5 \\
\hline \hline & \multicolumn{6}{c}{Models} & \\
 \hline
 M68 & $-$7.34 $\pm$ 0.05 & 0.00 $\pm$ 0.02 & 0.60 $\pm$ 0.02 & 0.42 $\pm$ 0.01 & 0.87 $\pm$ 0.02 & 0.57 $\pm$ 0.07 & 1.79 $\pm$ 0.07\\
 M15 & $-$9.18 $\pm$ 0.02 & 0.00 $\pm$ 0.01 & 0.60 $\pm$ 0.01 & 0.420 $\pm$ 0.003 & 0.89 $\pm$ 0.01 & 0.59 $\pm$ 0.02 & 1.58 $\pm$ 0.03\\
 M30 & $-$7.43 $\pm$ 0.05 & $-$0.01 $\pm$ 0.01 & 0.59 $\pm$ 0.02 & 0.41 $\pm$ 0.01 & 0.87 $\pm$ 0.02 & 0.54 $\pm$ 0.05 & 1.76 $\pm$ 0.07\\
 \hline \hline
 \end{tabular}
 \end{footnotesize}
 \end{center}
 \end{table*}

%% file: 4197.tab2.tex
 \begin{table*}[t]
 \small
 \begin{footnotesize}
 \caption {Integrated colours in the \emph{standard} \emph{UBVRIJHK} photometric filters for our \emph{standard} model with $M_V^{tot}=-8$ and Z=0.0002.}
\label{standardcol}
 \begin{tabular}{crrrrrrr}
 \hline \hline
 Age ($Myr$) & $M_V^{tot}$ & $U-B$ & $B-V$ & $V-R$ & $V-I$ & $V-J$ & $V-K$ \\
 \hline
     50. & $-$8.0 $\pm$ 0.1 & $-$0.56 $\pm$ 0.03 &   $-$0.10 $\pm$ 0.08 & $-$0.02 $\pm$ 0.08&    0.0 $\pm$ 0.2&      0.0  $\pm$ 0.3 & 0.1 $\pm$ 0.5\\
    100. & $-$8.0 $\pm$ 0.1 & $-$0.39 $\pm$ 0.03 &      0.00 $\pm$ 0.07 &   0.06 $\pm$ 0.07 &    0.2 $\pm$ 0.2 &     0.4 $\pm$ 0.3 & 0.7 $\pm$ 0.5\\
    150. & $-$8.0 $\pm$ 0.1 & $-$0.31 $\pm$ 0.02 &      0.01 $\pm$ 0.04 &   0.06 $\pm$ 0.04 &    0.2 $\pm$ 0.1 &     0.4 $\pm$ 0.2 & 0.7 $\pm$ 0.3\\
    200. & $-$8.00 $\pm$ 0.05 & $-$0.22 $\pm$ 0.01 &    0.09 $\pm$ 0.05 &   0.14 $\pm$ 0.05 &    0.4 $\pm$ 0.1 &     0.7 $\pm$ 0.2 & 1.2 $\pm$ 0.3\\
    300. & $-$8.01 $\pm$ 0.09 & $-$0.12 $\pm$ 0.02 &    0.14 $\pm$ 0.07 &   0.16 $\pm$ 0.07 &    0.4 $\pm$ 0.1 &     0.8 $\pm$ 0.2 & 1.3 $\pm$ 0.4\\
    400. & $-$8.06 $\pm$ 0.08 & $-$0.05 $\pm$ 0.01 &    0.16 $\pm$ 0.04 &   0.16 $\pm$ 0.04 &    0.41 $\pm$ 0.09 &   0.8 $\pm$ 0.2 & 1.3 $\pm$ 0.2\\
    500. & $-$8.02 $\pm$ 0.05 & $-$0.01 $\pm$ 0.01 &    0.19 $\pm$ 0.03 &   0.18 $\pm$ 0.03 &    0.44 $\pm$ 0.07 &   0.8 $\pm$ 0.1 & 1.3 $\pm$ 0.2\\
    600. & $-$8.06 $\pm$ 0.08 &    0.00 $\pm$ 0.01 &    0.26 $\pm$ 0.03 &   0.22 $\pm$ 0.03 &    0.52 $\pm$ 0.06 &   1.0 $\pm$ 0.1 & 1.4 $\pm$ 0.2\\
    800. & $-$8.03 $\pm$ 0.06 &    0.01 $\pm$ 0.01 &    0.34 $\pm$ 0.04 &   0.27 $\pm$ 0.03 &    0.62 $\pm$ 0.07 &   1.1 $\pm$ 0.1 & 1.7 $\pm$ 0.2\\
   1000. & $-$7.98 $\pm$ 0.04 &    0.00 $\pm$ 0.01 &    0.36 $\pm$ 0.02 &   0.28 $\pm$ 0.01 &    0.63 $\pm$ 0.03 &   1.13 $\pm$ 0.06 & 1.6 $\pm$ 0.1\\
   1100. & $-$8.02 $\pm$ 0.06 &    0.00 $\pm$ 0.01 &    0.39 $\pm$ 0.03 &   0.30 $\pm$ 0.02 &    0.66 $\pm$ 0.04 &   1.16 $\pm$ 0.07 & 1.68 $\pm$ 0.09\\
   1500. & $-$8.03 $\pm$ 0.04 &    0.01 $\pm$ 0.01 &    0.41 $\pm$ 0.02 &   0.30 $\pm$ 0.01 &    0.67 $\pm$ 0.02 &   1.17 $\pm$ 0.04 & 1.70 $\pm$ 0.07\\
   1700. & $-$8.00 $\pm$ 0.04 &    0.02 $\pm$ 0.01 &    0.42 $\pm$ 0.02 &   0.31 $\pm$ 0.01 &    0.68 $\pm$ 0.03 &   1.20 $\pm$ 0.06 & 1.73 $\pm$ 0.08\\
   2000. & $-$7.97 $\pm$ 0.03 &    0.03 $\pm$ 0.01 &    0.44 $\pm$ 0.01 &   0.32 $\pm$ 0.01 &    0.71 $\pm$ 0.02 &   1.24 $\pm$ 0.04 & 1.78 $\pm$ 0.05\\
   3000. & $-$7.98 $\pm$ 0.05 &    0.03 $\pm$ 0.01 &    0.49 $\pm$ 0.02 &   0.35 $\pm$ 0.01 &    0.75 $\pm$ 0.02 &   1.29 $\pm$ 0.03 & 1.83 $\pm$ 0.05\\
   4000. & $-$8.04 $\pm$ 0.06 &    0.01 $\pm$ 0.01 &    0.53 $\pm$ 0.02 &   0.37 $\pm$ 0.01 &    0.79 $\pm$ 0.02 &   1.34 $\pm$ 0.03 & 1.89 $\pm$ 0.05\\
   5000. & $-$8.01 $\pm$ 0.05 &    0.00 $\pm$ 0.01 &    0.55 $\pm$ 0.02 &   0.38 $\pm$ 0.01 &    0.80 $\pm$ 0.02 &   1.35 $\pm$ 0.04 & 1.89 $\pm$ 0.05\\
   6000. & $-$8.02 $\pm$ 0.03 & $-$0.02 $\pm$ 0.01 &    0.57 $\pm$ 0.01 &   0.39 $\pm$ 0.01 &    0.83 $\pm$ 0.01 &   1.39 $\pm$ 0.02 & 1.94 $\pm$ 0.03\\
   7000. & $-$8.03 $\pm$ 0.04 & $-$0.02 $\pm$ 0.01 &    0.60 $\pm$ 0.01 &   0.41 $\pm$ 0.01 &    0.85 $\pm$ 0.01 &   1.43 $\pm$ 0.02 & 1.99 $\pm$ 0.03\\
   8000. & $-$8.00 $\pm$ 0.03 & $-$0.02 $\pm$ 0.01 &    0.61 $\pm$ 0.01 &   0.41 $\pm$ 0.01 &    0.86 $\pm$ 0.01 &   1.44 $\pm$ 0.02 & 2.00 $\pm$ 0.03\\
   9000. & $-$7.99 $\pm$ 0.02 & $-$0.01 $\pm$ 0.01 &    0.61 $\pm$ 0.01 &   0.41 $\pm$ 0.01&    0.87 $\pm$ 0.01 &   1.45 $\pm$ 0.02 & 2.02 $\pm$ 0.02\\
  10000. & $-$7.94 $\pm$ 0.03 & $-$0.00 $\pm$ 0.01 &    0.61 $\pm$ 0.01 &   0.42 $\pm$ 0.01 &    0.87 $\pm$ 0.01 &   1.47 $\pm$ 0.02 & 2.04 $\pm$ 0.03\\
  11000. & $-$8.01 $\pm$ 0.04 &    0.00 $\pm$ 0.01 &    0.60 $\pm$ 0.01 &   0.42 $\pm$ 0.01 &    0.87 $\pm$ 0.01 &   1.46 $\pm$ 0.02 & 2.04 $\pm$ 0.02\\
  12000. & $-$7.99 $\pm$ 0.04 &    0.00 $\pm$ 0.01 &    0.61 $\pm$ 0.01 &   0.42 $\pm$ 0.01 &    0.88 $\pm$ 0.01 &   1.48 $\pm$ 0.02 & 2.06 $\pm$ 0.03\\
  13000. & $-$7.95 $\pm$ 0.05 &    0.01 $\pm$ 0.01 &    0.62 $\pm$ 0.02 &   0.43 $\pm$ 0.01 &    0.90 $\pm$ 0.02 &   1.51 $\pm$ 0.03 & 2.10 $\pm$ 0.04\\
  14000. & $-$8.00 $\pm$ 0.04 &    0.00 $\pm$ 0.01 &    0.63 $\pm$ 0.01 &   0.43 $\pm$ 0.01 &    0.90 $\pm$ 0.01 &   1.52 $\pm$ 0.02 & 2.11 $\pm$ 0.03\\
  15000. & $-$8.01 $\pm$ 0.04 & $-$0.02 $\pm$ 0.01 &    0.62 $\pm$ 0.02 &   0.43 $\pm$ 0.01 &    0.90 $\pm$ 0.01 &   1.51 $\pm$ 0.03 & 2.10 $\pm$ 0.04\\
 \hline \hline
 \end{tabular}
 \end{footnotesize}
 \end{table*}

%% file: 4197.tab3.tex
\begin{table*}
\small
\caption{As in Table \ref{standardcol} but for WFPC2 (first
three columns), and NICMOS1 (last two columns) photometric bands.
} \label{standardHST}
\begin{center}
\begin{tabular}{crrrrr}
\hline \hline
 Age ($Myr$) & $V-439W(2)$ & $V-F555W(2) $ & $V-F814W(2) $ & $V-F110W(1) $ & $V-F160W(1) $ \\
\hline
    50. & 0.09 $\pm$ 0.08      &     0.009 $\pm$    0.005 &     0.0 $\pm$   0.2  &    0.0 $\pm$    0.3 &    0.0 $\pm$    0.5\\
   100. &$-$0.01 $\pm$    0.08 &     0.002 $\pm$    0.005 &     0.2  $\pm$  0.2  &    0.3 $\pm$    0.3 &    0.6 $\pm$    0.5  \\
   150. &$-$0.02 $\pm$    0.03 &     0.000 $\pm$    0.002 &     0.2 $\pm$   0.1 &     0.3 $\pm$    0.1 &    0.6 $\pm$    0.3  \\
   200. &$-$0.11 $\pm$    0.05 &   $-$0.006 $\pm$    0.003 &    0.3 $\pm$   0.1 &     0.6 $\pm$    0.2 &    1.1 $\pm$    0.3  \\
   300. &$-$0.15 $\pm$    0.07 &   $-$0.009 $\pm$    0.004 &    0.4 $\pm$   0.1 &     0.6 $\pm$    0.2 &    1.2 $\pm$    0.4  \\
   400. &$-$0.18 $\pm$    0.05 &   $-$0.011 $\pm$    0.003 &    0.4 $\pm$   0.09 &     0.6 $\pm$    0.1 &   1.2 $\pm$    0.2  \\
   500. &$-$0.21 $\pm$    0.04 &   $-$0.013 $\pm$    0.002 &    0.42 $\pm$   0.07 &    0.7 $\pm$    0.1 &   1.2 $\pm$    0.2  \\
   600. &$-$0.28 $\pm$    0.04 &   $-$0.018 $\pm$    0.002 &    0.50 $\pm$   0.06 &    0.8 $\pm$    0.1 &   1.3 $\pm$    0.1 \\
   800. &$-$0.38 $\pm$    0.04 &   $-$0.023 $\pm$    0.002 &    0.60 $\pm$   0.07 &    0.9 $\pm$    0.1 &   1.5 $\pm$    0.1 \\
  1000. &$-$0.40 $\pm$    0.02 &   $-$0.024 $\pm$    0.001 &    0.61 $\pm$   0.03 &    0.93 $\pm$   0.05 &   1.53 $\pm$    0.08\\
  1100. &$-$0.43 $\pm$    0.03 &   $-$0.025 $\pm$    0.001 &    0.63 $\pm$   0.04 &    0.99 $\pm$   0.06 &   1.57 $\pm$     0.10\\
  1500. &$-$0.45 $\pm$    0.02 &   $-$0.026 $\pm$    0.001 &    0.64 $\pm$   0.02 &    0.97 $\pm$   0.04 &   1.58 $\pm$    0.06\\
  1700. &$-$0.46 $\pm$    0.02 &   $-$0.027 $\pm$    0.001 &    0.66 $\pm$   0.03 &    0.99 $\pm$   0.05 &   1.61 $\pm$    0.08\\
  2000. &$-$0.49 $\pm$    0.01 &   $-$0.028 $\pm$    0.001 &    0.68 $\pm$   0.02 &   1.03 $\pm$    0.03 &   1.66 $\pm$    0.05\\
  3000. &$-$0.55 $\pm$    0.02 &   $-$0.031 $\pm$    0.001 &    0.72 $\pm$   0.02 &   1.07 $\pm$    0.03 &   1.71 $\pm$    0.05\\
  4000. &$-$0.60 $\pm$    0.02 &   $-$0.033 $\pm$    0.001 &    0.76 $\pm$   0.02 &   1.12 $\pm$    0.03 &   1.77 $\pm$    0.04\\
  5000. &$-$0.61 $\pm$    0.02 &   $-$0.033 $\pm$    0.001 &    0.76 $\pm$   0.02 &   1.11 $\pm$    0.03 &   1.75 $\pm$    0.04\\
  6000. &$-$0.65 $\pm$    0.01 &   $-$0.035 $\pm$    0.001 &    0.89 $\pm$   0.01 &   1.17 $\pm$    0.02 &    1.82 $\pm$   0.02\\
  7000. &$-$0.67 $\pm$    0.02 &   $-$0.035 $\pm$    0.001 &    0.81 $\pm$   0.02 &   1.18 $\pm$    0.03 &    1.84 $\pm$   0.04\\
  8000. &$-$0.69 $\pm$    0.01 &   $-$0.036 $\pm$    0.001 &    0.83 $\pm$   0.01 &   1.21 $\pm$    0.02 &   1.88 $\pm$    0.03\\
  9000. &$-$0.70 $\pm$    0.02 &   $-$0.036 $\pm$    0.001 &    0.83 $\pm$   0.01 &   1.22 $\pm$    0.02 &   1.90 $\pm$    0.03\\
 10000. &$-$0.70 $\pm$    0.01 &   $-$0.036 $\pm$    0.001 &    0.84 $\pm$   0.01 &   1.23 $\pm$    0.02 &   1.92 $\pm$    0.03\\
 11000. &$-$0.69 $\pm$    0.01 &   $-$0.035 $\pm$    0.001 &    0.84 $\pm$   0.02 &   1.23 $\pm$    0.01 &   1.91 $\pm$    0.02\\
 12000. &$-$0.69 $\pm$    0.01 &   $-$0.035 $\pm$    0.001 &    0.85 $\pm$   0.01 &   1.25 $\pm$    0.02 &   1.94 $\pm$    0.03\\
 13000. &$-$0.71 $\pm$    0.02 &   $-$0.036 $\pm$    0.001 &    0.86 $\pm$   0.01 &   1.27 $\pm$    0.02 &   1.97 $\pm$    0.03\\
 14000. &$-$0.72 $\pm$    0.01 &   $-$0.036 $\pm$    0.001 &    0.87 $\pm$   0.01 &   1.28 $\pm$    0.02 &   1.98 $\pm$    0.03\\
 15000. &$-$0.72 $\pm$    0.02 &   $-$0.036 $\pm$    0.001 &    0.87 $\pm$   0.02 &   1.28 $\pm$    0.02 &   1.97 $\pm$    0.04\\
\hline \hline
\end{tabular}
\end{center}
\end{table*}

%% file: 4197.tab4.tex
\begin{table*}
 \caption{The physical ingredients adopted in the models plotted in
   Fig.\ref{fig:comparison}.
   The quoted values for the overshooting (OS)
   efficiency, expressed in unit of pressure-scale-height above the
   Schwarzschild convective border, are only indicative because different sets
   of evolutionary tracks adopt different prescriptions for the overshooting
   mechanisms and for the minimum mass at which the overshooting is fully
   efficient (see the related papers for more details); Y=YES; N=NO.
   WD indicates if white dwarfs are included in the
   calculations; ${\cal M}/L$ if mass--to--light ratio values
   are available. In the paper by Kurth et al. (1999) the
   normalization procedure is not described. In the paper by  Anders \&
   Fritze-v. Alvensleben (2003), the SSPs have been
   evolved to the selected ages starting
   with an initial stellar mass of $10^6$M$_{\odot}$;
   the adopted value for $M_{up}$ is not specified.
   $^1$ Detailed information on the adopted
   evolutionary tracks can be found in the paper quoted in Col. 1. $^2$ Latest
   version available on the web, see Maraston (2005). $^3$ Sinthetic models
   are also calculated for Salpeter (1955) IMF. $^4$ See discussion in Maraston (2005). $^5$
   Carbon stars included.  
    }
 \label{table:ingredients} \tiny
 \begin{tabular}{clllllllll}
 \hline \hline
 Authors &Evolutionary &Stellar Spectral &IMF & Normalization &Thermal Pulses & OS & $M_{up}$ & WD &${\cal M}/L$\\
                 &Tracks &Library & & & &[H$_{\mathrm p}$] & [M$_{\odot}$] & & \\
 \hline
& & & & & & & &\\
 Kurth et al. & Padua Lib. 1994$^{1}$ & Lejeune et al. & Salpeter & --- & N & $\approx$0.2 & 5 & N & N \\
    (1999) & + Chabrier \& Baraffe & (1997, 1998) & (1955) &  & & & & \\
   & (1997) &  & & &  & & & &\\
 Brocato et al. & Frascati Lib. & Kurucz (1993) & Scalo & N. of stars & Groenewegen \& & N & 5 & Y &N\\
 (2000) & 1991-2000$^{1}$ & Castelli et al. (1997) & (1986) & \& $M_V^{tot}$ & de Jong (1993) & & & &\\
 & & & & & & & \\
 Girardi et al. & Padua Lib. & Kurucz (1992) & Salpeter &
 $1M_{\odot}$ & Groenewegen \& & $\approx$0.25 & 5 & Y &N\\
     (2000) & 1994-2000$^{1}$ & &(1955) & & de Jong (1993) & & & & \\
                & & & & & + Marigo (1998) & & & &\\
 & & & & & & &\\
 Zhang et al. & Pols et al. (1998) & Lejeune et al. & Kroupa et al. &
 $1M_{\odot}$ &  Vassiliadis \& Wood & 0.22$\div$0.4 & 5 & Y & N\\
 (2002) & & (1997, 1998) & (1993) & & (1993) & & & & \\
  & & & & & & & &\\
Anders \& & Padua Lib. 1994-2000$^{1}$ & Lejeune et al. & Salpeter & Initial  & Groenewegen \& & $\approx$0.25 & - & Y & Y \\
 Fritze-v. Alvensleben & + Chabrier \& Baraffe & (1997, 1998) & (1955) & cluster mass & de Jong (1993)& & & & \\
 (2003)  & (1997) &  & & $1.6\cdot 10^9$M$_{\odot}$ & & & & &\\
  & & & & & & & &\\
 Bruzual \& Charlot & Padua Lib. 1994$^{1}$ & Westera et al. & Chabrier &
 $1M_{\odot}$& Vassiliadis \& Wood$^5$ & $\approx$0.2 & 5 & Y & Y\\
 (2003) & + Baraffe (1998) & (2002) &(2003) & & (1993) & & & & \\
  & & & & & & & &\\
Maraston & Cassisi et al. & Lejeune et al. & Kroupa    & $1M_{\odot}$   & Renzini & N & 8.5 & Y & Y\\
   (2005)& (2000)$^1$     & (1998)$^2$     &(2001)$^3$ & & (1992)$^{4,5}$ & & & &\\
 & & & & & & & &\\
Present work & Pisa Lib. 2004 & Castelli (1999) & Kroupa & $M_V^{tot}$ & Wagenhuber & N & 6.5 & Y & Y\\
                       & + Baraffe et al.& &(2002) & & \& Groenewegen & & & &\\
 & (1997) & & & & (1998) & & & &\\
& & & & & & & &\\
 \hline \hline
 \end{tabular}
 \end{table*}

%% file: 4197.tab5.tex
\begin{table*}[!tb]
\caption{Mass--to--light ratios for selected ages in several 
photometric passbands from our {\em standard} model with different
assumptions of $M_{up}$.}
\label{tab:Mup}
\begin{center}
\begin{tabular}{lccccccc}
\hline \hline
 & $M/L_V$ & $M/L_U$ & $M/L_B$ & $M/L_R$ & $M/L_I$ & $M/L_J$  &  $M/L_K$\\
\hline
$M_{up}=5\,M_{\odot}$ & & & & & & &\\
\hline
$500\,Myr$ & $0.21  \pm 0.01$ & $0.126  \pm  0.003$ & $0.139  \pm  0.004$ & $0.25  \pm  0.02$ & $0.28  \pm  0.04$ & $0.28 \pm   0.06$ & $0.26   \pm    0.08$ \\
$800\,Myr$ &$  0.28 \pm   0.02 $&$0.194    \pm   0.007 $&$0.210    \pm   0.008 $&$0.31    \pm   0.03 $&$0.31    \pm    0.04$&$0.29    \pm   0.05  $&$0.25    \pm   0.06 $\\
$3\,Gyr$  &$0.66   \pm   0.03 $&$0.55    \pm   0.01 $&$0.58    \pm   0.02 $&$0.67    \pm   0.03 $&$0.64    \pm   0.03$&$0.58    \pm   0.04  $&$0.50    \pm   0.04 $\\
$10\,Gyr$  &1.55$   \pm   0.04 $&$1.45    \pm   0.02 $&$1.58    \pm   0.03 $&$1.47    \pm   0.04 $&$1.33    \pm0.05   $&$1.15    \pm   0.05  $&$0.95    \pm   0.05 $\\
\hline
$M_{up}=6.5\,M_{\odot}$ & & & & & & &\\
\hline
$500\,Myr$ & 0.22 $\pm$ 0.01 & 0.127 $\pm$ 0.003 & 0.140 $\pm$ 0.004 & 0.26 $\pm$ 0.02 & 0.28 $\pm$ 0.03 & 0.29 $\pm$ 0.06 & 0.26 $\pm$ 0.08\\
$800\,Myr$ & 0.28 $\pm$ 0.01 & 0.191 $\pm$ 0.005 & 0.207 $\pm$ 0.006 & 0.30 $\pm$ 0.02 & 0.30 $\pm$ 0.03 & 0.28 $\pm$ 0.05 & 0.25 $\pm$ 0.06\\
$3\,Gyr$   & 0.66 $\pm$ 0.03 & 0.55 $\pm$ 0.01 & 0.58 $\pm$ 0.02 & 0.67 $\pm$ 0.03 & 0.64 $\pm$ 0.04 & 0.58 $\pm$ 0.04 & 0.49 $\pm$ 0.04\\
$10\,Gyr$  & 1.58 $\pm$ 0.05 & 1.47 $\pm$ 0.02 & 1.60 $\pm$ 0.03 & 1.49 $\pm$ 0.05 & 1.36 $\pm$ 0.06 & 1.17 $\pm$ 0.06 & 0.97 $\pm$ 0.06\\
\hline
$M_{up}=10\,M_{\odot}$ & & & & & & &\\
\hline
$500\,Myr$  &$0.22    \pm   0.01 $&$0.130    \pm   0.002 $&$0.143    \pm  0.003 $&$ 0.26    \pm   0.02 $&$0.28    \pm   0.04 $&$0.28    \pm   0.06  $&$0.26    \pm   0.08 $\\
$800\,Myr$  &$0.29    \pm   0.01 $&$0.199    \pm   0.005 $&$0.214    \pm   0.005 $&$ 0.31   \pm   0.02 $&$0.31    \pm   0.03 $&$0.29    \pm   0.05  $&$0.25    \pm   0.05 $\\
$3\,Gyr$    &$0.68    \pm   0.03 $&$0.56     \pm   0.01  $&$0.60     \pm   0.02  $&$ 0.69   \pm   0.04 $&$0.66    \pm   0.04 $&$ 0.60   \pm   0.04  $&$0.51    \pm   0.05 $\\
$10 \,Gyr$  &$1.58   \pm   0.05 $&$1.48    \pm   0.02 $&$1.61    \pm   0.04 $&$1.49    \pm   0.06 $&$1.35    \pm   0.06 $&$1.16    \pm   0.06  $&$0.96    \pm   0.06 $\\
\hline
\hline
\end{tabular}
\end{center}
\end{table*}

%% file: 4197.tab6.tex
\begin{table*}[t!]
 \caption{Theoretical ${\cal M}/L$ as a function of
 the age in different photometric bands (solar units) obtained
 for \emph{standard} models with $M_V^{tot}=-8$.}
 \label{standardML}
 \small
 \begin{center}
 \begin{tabular}{ccccccccc}
 \hline \hline
  Age ($Myr$)  &    ${\cal M}/L_V$  &    ${\cal M}/L_U$  &    ${\cal M}/L_B$  &    ${\cal M}/L_R$  &    ${\cal M}/L_I$  &    ${\cal M}/L_J$  &    ${\cal M}/L_K$\\
 \hline
     50.  &    0.12 $\pm$ 0.01  &    0.032 $\pm$ 0.003  &    0.06 $\pm$ 0.01  &      0.17 $\pm$ 0.02  &    0.24 $\pm$ 0.05  &    0.4  $\pm$ 0.1   &    0.5  $\pm$ 0.2\\
    100.  &    0.13 $\pm$ 0.02  &    0.046 $\pm$ 0.003  &    0.07 $\pm$ 0.01  &      0.18 $\pm$ 0.04  &    0.22 $\pm$ 0.07  &    0.3  $\pm$ 0.1   &    0.3 $\pm$ 0.2\\
    150.  &    0.16 $\pm$ 0.01  &    0.061 $\pm$ 0.003  &    0.088 $\pm$ 0.005&      0.21 $\pm$ 0.02  &    0.26 $\pm$ 0.04  &    0.33 $\pm$ 0.08  &    0.3 $\pm$ 0.1\\
    200.  &    0.16 $\pm$ 0.01  &    0.070 $\pm$ 0.002  &    0.092 $\pm$ 0.003&      0.19 $\pm$ 0.01  &    0.22 $\pm$ 0.03  &    0.23 $\pm$ 0.05  &    0.21 $\pm$ 0.06\\
    300.  &    0.18 $\pm$ 0.01  &    0.091 $\pm$ 0.004  &    0.110 $\pm$ 0.005&      0.22 $\pm$ 0.03  &    0.24 $\pm$ 0.05  &    0.25 $\pm$ 0.08  &    0.24 $\pm$ 0.10\\
    400.  &    0.20 $\pm$ 0.01  &    0.109 $\pm$ 0.003  &    0.124 $\pm$ 0.005&      0.24 $\pm$ 0.02  &    0.26 $\pm$ 0.04  &    0.27 $\pm$ 0.05  &    0.25 $\pm$ 0.06\\
    500.  &    0.22 $\pm$ 0.01  &    0.127 $\pm$ 0.003  &    0.140 $\pm$ 0.004&      0.25 $\pm$ 0.02  &    0.28 $\pm$ 0.03  &    0.29 $\pm$ 0.06  &    0.26 $\pm$ 0.08\\
    600.  &    0.23 $\pm$ 0.02  &    0.148 $\pm$ 0.005  &    0.16 $\pm$ 0.01  &      0.27 $\pm$ 0.02  &    0.28 $\pm$ 0.03  &    0.28 $\pm$ 0.05  &    0.25 $\pm$ 0.06\\
    800.  &    0.28 $\pm$ 0.01  &    0.191 $\pm$ 0.005  &    0.21 $\pm$ 0.01  &      0.30 $\pm$ 0.02  &    0.30 $\pm$ 0.03  &    0.28 $\pm$ 0.05  &    0.25 $\pm$ 0.05\\
   1000.  &    0.33 $\pm$ 0.01  &    0.23 $\pm$ 0.01    &    0.25 $\pm$ 0.01  &      0.35 $\pm$ 0.01  &    0.35 $\pm$ 0.02  &    0.33 $\pm$ 0.02  &    0.29 $\pm$ 0.03\\
   1100.  &    0.34 $\pm$ 0.02  &    0.24 $\pm$ 0.01    &    0.26 $\pm$ 0.01  &      0.36 $\pm$ 0.03  &    0.36 $\pm$ 0.03  &    0.33 $\pm$ 0.04  &    0.29 $\pm$ 0.04\\
   1500.  &    0.43 $\pm$ 0.02  &    0.32 $\pm$ 0.01    &    0.35 $\pm$ 0.01  &      0.46 $\pm$ 0.02  &    0.45 $\pm$ 0.03  &    0.42 $\pm$ 0.03  &    0.37 $\pm$ 0.03\\
   1700.  &    0.47 $\pm$ 0.02  &    0.36 $\pm$ 0.01    &    0.38 $\pm$ 0.01  &      0.49 $\pm$ 0.03  &    0.48 $\pm$ 0.03  &    0.45 $\pm$ 0.04  &    0.39 $\pm$ 0.05\\
   2000.  &    0.52 $\pm$ 0.01  &    0.40 $\pm$ 0.01    &    0.43 $\pm$ 0.01  &      0.53 $\pm$ 0.02  &    0.52 $\pm$ 0.02  &    0.47 $\pm$ 0.03  &    0.40 $\pm$ 0.03\\
   3000.  &    0.66 $\pm$ 0.03  &    0.54 $\pm$ 0.01    &    0.59 $\pm$ 0.02  &      0.67 $\pm$ 0.03  &    0.64 $\pm$ 0.04  &    0.58 $\pm$ 0.04  &    0.49 $\pm$ 0.04\\
   4000.  &    0.81 $\pm$ 0.04  &    0.69 $\pm$ 0.02    &    0.74 $\pm$ 0.03  &      0.80 $\pm$ 0.05  &    0.75 $\pm$ 0.05  &    0.68 $\pm$ 0.05  &    0.57 $\pm$ 0.05\\
   5000.  &    0.99 $\pm$ 0.05  &    0.84 $\pm$ 0.02    &    0.92 $\pm$ 0.03  &      0.97 $\pm$ 0.06  &    0.91 $\pm$ 0.06  &    0.82 $\pm$ 0.07  &    0.70 $\pm$ 0.07\\
   6000.  &    1.12 $\pm$ 0.03  &    0.97 $\pm$ 0.01    &    1.08 $\pm$ 0.02  &      1.09 $\pm$ 0.03  &    1.01 $\pm$ 0.03  &    0.89 $\pm$ 0.03  &    0.76 $\pm$ 0.03\\
   7000.  &    1.22 $\pm$ 0.04  &    1.09 $\pm$ 0.02    &    1.21 $\pm$ 0.03  &      1.17 $\pm$ 0.05  &    1.08 $\pm$ 0.05  &    0.94 $\pm$ 0.05  &    0.80 $\pm$ 0.05\\
   8000.  &    1.36 $\pm$ 0.04  &    1.22 $\pm$ 0.02    &    1.36 $\pm$ 0.03  &      1.30 $\pm$ 0.05  &    1.18 $\pm$ 0.05  &    1.03 $\pm$ 0.05  &    0.86 $\pm$ 0.05\\
   9000.  &    1.45 $\pm$ 0.03  &    1.33 $\pm$ 0.02    &    1.47 $\pm$ 0.02  &      1.38 $\pm$ 0.04  &    1.26 $\pm$ 0.04  &    1.09 $\pm$ 0.04  &    0.91 $\pm$ 0.04\\
  10000.  &    1.58 $\pm$ 0.05  &    1.47 $\pm$ 0.02    &    1.60 $\pm$ 0.03  &      1.49 $\pm$ 0.05  &    1.36 $\pm$ 0.06  &    1.17 $\pm$ 0.06  &    0.97 $\pm$ 0.06\\
  11000.  &    1.69 $\pm$ 0.06  &    1.58 $\pm$ 0.04    &    1.71 $\pm$ 0.05  &      1.61 $\pm$ 0.06  &    1.46 $\pm$ 0.06  &    1.25 $\pm$ 0.06  &    1.04 $\pm$ 0.05\\
  12000.  &    1.80 $\pm$ 0.07  &    1.69 $\pm$ 0.04    &    1.83 $\pm$ 0.06  &      1.70 $\pm$ 0.07  &    1.54 $\pm$ 0.07  &    1.31 $\pm$ 0.07  &    1.08 $\pm$ 0.07\\
  13000.  &    1.86 $\pm$ 0.09  &    1.78 $\pm$ 0.05    &    1.93 $\pm$ 0.07  &      1.74 $\pm$ 0.09  &    1.57 $\pm$ 0.09  &    1.33 $\pm$ 0.09  &    1.08 $\pm$ 0.08\\
  14000.  &    1.99 $\pm$ 0.07  &    1.92 $\pm$ 0.04    &    2.09 $\pm$ 0.06  &      1.86 $\pm$ 0.07  &    1.67 $\pm$ 0.07  &    1.41 $\pm$ 0.07  &    1.15 $\pm$ 0.07\\
  15000.  &    2.13 $\pm$ 0.07  &    2.04 $\pm$ 0.03    &    2.25 $\pm$ 0.06  &      1.99 $\pm$ 0.08  &    1.78 $\pm$ 0.09  &    1.51 $\pm$ 0.09  &    1.24 $\pm$ 0.09\\
 \hline \hline
 \end{tabular}
 \end{center}
 \end{table*}

%% file: 4197.bbl
\begin{thebibliography}{78}
\expandafter\ifx\csname
natexlab\endcsname\relax\def\natexlab#1{#1}\fi
\expandafter\ifx\csname url\endcsname\relax
  \def\url#1{{\tt #1}}\fi
\expandafter\ifx\csname
urlprefix\endcsname\relax\def\urlprefix{URL }\fi


\bibitem[{{Allard} et~al.(1997){Allard}, {Hauschildt}, {Alexander}, \&
  {Starrfield}}]{Allard+97}
{Allard} F., {Hauschildt} P.H., {Alexander} D.R., {Starrfield} S.,
1997, \araa,
  35, 137

\bibitem[{{Alves} et~al.(2000){Alves}, {Bond}, \& {Livio}}]{Alves+00}
{Alves} D.R., {Bond} H.E., {Livio} M., 2000, \aj, 120, 2044

\bibitem[{{Anders} \& {Fritze-v.Alvensleben} (2003)}]{Anders+03}
{Anders} P., {Fritze-v.~Alvensleben} U., 2003, \aap, 401, 1063

\bibitem[{{Angeletti} et~al.(1980){Angeletti}, {Dolcetta}, \&
  {Giannone}}]{Angeletti+80}
{Angeletti} L., {Dolcetta} R., {Giannone} P., 1980, \apss, 69, 45

\bibitem[{{Bailyn} et~al.(1989){Bailyn}, {Grindlay}, {Cohn}, {Lugger}, {Stetson},
 {Hesser}}]{Bailyn+89} {Bailyn}, C.D., {Grindlay}, J.E., {Cohn}, H., {Lugger}, P.M., {Stetson},
 P.B., {Hesser} J.E. 1989, \aj, 98, 882

\bibitem[{{Baraffe} et~al.(1997){Baraffe}, {Chabrier}, {Allard}, \&
  {Hauschildt}}]{Baraffe+97}
{Baraffe} I., {Chabrier} G., {Allard} F., {Hauschildt} P.H., 1997,
\aap, 327,
  1054

\bibitem[{{Barbaro} \& {Bertelli}(1977)}]{Barbaro&Bertelli77}
{Barbaro} C., {Bertelli} C., 1977, \aap, 54, 243

\bibitem[{{Barmina} et~al.(2002){Barmina}, {Girardi}, \& {Chiosi}}]{Barmina+02}
{Barmina} R., {Girardi} L., {Chiosi} C., 2002, \aap, 385, 847

\bibitem[{{Baud} \& {Habing}(1983)}]{Baud&Habing83}
{Baud} B., {Habing} H.J., 1983, \aap, 127, 73

\bibitem[{{Bergeron} et~al.(1995){Bergeron}, {Saumon}, \&
  {Wesemael}}]{Bergeron+95}
{Bergeron} P., {Saumon} D., {Wesemael} F., 1995, \apj, 443, 764

\bibitem[{{Bessell} et~al.(1998){Bessell}, {Castelli}, \& {Plez}}]{Bessel+98}
{Bessell} M.S., {Castelli} F., {Plez} B., 1998, \aap, 333, 231

\bibitem[{{Blocker} et~al.(1997){Blocker}, {Herwig}, {Driebe}, {Bramkamp}, \&
  {Schonberner}}]{Blocker&Schon97}
{Blocker} T., {Herwig} F., {Driebe} T., {Bramkamp} H.,
{Schonberner} D., 1997,
  In: IAU Symp. 180: Planetary Nebulae, 389

\bibitem[{{Boily} et~al.(2005){Boily}, {Lan{\c c}on}, {Deiters}, \&
  {Heggie}}]{Boily+05}
{Boily} C.M., {Lan{\c c}on} A., {Deiters} S., {Heggie} D.C., 2005,
\apjl, 620,
  L27

\bibitem[{{Bombaci} et~al.(2004){Bombaci}, {Parenti}, \& {Vida{\~
  n}a}}]{Bombaci+04}
{Bombaci} I., {Parenti} I., {Vida{\~ n}a} I., 2004, \apj, 614, 314

\bibitem[{{Bressan} et~al.(1994){Bressan}, {Chiosi}, \& {Fagotto}}]{Bressan+94}
{Bressan} A., {Chiosi} C., {Fagotto} F., 1994, \apjs, 94, 63

\bibitem[{{Brocato} et~al.(1990){Brocato}, {Caputo}, {di Giorgio},
  {Santolamazza}, \& {Richichi}}]{Brocato+90}
{Brocato} E., {Caputo} F., {di Giorgio} A.M., {Santolamazza} P.,
{Richichi} A., 1990, Memorie della Societa Astronomica Italiana,
61, 137

\bibitem[{{Brocato} et~al.(1990b){Brocato}, {Matteucci},
{Mazzitelli},\& {Tornamb\'e}}]{Brocato+90b} {Brocato} E.,
{Matteucci} F., {Mazzitelli} I., {Tornamb\'e} A., 1990b, \apj,
349, 458

\bibitem[{{Brocato} et~al.(1999){Brocato}, {Castellani}, {Raimondo}, \&
  {Romaniello}}]{Brocato+99}
{Brocato} E., {Castellani} V., {Raimondo} G., {Romaniello} M.,
1999, \aaps,
  136, 65

\bibitem[{{Brocato} et~al.(2000){Brocato}, {Castellani}, {Poli}, \&
  {Raimondo}}]{Brocato+00}
{Brocato} E., {Castellani} V., {Poli} F.M., {Raimondo} G., 2000,
\aaps, 146, 91

\bibitem[{{Brocato} et~al.(2003){Brocato}, {Castellani}, {Di Carlo},
  {Raimondo}, \& {Walker}}]{Brocato+03}
{Brocato} E., {Castellani} V., {Di Carlo} E., {Raimondo} G.,
{Walker} A.R.,
  2003, \aj, 125, 3111

\bibitem[{{Bruzual} \& {Charlot}(2003)}]{Bruzual&Charlot03}
{Bruzual} G., {Charlot} S., 2003, \mnras, 344, 1000

\bibitem[{{Burstein} et~al.(1984)}]{Burstein+84}
{Burstein} D., {Faber} S. M., {Gaskell} C. M., {Krumm} N., 1984,
\apj, 287,586

\bibitem[{{Buzzoni}(1993)}]{Buzzoni93}
{Buzzoni} A., 1993, \aap, 275, 433

\bibitem[{{Cardelli} et~al.(1989){Cardelli}, {Clayton}, \&
  {Mathis}}]{Cardelli+89}
{Cardelli} J.A., {Clayton} G.C., {Mathis} J.S., 1989, \apj, 345,
245

\bibitem[{{Cariulo} et~al.(2004){Cariulo}, {Degl'Innocenti}, \&
 {Castellani}}]{Cariulo+04}
{Cariulo} P., {Degl'Innocenti} S., {Castellani} V., 2004, \aap,
421, 1121

\bibitem[{{Carretta} et~al.(2000){Carretta}, {Gratton}, {Clementini}, \& {Fusi
  Pecci}}]{Carretta+00}
{Carretta} E., {Gratton} R.G., {Clementini} G., {Fusi Pecci} F.,
2000, \apj,  533, 215

\bibitem[{{Castellani} et~al.(1985){Castellani}, {Chieffi}, {Tornambe}, \&
  {Pulone}}]{Castellani+85}
{Castellani} V., {Chieffi} A., {Tornambe} A., {Pulone} L., 1985,
\apj, 296, 204

\bibitem[{{Castellani} et~al.(2003){Castellani}, {Degl'Innocenti}, {Marconi}, \&
  {Prada Moroni}}]{Castellani+03} {Castellani} V., {Degl'Innocenti} S., {Marconi} M.,
{Prada Moroni} P.G., {Sestito} P., 2003, \aap 404, 645

\bibitem[{{Castelli}(1999)}]{Castelli99}
{Castelli} F., 1999, \aap, 346, 564

\bibitem[{{Castelli} et~al.(1997){Castelli}, {Gratton}, \&
  {Kurucz}}]{Castelli+97}
{Castelli} F., {Gratton} R.G., {Kurucz} R.L., 1997, \aap, 318, 841

\bibitem[{{Cervi{\~n}o} \& {Valls-Gabaud}(2003)}]{Cervino+03}
{Cervi{\~n}o}, M. and {Valls-Gabaud}, D., 2003, \mnras, 338, 481

\bibitem[{{Chabrier} \& {Baraffe}(1997)}]{Chabrier&Baraffe97}
{Chabrier} G., {Baraffe} I., 1997, \aap, 327, 1039

\bibitem[{{Chabrier} \& {Mera}(1997)}]{Chabrier&Mera97}
{Chabrier} G., {Mera} D., 1997, \aap, 328, 83

\bibitem[{{Charlot} \& {Bruzual}(1991)}]{Charlot&Bruzual91}
{Charlot} S., {Bruzual} G., 1991, \apj, 367, 126

\bibitem[{{Charlot} et~al.(1996){Charlot}, {Worthey}, \&
  {Bressan}}]{Charlot+96} {Charlot} S., {Worthey} G., {Bressan} A., 1996, \apj, 457, 625

\bibitem[{{Chiosi} et~al.(1988){Chiosi}, {Bertelli}, \& {Bressan}}]{Chiosi+88}
{Chiosi} C., {Bertelli} G., {Bressan} A., 1988, \aap, 196, 84

\bibitem[{{Ciacio} et~al.(1997){Ciacio}, {Degl'Innocenti}, \& {Ricci}}]{Ciacio+97}
{Ciacio} F., {Degl'Innocenti} S., {Ricci} B., 1997, A\&AS 123, 449

\bibitem[{{Cote} et~al.(1998){Cote}, {Marzke}, \& {West}}]{Cote+98}
{Cote} P., {Marzke} R.O., {West} M.J., 1998, \apj, 501, 554

\bibitem[{{De Marchi} \& {Paresce} (1994){De Marchi}, \& {Paresce}}]{DeMarchi+94}
{De Marchi} G., {Paresce} F. 1994, \apj, 422, 597

\bibitem[{{Di Criscienzo} et~al.(2004){Di Criscienzo}, {Marconi}, \&
  {Caputo}}]{Dicriscienzo+04}
{Di Criscienzo} M., {Marconi} M., {Caputo} F., 2004, \apj, 612,
1092

\bibitem[{{Dominguez} et~al.(1999){Dominguez}, {Chieffi}, {Limongi}, \&
  {Straniero}}]{Dominguez+99}
{Dominguez} I., {Chieffi} A., {Limongi} M., {Straniero} O., 1999,
\apj, 524,
  226

\bibitem[{{Fagiolini}(2004)}]{Fagiolini04}
{Fagiolini} M., 2004, Diploma Thesis, University of Pisa, Italy

\bibitem[{{Fellhauer} et~al.(2003){Fellhauer}, {Lin}, {Bolte}, {Aarseth}, \&
  {Williams}}]{Fellhauer+03}
{Fellhauer} M., {Lin} D.N.C., {Bolte} M., {Aarseth} S.J.,
{Williams} K.A.,
  2003, \apjl, 595, L53

\bibitem[{{Ferraro} et~al.(1999){Ferraro}, {Messineo}, {Fusi Pecci}
  et~al.}]{Ferraro+99}
{Ferraro} F.R., {Messineo} M., {Fusi Pecci} F., et~al., 1999, \aj,
118, 1738

\bibitem[{{Girardi} \& {Bertelli}(1998)}]{Gir&Ber98}
{Girardi} L., {Bertelli} L., 1998, \mnras,300, 533

\bibitem[{{Girardi} et~al.(2000){Girardi}, {Bressan}, {Bertelli}, \&
  {Chiosi}}]{Girardi+00}
{Girardi} L., {Bressan} A., {Bertelli} G., {Chiosi} C., 2000,
\aaps, 141, 371

\bibitem[{{Harris}(1996)}]{Harris96}
{Harris} W.E., 1996, VizieR Online Data Catalog, 7195, 0

\bibitem[{{Harris}(2003)}]{Harris03}
{Harris} W.E., 2003, In: A Decade of Hubble Space Telescope
Science, 78--100

\bibitem[{{Hurley} \& {Shara}(2003)}]{Hurley&Shara03}
{Hurley} J.R., {Shara} M.M., 2003, \apj, 589, 179

\bibitem[{{Kron} \& {Mayall}(1960)}]{Kron+60}
{Kron} G.E., {Mayall} N.U., 1960, \aj, 65, 581

\bibitem[{{Kroupa}(2002)}]{Kroupa02}
{Kroupa} P., 2002, Science, 295, 82

\bibitem[{{Kurth} et~al.(1999){Kurth}, {Fritze-v.~Alvensleben}, \&
  {Fricke}}]{Kurth+99}
{Kurth} O.M., {Fritze-v.~Alvensleben} U., {Fricke} K.J., 1999,
\aaps, 138, 19

\bibitem[{{Lattanzio} \& Wood (2003){Lattanzio}, \& {Wood}}]{Lattanzio&Wood03}
{Lattanzio} J. C., {Wood} P. R., 2003, in Asymptotic giant branch
stars, by Harm J. Habing and Hans Olofsson. Astronomy and
astrophysics library, New York, Berlin: Springer, 2003, p. 23
(2003)

\bibitem[{{Larsen}(2000)}]{Larsen00}
{Larsen} S.S., 2000, \mnras, 319, 893

\bibitem[{{Lee} et~al.(1994){Lee}, {Demarque}, \& {Zinn}}]{Lee+94}
{Lee} Y., {Demarque} P., {Zinn} R., 1994, \apj, 423, 248

\bibitem[{{Maeder} \& {Zahn}(1998)}]{Maeder&Zahn98}
{Maeder} A., {Zahn} J., 1998, \aap, 334, 1000

\bibitem[{{Maraston}(1998)}]{Maraston98}
{Maraston} C., 1998, \mnras, 300, 872

\bibitem[{{Maraston}(2005)}]{Maraston05}
{Maraston} C., 2005, \mnras, 362, 799

\bibitem[{{Marigo} et~al.(1999)}]{Marigo+99} {Marigo}, P., {Girardi}, L., \& {Bressan}, A. 1999, \aap, 344, 123

\bibitem[{{Matteucci} et~al.(2002){Matteucci}, {Ripepi}, {Brocato}, \&
  {Castellani}}]{Matteucci+02}
{Matteucci} A., {Ripepi} V., {Brocato} E., {Castellani} V., 2002,
\aap, 387,
  861

\bibitem[{{McNamara} et~al.(2004){McNamara}, {Harrison}, \&
  {Baumgardt}}]{McNamara+04}
{McNamara} B.J., {Harrison} T.E., {Baumgardt} H., 2004, \apj, 602,
264

\bibitem[{{Origlia} \& {Leitherer}(2000)}]{Origlia&Leitherer00}
{Origlia} L., {Leitherer} C., 2000, \aj, 119, 2018

\bibitem[{{Palacios} et~al.(2003){Palacios}, {Talon}, {Charbonnel}, \&
  {Forestini}}]{Palacios+03}
{Palacios} A., {Talon} S., {Charbonnel} C., {Forestini} M., 2003,
\aap, 399,  603


\bibitem[{{Pasquali} et~al.(2004){Pasquali}, {De Marchi}, {Pulone}, \&
  {Brigas}}]{Pasquali+04}
{Pasquali} A., {De Marchi} G., {Pulone} L., {Brigas} M. S., 2004,
\aap, 428,  469

\bibitem[{{Pease}(1928)}]{Pease28}
{Pease} F.G., 1928, \pasp, 40, 342

\bibitem[{{Piotto} et~al.(2002){Piotto}, {King}, {Djorgovski}
  et~al.}]{Piotto+02}
{Piotto} G., {King} I.R., {Djorgovski} S.G., et~al., 2002, \aap,
391, 945

\bibitem[{{Pols} et~al.(1998){Pols}, {Schroder}, {Hurley}, {Tout}, \&
  {Eggleton}}]{Pols+98}
{Pols} O.R., {Schroder} K., {Hurley} J.R., {Tout} C.A., {Eggleton}
P.P., 1998,
  \mnras, 298, 525

\bibitem[{{Prada Moroni} \& {Straniero}(2005)}]{PradaMoroni05}
{Prada Moroni} P.G., {Straniero} O., 2005, \apj, in publication

\bibitem[{{Pryor} \& {Meylan}(1993)}]{Pryor&Meylan93}
{Pryor} C., {Meylan} G., 1993, In: ASP Conf. Ser. 50: Structure
and Dynamics of
  Globular Clusters, 357

\bibitem[{{Raimondo} et~al.(2005){Raimondo}, {Brocato}, {Cantiello}, \&
  {Capaccioli}}]{Raimondo+05}
{Raimondo} G., {Brocato} E., {Cantiello} M., {Capaccioli} M.,
2005, \aj, 130, 2625

\bibitem[{{Reed}(1985)}]{Reed85}
{Reed} B.C., 1985, \pasp, 97, 120

\bibitem[{{Reed} et~al.(1988){Reed}, {Hesser}, \& {Shawl}}]{Reed+88}
{Reed} B.C., {Hesser} J.E., {Shawl} S.J., 1988, \pasp, 100, 545

\bibitem[{{Reimers}(1975)}]{Reimers75}
{Reimers} D., 1975, Memoires of the Societe Royale des Sciences de
Liege, 8, 369

\bibitem[{{Renzini} \& {Buzzoni}(1986)}]{Renzini&Buzzoni86}
{Renzini}, A., \& {Buzzoni}, A. 1986,
    in Proceedings of the Fourth Workshop, Erice,
    Italy, March 12-22, 1985, Spectral evolution of galaxies
    (Dordrecht, D. Reidel Publishing Co.), 195

\bibitem[{{Renzini} \& {Voli}(1981)}]{Renzini&Voli81}
{Renzini} A., {Voli} M., 1981, \aap, 94, 175

\bibitem[{{Rey} et~al.(2001){Rey}, {Yoon}, {Lee}, {Chaboyer}, \&
  {Sarajedini}}]{Rey+01}
{Rey} S., {Yoon} S., {Lee} Y., {Chaboyer} B., {Sarajedini} A.,
2001, \aj, 122,
  3219

\bibitem[{{Richard} et~al.(2002){Richard}, {Michaud}, \& {Richer}}]{Richard+02}
{Richard} O., {Michaud} G., {Richer} J., 2002, \apj, 580, 1100

\bibitem[{{Richer} et~al.(1998){Richer}, {Michaud}, {Rogers}
  et~al.}]{Richer+98}
{Richer} J., {Michaud} G., {Rogers} F., et~al., 1998, \apj, 492,
833

\bibitem[{{Rosenberg} et~al.(2000){Rosenberg}, {Aparicio}, {Saviane}
  et~al.}]{Rosenberg+00}
{Rosenberg} A., {Aparicio} A., {Saviane} I., et~al., 2000, \aaps,
145, 451

\bibitem[{{Salaris} et~al.(2000){Salaris}, {Garc{\'{\i}}a-Berro}, {Hernanz},
  {Isern}, \& {Saumon}}]{Salaris+00}
{Salaris} M., {Garc{\'{\i}}a-Berro} E., {Hernanz} M., {Isern} J.,
{Saumon} D.,
  2000, \apj, 544, 1036

\bibitem[{{Sandquist} et~al.(1999){Sandquist}, {Bolte}, {Langer}, {Hesser}, \&
  {Mendes de Oliveira}}]{Sandquist+99}
{Sandquist} E.L., {Bolte} M., {Langer} G.E., {Hesser} J.E.,
{Mendes de
  Oliveira} C., 1999, \apj, 518, 262

\bibitem[{{Santos} \& {Frogel}(1997)}]{Santos&Frogel97}
{Santos} J.F.C., {Frogel} J.A., 1997, \apj, 479, 764

\bibitem[{{Saumon} \& {Jacobson}(1999)}]{Saumon&Jacobson99}
{Saumon} D., {Jacobson} S.B., 1999, \apjl, 511, L107

\bibitem[{{Schulz} et~al.(2002){Schulz}, {Fritze-v.Alvensleben}, {M$\ddot{\mathrm{o}}$ller}, \&
  {Fricke}}]{Schulz+02}
{Schulz} J., {Fritze-v.~Alvensleben} U., {M$\ddot{\mathrm{o}}$ller} C.S., {Fricke} K.J.,
2002, \aap, 392, 1

\bibitem[{{Spitzer}(1987)}]{Spitzer87}
{Spitzer} L., 1987, {Dynamical evolution of globular clusters},
Princeton, NJ,  Princeton University Press, 1987, 191 p.

\bibitem[{{Stetson}(1994)}]{Stetson94}
{Stetson} P.B., 1994, PASP, 106, 250

\bibitem[{{Straniero} et~al.(2003)}]{Straniero+03}
{Straniero} O., {Domínguez}, I., {Cristallo}, R., {Gallino}, R.
2003, PASA, 20, 389

\bibitem[{{Thoul} et~al.(1994){Thoul}, {Bahcall}, \& {Loeb}}]{Thoul+94}
{Thoul} A.A., {Bahcall} J.N., {Loeb} A., 1994, \apj, 421, 828

\bibitem[{{Tinsley}(1972)}]{Tinsley72}
{Tinsley} B.M., 1972, \aap, 20, 383

\bibitem[{{Tornambe} \& {Chieffi}(1986)}]{Tornambe&Chieffi86}
{Tornambe} A., {Chieffi} A., 1986, \mnras, 220, 529

\bibitem[{{Vazdekis}(1999)}]{Vazdekis99}
{Vazdekis} A., 1999, \apj, 513, 224

\bibitem[{{Vesperini} \& {Heggie}(1997)}]{Vesperini&Heggie97}
{Vesperini} E., {Heggie} D.C., 1997, \mnras, 289, 898

\bibitem[{{Wagenhuber} \& {Groenewegen}(1998)}]{Wagenhuber&Groenewegen98}
{Wagenhuber} J., {Groenewegen} M.A.T., 1998, \aap, 340, 183

\bibitem[{{Walker}(1994)}]{Walker94}
{Walker} A.R., 1994, \aj, 108, 555

\bibitem[{{West} et~al.(2004){West}, {C{\^ o}t{\' e}}, {Marzke}, \& {Jord{\'
  a}n}}]{West+04}
{West} M.J., {C{\^ o}t{\' e}} P., {Marzke} R.O., {Jord{\' a}n} A.,
2004, \nat,
  427, 31

\bibitem[{{Worthey}(1994)}]{Worthey94}
{Worthey} G., 1994, \apjs, 95, 106

\bibitem[{{Yi}(2003)}]{Yi03}
{Yi} S.K., 2003, \apj, 582, 202

\bibitem[{{Zhang} et~al.(2002){Zhang}, {Han}, {Li}, \& {Hurley}}]{Zhang+02}
{Zhang} F., {Han} Z., {Li} L., {Hurley} J.R., 2002, \mnras, 334,
883

\bibitem[{{Zhang} et~al.(2005){Zhang}, {Li}, \& {Han}}]{Zhang+05}
{Zhang} F., {Li} L., {Han} Z. 2005, \mnras, 364, 503


 \end{thebibliography}
